\documentclass[traditabstract]{aa}

\usepackage{graphicx}
\usepackage{natbib}
\usepackage{amssymb}
\usepackage{amsmath}
\usepackage{rotating}
\usepackage{supertabular}
\usepackage[bookmarks=true, bookmarksnumbered=true, breaklinks=true, colorlinks=true, linkcolor=BrickRed, citecolor=Blue, urlcolor=Blue, filecolor=Blue]{hyperref}
\usepackage[all]{hypcap}
\usepackage{txfonts}
\usepackage{fixltx2e}
\usepackage{multirow}
\usepackage{pifont}
\usepackage[usenames,dvipsnames]{color}
\usepackage[all]{hypcap}
\def\apjl{ApJ}%
\def\apjs{ApJS}%
\def\ao{Appl.\ Opt.}%
\def\aap{A\&A}%

\newcommand{\ph}{\phantom{$-$}}
\newcommand{\pst}{\phantom{*}}
\newcommand{\phz}{\phantom{0}}

\newcommand{\oneDAV}{$\langle\mathrm{3D}\rangle$}

\newcommand{\marcs}{\textsc{marcs}}
\def\urltilda{\kern -.15em\lower .7ex\hbox{\~{}}\kern .04em}

\defcitealias{AG89}{AG89}
\defcitealias{GS98}{GS98}
\defcitealias{AGS05}{AGS05}
\defcitealias{AGSS}{AGSS09}
\defcitealias{AGSS_NaCa}{Paper I}
\defcitealias{AGSS_heavy}{Paper III}
\defcitealias{HM}{HM}
\defcitealias{CB62}{CB}
\defcitealias{Greenlee79}{GW}

\newcommand{\altsci}{Sc\,\textsc{i}}
\newcommand{\altscii}{Sc\,\textsc{ii}}
\newcommand{\alttii}{Ti\,\textsc{i}}
\newcommand{\alttiii}{Ti\,\textsc{ii}}
\newcommand{\altvi}{V\,\textsc{i}}
\newcommand{\altvii}{V\,\textsc{ii}}
\newcommand{\altcri}{Cr\,\textsc{i}}
\newcommand{\altcrii}{Cr\,\textsc{ii}}
\newcommand{\altmni}{Mn\,\textsc{i}}
\newcommand{\altmnii}{Mn\,\textsc{ii}}
\newcommand{\altfei}{Fe\,\textsc{i}}
\newcommand{\altfeii}{Fe\,\textsc{ii}}
\newcommand{\altcoi}{Co\,\textsc{i}}
\newcommand{\altcoii}{Co\,\textsc{ii}}
\newcommand{\altnii}{Ni\,\textsc{i}}
\newcommand{\altniii}{Ni\,\textsc{ii}}
\newcommand{\scani}{\altsci\ }
\newcommand{\scanii}{\altscii\ }
\newcommand{\tii}{\alttii\ }
\newcommand{\tiii}{\alttiii\ }
\newcommand{\vi}{\altvi\ }
\newcommand{\vii}{\altvii\ }
\newcommand{\cri}{\altcri\ }
\newcommand{\crii}{\altcrii\ }
\newcommand{\mni}{\altmni\ }
\newcommand{\mnii}{\altmnii\ }
\newcommand{\fei}{\altfei\ }
\newcommand{\feii}{\altfeii\ }
\newcommand{\coi}{\altcoi\ }

\newcommand{\nii}{\altnii\ }
\newcommand{\niii}{\altniii\ }

\begin{document}

\title{The elemental composition of the Sun}
\subtitle{II. The iron group elements Sc to Ni}

\titlerunning{Solar abundances II. The iron group (Sc to Ni)}
\authorrunning{Scott et al.}

\author{Pat Scott\inst{1}
\and
Martin Asplund\inst{2}
\and
Nicolas Grevesse\inst{3,4}
\and
Maria Bergemann\inst{5}
\and
A.~Jacques Sauval\inst{6}}

\institute{Department of Physics, Imperial College London, Blackett Laboratory, Prince Consort Road, London SW7 2AZ, UK\\
\email{p.scott@imperial.ac.uk}
\and
Research School of Astronomy and Astrophysics, Australian National University, Cotter Rd., Weston Creek, ACT 2611, Australia\\
\email{martin.asplund@anu.edu.au}
\and
Centre Spatial de Li{\`e}ge, Universit{\'e} de Li{\`e}ge, avenue Pr{\'e} Aily, B-4031 Angleur-Li{\`e}ge, Belgium
\and
Institut d'Astrophysique et de G{\'e}ophysique, Universit{\'e} de Li{\`e}ge, All{\'e}e du 6 ao{\^u}t, 17, B5C, B-4000 Li{\`e}ge, Belgium\\
\email{nicolas.grevesse@ulg.ac.be}
\and
Institute of Astronomy, University of Cambridge, Madingley Road, CB3 0HA, Cambridge, UK\\
\email{mbergema@ast.cam.ac.uk}
\and
Observatoire Royal de Belgique, avenue circulaire, 3, B-1180 Bruxelles, Belgium\\
\email{jacques.sauval@oma.be}
}

\date{Received 1 May 2014 / Accepted 1 Sep 2014}

\abstract{We redetermine the abundances of all iron group nuclei in the Sun, based on neutral and singly-ionised lines of Sc, Ti, V, Mn, Fe, Co and Ni in the solar spectrum.  We employ a realistic 3D hydrodynamic model solar atmosphere, corrections for departures from local thermodynamic equilibrium (NLTE), stringent line selection procedures and high quality observational data.  We have scoured the literature for the best quality oscillator strengths, hyperfine constants and isotopic separations available for our chosen lines.  We find $\log \epsilon_\mathrm{Sc}=3.16\pm0.04$, $\log \epsilon_\mathrm{Ti}=4.93\pm0.04$, $\log \epsilon_\mathrm{V}=3.89\pm0.08$, $\log \epsilon_\mathrm{Cr}=5.62\pm0.04$, $\log \epsilon_\mathrm{Mn}=5.42\pm0.04$, $\log \epsilon_\mathrm{Fe}=7.47\pm0.04$, $\log \epsilon_\mathrm{Co}=4.93\pm0.05$ and $\log \epsilon_\mathrm{Ni}=6.20\pm0.04$. Our uncertainties factor in both statistical and systematic errors (the latter estimated for possible errors in the model atmospheres and NLTE line formation). The new abundances are generally in good agreement with the CI meteoritic abundances but with some notable exceptions.  This analysis constitutes both a full exposition and a slight update of the preliminary results we presented in \citeauthor*{AGSS} (\citeyear{AGSS}), including full line lists and details of all input data we employed.}

\maketitle
%
%________________________________________________________________

\section{Introduction}
\label{intro}
Cosmic abundances of the transition metals Sc -- Ni ($21\le Z\le 28$) tend to form a `peak' around iron.
This behaviour approximately tracks the variation in average binding energy per nucleon with $Z$, and reflects the
predominantly common origin of iron peak nuclei in core-collapse and thermonuclear supernovae \citep[e.g.][]{Pagel97}.  Variations of abundances within the group
provide information on nuclear physics and the physical environments in which the elements were processed.  To compare such analyses with theories
of stellar structure and evolution, galactic chemical evolution, supernova nucleosynthesis and the formation history
of the solar system, accurate solar abundances of the iron group elements are required.  In this paper, we present a
reanalysis of the solar composition of the iron peak elements Sc, Ti, V, Cr, Mn, Fe, Co and Ni, using a realistic 3D hydrodynamic
solar model atmosphere.

This paper is part of a series detailing, and updating, the chemical composition of the Sun presented in \citet[][hereafter \citetalias{AGSS}]{AGSS}.  This paper covers the iron group nuclei Sc -- Ni.   \citet[][hereafter \citetalias{AGSS_NaCa}]{AGSS_NaCa} deals with the intermediate-mass elements Na -- Ca, whereas \citet[hereafter \citetalias{AGSS_heavy}]{AGSS_heavy} is devoted to the heavy elements Cu -- Th. Later studies will describe the analysis of the light elements C, N and O, as well as summarise and compare the solar photospheric abundances with the meteoritic evidence, indications from helioseismology and the solar neighbourhood. In an earlier series of papers (\citealt{AspI}; \citealt{AspII}; \citealt{AspIII}; \citealt{2001ApJ...556L..63A}, \citealt{2002ApJ...573L.137A}, \citealt{AspIV}; \citealt{AspV}; \citealt{AspVI}; \citealt{ScottVII}; \citealt{AGS05}, hereafter \citetalias{AGS05}; \citealt{2008A&A...490..817M}, \citealt{Scott09Ni}), we examined the abundances of all elements up to Ca, as well as Fe and Ni, using a predecessor of the current 3D solar model atmosphere.  The only 3D solar analyses of any iron group elements to date have been of nickel \citep{Scott09Ni} and iron itself (\citealt{1994A&A...291..635A, AspI, AspII, Caffau11}; also the 1D calculations of \citealt{MB_fe} based on an averaged 3D model).

In Sect.~\ref{previouswork} we summarise the current state of knowledge about the solar abundances of the iron peak elements.  We describe the observational data we employ in Sect.~\ref{observations}, then give brief recapitulations of our solar model atmosphere, line synthesis code (Sect.~\ref{model}) and abundance calculations (Sect.~\ref{calculations}).  In Sect.~\ref{atomicdata} we justify our selection of atomic data, spectral lines and non-LTE (NLTE) corrections.  Our results are presented in Sect.~\ref{results}, discussed in Sect.~\ref{discussion}, compared to previous compilations in Sect.~\ref{compilations} and summarised in Sect.~\ref{conclusions}.

\section{Previous solar analyses of the iron group}
\label{previouswork}

\textit{Scandium:} Earlier reference compilations of the solar composition (\citealt{GS98}, hereafter \citetalias{GS98}; \citetalias{AGS05}) 
included the scandium abundance $\log \epsilon_\mathrm{Sc}=3.05\pm0.08$ from \citet{Youssef89},
derived using spectrum synthesis of Sc\,\textsc{ii} lines and the \citeauthor{HM} (\citeyear{HM}: hereafter \citetalias{HM}) model atmosphere.
A more recent study, adopted as the standard in \citet{Grevesse07}, was that of \citet{Neuforge93}, who used
the \citetalias{HM} model with both Sc\,\textsc{i} and Sc\,\textsc{ii} lines.  Despite using accurate oscillator
strengths (the same values as we use in this paper in fact), abundance scatter was high
($\log \epsilon_\mathrm{Sc}=3.14\pm0.12$ for Sc\,\textsc{i}, $3.20\pm0.07$ for Sc\,\textsc{ii}).
The dominant ionisation stage of scandium is Sc\,\textsc{ii}, so one expects Sc\,\textsc{ii} lines to return
the most reliable abundances.  However, the Sc\,\textsc{ii} results could not be reconciled with the meteoritic
value, nor could close agreement between ionisation stages be claimed.  \citet{Zhang08} performed
a detailed analysis of NLTE effects on Sc\,\textsc{i} and Sc\,\textsc{ii} lines in the Sun, finding
large NLTE corrections to abundances from Sc\,\textsc{i} lines.  This finally reconciled abundances
from the neutral and once-ionised species: with the MAFAGS-ODF 1D model atmosphere (based on opacity
distribution functions: ODF; \citealt{Fuhrmann97}), \citeauthor{Zhang08} found $\log \epsilon_\mathrm{Sc}=3.07$--$3.13$, depending upon
the oscillator strengths adopted.

\textit{Titanium:} \citet{Blackwell87} presented a thorough study of the solar abundance of Ti using a large number of
Ti\,\textsc{i} lines with both the \citetalias{HM} and \textsc{marcs} \citep{MARCS75} solar photospheric models (see Sect.~\ref{model}).
These results were corrected by \citet{Grevesse89} for a systematic shift in Ti\,\textsc{i} oscillator strengths
of $0.056$\,dex (see Sect.~\ref{Tigfs}), resulting in $\log \epsilon_\mathrm{Ti}=4.99\pm0.04$ with the \citetalias{HM}
 model.  With the \textsc{marcs} model, the result would have been $0.10$\,dex smaller.  Using Ti\,\textsc{ii}
 (the dominant species) instead, along with the \citetalias{HM} model, \citet{Bizzarri93} found
 $\log \epsilon_\mathrm{Ti}=5.04\pm0.04$, in very good agreement with the Ti\,\textsc{i} result.  The first NLTE analysis of the solar Ti abundance was performed by \citet{Bergemann11}, who found strong NLTE effects in \tii line formation, and 
a severe dependence upon the adopted solar model atmosphere and rates of inelastic collisions with H\,\textsc{i} and $e^{-}$.  From \tiii lines, \citet{Bergemann11} found a mean abundance of $\log \epsilon_\mathrm{Ti}=4.93$--$4.98$, depending mostly on the adopted oscillator strengths.  The most recent results are from \citet{Lawler13} and \citet{Wood13}, who found $\log \epsilon_\mathrm{Ti}=4.97\pm0.04$ and $\log \epsilon_\mathrm{Ti}=4.98\pm0.03$ from \tii and \tiii lines respectively, using spectrum synthesis with the \citetalias{HM} model and new laboratory oscillator strengths.

\textit{Vanadium:} The most recent derivations of the solar abundance of vanadium are due to \citet[][ with
V\,\textsc{i} and the \citetalias{HM} model: $\log \epsilon_\mathrm{V}=3.99\pm0.01$]{Whaling85} and \citet[][ with
V\,\textsc{i}, V\,\textsc{ii} and the \citetalias{HM} model: $\log
\epsilon_\mathrm{V}=4.00\pm0.02$]{Biemont89}.  The latter is the
previously-adopted reference abundance (\citetalias{GS98}; \citetalias{AGS05};
\citealt{Grevesse07}), giving more weight to the V\,\textsc{i} data,
which is derived from a much larger number of lines than the
V\,\textsc{ii} result.  The error estimate given is probably
unrealistically low however, as we describe in Sect.~\ref{Vgfs}.  Both these studies assumed that V lines form in LTE.
During the refereeing phase of our paper, we became aware of a recent determination of the solar V abundance using newly determined experimental transition probabilities for V\,\textsc{ii} \citep{Wood14V2}. Using spectrum synthesis with the HM model for a set of 15 often heavily blended V\,\textsc{ii} lines, they estimated $\log \epsilon_\mathrm{V}=3.95\pm0.01$ ($\sigma=0.05$\,dex). 

\textit{Chromium:} Recent compilations (\citetalias{AGS05}; \citealt{Grevesse07})
recommended a Cr abundance of $\log
\epsilon_\mathrm{Cr}=5.64\pm0.10$, derived from two papers.  Using
various $gf$-values available at the time, \citet{Biemont78c} found
$\log \epsilon_\mathrm{Cr}=5.67\pm0.03$ with Cr\,\textsc{i} lines
and the \citetalias{HM} model, and $\log
\epsilon_\mathrm{Cr}=5.64\pm0.03$ using the VAL \citep{VAL76} model.
\citet{Blackwell87}, using the accurate $gf$-values measured at
Oxford and different solar spectra, found $\log
\epsilon_\mathrm{Cr}=5.68\pm0.06$ with the \citetalias{HM} model. As
for Ti\,\textsc{i}, with the \textsc{marcs} model this
would have been $0.10$\,dex smaller.  \citet{Sobeck07}
measured new $gf$-values for Cr\,\textsc{i} lines (see
Sect.~\ref{Crgfs}) and used them to revise the solar abundance assuming LTE. When
two highly discrepant outlying lines are removed,
the results are $\log \epsilon_\mathrm{Cr}=5.64\pm0.05$ with the
\citetalias{HM} model and $\log
\epsilon_\mathrm{Cr}=5.53\pm0.05$ with \textsc{marcs}.
\citeauthor{Sobeck07} also used a small number of Cr\,\textsc{ii}
lines with $gf$-values from \citet{Nilsson06}. These lines lead to
higher abundances and much larger dispersions: $\log
\epsilon_\mathrm{Cr}=5.77\pm0.13$ (\citetalias{HM}) and $\log
\epsilon_\mathrm{Cr}=5.67\pm0.13$ (\textsc{marcs}).  \citet{Bergemann10} investigated
NLTE effects in solar Cr line formation for the first time, finding corrections of order $+$0.05--0.10\,dex
to abundances from \cri with the 1D MAFAGS-ODF model.  Using \crii lines and 
$gf$-values from \citet{Nilsson06}, \citeauthor{Bergemann10} confirmed the high abundance 
and large scatter seen by \citet{Sobeck07}.  With the complete exclusion of inelastic collisions
with hydrogen, chosen so as to satisfy Cr ionisation balance for the Sun and a number of late-type stars, they also found 
a high abundance from \altcri: $\log \epsilon_\mathrm{Cr}=5.74\pm0.05$.

\textit{Manganese:} Previous reference solar manganese abundances (\citetalias{GS98}; \citetalias{AGS05}; \citealt{Grevesse07})
came from \citet{Booth84b}, who found $\log \epsilon_\mathrm{Mn}=5.39\pm0.03$ using the \citetalias{HM} model
and Mn\,\textsc{i} lines.  The derived abundance is almost 3$\sigma$ below the meteoritic value \citep{Lodders09}, quite a striking
discrepancy when one considers that agreement between photospheric and meteoritic values is typically quite good
 (cf.~\citealt{AG89}, hereafter \citetalias{AG89}; \citetalias{AGSS}; \citealt{Lodders09}).  The errors on the photospheric value have probably been
underestimated however, as revealed by a detailed investigation of Mn\,\textsc{i} oscillator strengths and line
selection (Sect.~\ref{Mngfs}).  \citet{Bergemann07}  made a detailed NLTE analysis of a large number
of Mn\,\textsc{i} lines in the solar flux spectrum, showing that NLTE abundance corrections are of order $+0.08$\,dex
for solar lines.  Their analysis with the MAFAGS-ODF model produced an abundance of
$\log \epsilon_\mathrm{Mn}=5.36\pm0.10$.  This work was subsequently revised with improved oscillator strengths by
\citet{BW07}, giving $\log \epsilon_\mathrm{Mn}=5.37\pm0.05$ with the same model.  Using the \citetalias{HM} model,
the result was $\log \epsilon_\mathrm{Mn}=5.46\pm0.08$, in reasonable agreement with the meteoritic value but exhibiting an
uncomfortably high scatter.

\textit{Iron:} \citet{GS99} and \citet{AspII} summarised the long and well-known debate as to whether the solar abundance of Fe is equal to or higher than seen in meteorites.  Discrepant results in older studies of the solar Fe abundance using 1D solar models appeared to be due to differences in the adopted $gf$-values, equivalent widths, microturbulent velocities and collisional damping parameters, as well as differences in computer codes. \citet{GS99} succeeded in reconciling LTE abundances from \fei and \feii lines by modifying the temperature structure of the \citetalias{HM} model, so as to remove the observed trend with excitation potential in abundances from \fei lines. 

The first pioneering work aimed at determining the solar Fe abundance using a 3D solar model that we are aware of was by \citet{1994A&A...291..635A}, who used two different 3D models (with what nowadays is obviously very modest numerical resolution and simplified radiative transfer). Perhaps not surprisingly, their derived Fe abundance showed a large difference between \fei ($\log \epsilon_\mathrm{Fe} \approx 7.0$) and \feii ($\log \epsilon_\mathrm{Fe} \approx 7.6$) lines when using either equivalent widths or line depths.
\citet{AspII} analysed \fei and \feii lines with a more realistic 3D model, albeit still in LTE. They found abundances from weak \fei lines to be
independent of the excitation energy, and in very good agreement with
both \feii results and the meteoritic abundance: $\log
\epsilon_\mathrm{Fe}=7.45\pm0.05$. Both \citet{GS99} and \citet{AspII} found that abundances derived from \feii lines, using $gf$-values available at the time, showed a very large scatter, 0.10\,dex.  \citet{Caffau11} analysed a set of \feii lines using a 3D solar model computed with the CO$^5$BOLD code \citet{COBOLD} and improved $gf$-values from
\cite{Melendez09}, finding $\log \epsilon_\mathrm{Fe}=7.52\pm0.06$.

\citet{2011A&A...528A..87M} and \citet{MB_fe} carried out NLTE calculations of Fe line formation, using the most up-to-date theoretical and experimental atomic data to construct their model atoms. \citet{2011A&A...528A..87M}, using the MAFAGS-OS\footnote{MAFAGS-OS models are successors to MAFAGS-ODF models by \citet{Fuhrmann97}, relying on opacity sampling instead of ODFs.} models, obtained $\log \epsilon_\mathrm{Fe}=7.56\pm0.09$ from \fei lines, and rather discrepant results from \feii lines ($7.41$--$7.56$\,dex depending on the adopted $gf$-values). \citet{MB_fe} also investigated NLTE Fe line formation with the \marcs\ and \oneDAV\ (Sect.~\ref{model}) model atmospheres, finding values fully consistent with the meteoritic abundance, and, in view of the small NLTE effects for the Sun, with the result of \citet{AspII}. With the \oneDAV\  model, they found a mean abundance of $\log \epsilon_\mathrm{Fe}=7.46\pm0.02$\,dex.

\textit{Cobalt:} The Co content of the Sun was derived by \citet{Cardon82} under the assumption of LTE using Co\,\textsc{i} lines, giving $\log \epsilon_\mathrm{Co}=4.92\pm0.08$ with the \citetalias{HM} model atmosphere. This was the reference value adopted by \citetalias{AG89}, \citetalias{GS98}, \citetalias{AGS05}, \citet{Grevesse07} and \citet{Lodders09}, although it only overlaps the meteoritic value because of the rather large errors.  Recently, \citet{Bergemann10Co} re-analysed a series of Co lines
in flux, taking into account departures from NLTE.  They found large NLTE corrections, of order $+$0.15\,dex. Using a MAFAGS-ODF solar photospheric model, they derived an NLTE Co abundance of 
$\log \epsilon_\mathrm{Co}=4.95\pm0.04$.

\textit{Nickel:} We recently provided a revised solar nickel abundance in the context of the Ni-blended forbidden
oxygen line at 630\,nm \citep{Scott09Ni}.  Using the 3D model of \citet{AspI}, that analysis gave
$\log \epsilon_\mathrm{Ni}=6.17\pm0.05$.  Here we update those results using an improved 3D solar model atmosphere \citepalias{AGSS,AGSS_NaCa}.  \citet{Wood14} found $\log \epsilon_\mathrm{Ni}=6.28\pm0.06$ by employing spectrum synthesis of \nii lines, the \citetalias{HM} model and new laboratory oscillator strengths.  The previous reference solar Ni abundance \citep{GS98, AGS05, Grevesse07} was $\log \epsilon_\mathrm{Ni}=6.25\pm0.09$, from an HM-based analysis of \nii by \citet{Biemont80}.

\section{Observations}
\label{observations}

We compared theoretical line profiles to the Fourier Transform Spectrograph (FTS) spectral intensity atlas of
\citet[][see also \citealt{Neckel99}]{Brault87} at solar disk-centre ($\mu=1$).
We removed the solar gravitational redshift of 633 m\,s$^{-1}$ from the observed spectrum, and convolved simulated
profiles with an instrumental sinc function of width $\Delta\sigma = \frac{c}{R}= 0.857$ km\,s$^{-1}$, reflecting
the FTS resolving power $R=350\,000$ \citep{Neckel99}.

Our adopted equivalent widths are the integrated values we previously obtained in full $\chi^2$-based profile fits, using the earlier version of the 3D model \citep{AspI} and the observed FTS spectrum of \citet{Brault87}.  We masked sections of profiles perturbed by nearby lines
from the fitting procedure.  We fitted local continua independently using nearby clear sections of the spectrum.  We were sure to use the same spectral regions to integrate both the observed and theoretical profiles. As a cross-check, we also directly measured the equivalent widths of all lines on two different disc-centre solar atlases: the FTS atlas mentioned
above \citep{Brault87}, and the atlas of \citet{LiegeAtlas} recorded with a classical double-pass spectrometer at the Jungfraujoch high-altitude station.  We noted excellent agreement between these two sets of measurements, and with the equivalent widths derived from the fitted 3D profiles (i.e. to within 1--2\%).  In order to ensure that our 1D and 3D abundances were derived consistently, for the 1D analyses we used the same equivalent widths as in the 3D analysis (i.e. those arising from the earlier 3D line profile fits).

\section{Solar model atmospheres and spectral line formation}
\label{model}

We use the improved 3D model atmosphere introduced in \citetalias{AGSS} and described in more detail in \citetalias{AGSS_NaCa}.  We carried out comparative calculations with four 1D models: \citetalias{HM}, \textsc{marcs} \citep{MARCS75, MARCS97, OSMARCS}, {\sc miss} \citep{MISS} and \oneDAV.  The \oneDAV\ model is a temporal average of the 3D model, contracted into the vertical dimension with horizontal averages taken over surfaces of common optical depth.  The reader is directed to \citetalias{AGSS_NaCa} for further details of these model atmospheres.

We obtain NLTE abundances by applying NLTE corrections to the values we obtained in LTE.\footnote{An NLTE abundance correction is defined as a difference in abundance required to equalise NLTE and LTE equivalent widths.} This is not strictly correct unless full 3D NLTE calculations are carried out; for computational reasons, this is not the case for any of the elements we investigate here. 3D NLTE line formation is still very challenging, and only very few such studies have been undertaken to date \citep[e.g.][]{AspIV, 2009A&A...508.1403P, 2013A&A...554A..96L}. Instead, we apply NLTE abundance corrections computed using 1D model atmospheres. For most elements (Ti, Cr, Mn, Fe, Co), we computed NLTE corrections for solar disk-centre intensity profiles of the selected lines, using the \citetalias{HM}, \textsc{marcs}, and \oneDAV\ models. For calculating 3D+NLTE abundances we adopt offsets computed with the \oneDAV\ model, which we expect to be a close approximation to the full 3D NLTE problem, given that radiative transfer proceeds primarily vertically. For {\sc miss} we adopt the offsets computed with the \citetalias{HM} model. We performed statistical equilibrium calculations with the DETAIL code \citep{Giddings81,Butler85}. For Sc, we rely on NLTE corrections from the literature, while V and Ni have not been exposed to a NLTE study. Our NLTE calculations are described in detail in Sect.~\ref{atomicdata}. 

We do not discuss the NLTE line formation in detail in this work, as this aspect has been extensively discussed previously \citep[e.g.][]{Bruls93,Bergemann07,Zhang08,Bergemann10Co,Bergemann10,Bergemann11,MB_fe}, along with descriptions of the adopted atomic models. In short, the Fe-group elements are predominantly singly-ionised in the solar atmosphere, and departures from LTE are significant only for the neutral species, which are overionised. NLTE effects on the lines of singly-ionised atoms are typically negligible. One remaining uncertainty in NLTE calculations is the unknown cross-sections for inelastic collisions between hydrogen and the element in question. In the absence of quantum mechanical calculations that still only exist for lighter elements, most NLTE studies rely on the classical and therefore uncertain formula of \citet{Drawin69}, which at best should be considered an order-of-magnitude estimate. Therefore, a scaling factor $S_{\rm H}$ for the Drawin cross-sections is used.  At least with iron lines and averaged 3D models, the unscaled \citeauthor{Drawin69} formula ($S_\mathrm{H}=1$) leads to ionisation balance and consistent inferred effective temperatures and surface gravities across a substantial sample of metal-poor stars \citep{MB_fe}.  Wherever possible, we therefore prefer to use $S_\mathrm{H}=1$ for iron-group elements; we do this for all elements where we calculate our own NLTE abundance corrections (Ti, Cr, Mn, Fe and Co).  For Sc, not having a model atom of our own to draw on, we must rely on results from the literature assuming $S_\mathrm{H}=0.1$ \citep{Zhang08}.  Indeed, this parameter remains quite uncertain, and will likely differ across lines, elements and stars. In the absence of detailed quantum mechanical calculations to rely on, or better, solar observations to guide our choices, our selection is by necessity somewhat arbitrary. However, we argue that it is a reasonable approach to adopt the same scaling factor $S_\mathrm{H}$ for all Fe-peak elements as empirically estimated for Fe. Reliable atomic physics computations are urgently needed for inelastic H collisions, not only for these but also other elements.

\section{Abundance calculations}
\label{calculations}

We derived abundances as per \citetalias{AGSS_NaCa}: by matching equivalent widths of simulated and observed line profiles, and including isotopic and hyperfine components in our calculations as blends.  

As in \citetalias{AGSS_NaCa}, the final uncertainties of our 3D+NLTE abundance results are the sum in quadrature of a systematic term and a statistical one.  We take the statistical term to be the standard error of the mean abundance.  We calculate the systematic term as the sum in quadrature of uncertainties due to the mean temperature structure (half the mean difference between the \oneDAV\ and HM results), atmospheric inhomogeneities (half the mean difference between the 3D and \oneDAV\ results), and departures from LTE (the greater of 0.03\,dex and half the mean NLTE correction).  

\section{Atomic data and line selection}
\label{atomicdata}

For each element and ionisation stage, we performed an extensive search of the atomic literature for the most
reliable oscillator strengths, hyperfine splitting constants, isotopic separations, wavelengths, excitation potentials, 
transition designations and partition functions. We preferred to make our own independent critical selection rather than relying on any existing 
compilation, though the NIST Atomic Transition Probability Bibliographical Database \citep{NISTbib} proved invaluable 
for this task.  We used the compilations of \citet{Martin88}, \citet{Fuhr88}, \citet{Doidge95} and especially 
\citet{Morton03} as guides and secondary comparators.

We extracted radiative broadening parameters from the Vienna Atomic Line Database \citep[VALD, ][]{VALD}.  We treated
collisional broadening of neutral lines via the Anstee-Barklem-O'Mara technique \citep{Anstee95, Barklem97, Barklem98}.
The broadening parameters $\sigma$ and $\alpha$ we used were previously calculated for many individual lines \citep{Barklem00}.  For others we interpolated within the tables of
\cite{Anstee95} or \cite{Barklem97}.  No such data exist for ionised iron-peak elements except iron itself \citep{2005A&A...435..373B}, so we employed
the classical \citet{Unsold} broadening recipe for such lines, with an enhancement factor of $2.0$.  The same is true for the small number of neutral lines that lie outside the Anstee-Barklem-O'Mara tables.  This scaling factor reflects the approximate proportionality typically seen between accurate modern broadening calculations and the \citet{Unsold} treatment, as observed over a large range of lines for which modern data are available. We note that most of the lines for which we have to resort to using scaled  \citet{Unsold} broadening are weak and thus insensitive to the adopted damping.

We typically only used a line if it had a $gf$-value available from
the source that we deemed most reliable. Each candidate line was
checked for blends, by inspection of the solar spectra
\citep{Brault87, LiegeAtlas} and the tables of
\citet{Moore66}. Line strengths were also checked in
\citet{Moore66}, and only lines weaker than \mbox{$\sim$60 m\AA}
were generally allowed; in some circumstances, these
requirements were relaxed slightly.\footnote{Even with the 3D model, weaker lines are to be preferred because they lie on the linear section of the curve of growth and are less sensitive to errors in the treatment of broadening or the atmospheric temperature structure, which is less certain in the higher parts of the atmosphere where stronger lines are formed.} The selected lines were assigned
a relative ranking from 1 to 3 based upon their appearance in the
observed spectrum, with rankings sometimes also adjusted to reflect
differences in uncertainties in atomic data.  These rankings were
used to weight the contribution of each line to mean abundances. 
Note that the rankings are only indications of relative merit
within a line list, so the same rank for lines of different species does not necessarily imply the same line quality.

Our adopted lines, oscillator strengths, NLTE corrections, equivalent widths, excitation potentials and derived abundances
for all elements are given in Table~\ref{table:lines}.  We provide isotopic and hyperfine splitting data separately in Table~\ref{table:hfs}.  The isotopic ratios given for individual elements are taken from \citetalias{AGSS}, but the original data are the terrestrial ratios recommended by \citet{IUPAC98}.  Our chosen partition functions are from Barklem \& Collet (in preparation), and our ionisation energies from NIST data tables.  These data are given in Table~\ref{table:partition}.

\subsection{Scandium}
\label{Scatomicdata}

Wavelengths, excitation potentials and transition identifications come from \citet{Kaufman88} for Sc\,\textsc{i}, and
from \citet{Johansson80} for Sc\,\textsc{ii}.  Scandium exhibits hyperfine but not isotopic structure, as it has just
one stable isotope \citep{IUPAC98}: $^{45}$Sc, with spin $I=\frac{7}{2}$.

\subsubsection{Oscillator strengths}
\label{Scgfs}

For both Sc\,\textsc{i} and Sc\,\textsc{ii}, we prefer the $gf$-values of \citet{Lawler89}.  These authors obtained
emission FTS branching fractions (BFs), which they set to an absolute scale using the time-resolved laser-induced
fluorescence (TRLIF) lifetimes of \citet{Marsden88}.  These techniques are currently the most accurate means available
for determining relative spectral intensities and radiative lifetimes respectively, and their combination is the most
reliable way of determining absolute atomic $gf$-values.  For Sc\,\textsc{ii}, accurate lifetimes are also available
from \citet{Vogel85}, where results are in excellent agreement with those of \citet{Marsden88}; 
using the former or the latter data would result in $gf$-values differing by less than
0.01 dex.  

Three very good solar lines (624.56, 630.07 and 632.08\,nm) were not measured by \citet{Lawler89}.  We derived $gf$-values for these lines from existing experimental (\citealt{CB62}, \citetalias{CB62}) and theoretical \citep{Kuruczweb} data, using the lifetimes of \citet{Vogel85} for normalisation.  However, the resulting scatter in the abundances from these lines (with all models) left us ultimately unconvinced as to the accuracy of the oscillator strengths, so we chose to discard these lines.

\subsubsection{Hyperfine structure}
\label{Schfs}

The HFS of Sc\,\textsc{i} has been studied extensively.  The atomic-beam magnetic-resonance technique (ABMR; also
known as laser-rf double resonance or ABMR-LIRF when detected using laser-induced resonance fluorescence) was employed
by \citet{Childs71} to give highly accurate data for the 3d4s$^2$ $^2$D$_{3/2,\,5/2}$ levels.
For the 3d4s4p levels, the data with lowest uncertainties are the FTS results of \citet{Aboussaid96}.  In some cases
\citet{Aboussaid96} provide more than one measurement for a given level; we take the average of these measurements,
weighted according to their uncertainties.  For the ($^3$F)4s $^2$F$_{5/2,\,7/2}$ levels, theoretical
results presented by \citet{Basar04} are the only data available.  \citet{Ertmer76} also presented ABMR data for the
($^3$F)4s $^4$F$_{3/2,\,9/2}$ levels.
We note that optogalvanic spectroscopy (OGS) data presented by \citet{Singh91} for the ($^1$D)4s $^2$D$_{3/2,\,5/2}$ levels are not
reliable, due to errors in their relative intensity formula pointed out by \citet{Aboussaid96}, and confirmed by
\citet{Bieron02} and \citet{Basar04}.  \citeauthor{Singh91} also measured the ($^3$F)4p $^4$G levels, though these
were not affected by this error; for these levels we thus adopt either the data of \citeauthor{Singh91} (
for $^4$G$_{5/2,\,11/2}$) or \citet[][for $^4$G$_{7/2,\,9/2}$]{Ertmer76}, based upon the
size of the quoted uncertainties in each case.

Work on the HFS of Sc\,\textsc{ii} is rather less common.  The most recent and accurate data that we could find come from
\citet{Villemoes92} and \citet{Mansour89}, the latter of whom employed the ultra-high-resolution ABMR technique.
We use the results of both these studies where available, adopting an average weighted according to the stated
uncertainties; in practice this means that the results of \citet{Mansour89} dominate due to their smaller error
 bars.  Where data are not available from both \citet{Villemoes92} and \citet{Mansour89}, we turn to each of these studies individually, followed by the experiments of \citet{Young88} and then \citet{Arnesen82}.  Apart from
the recent work by \citet{Zhang08}, previous determinations of the solar Sc abundance have not considered 
the effects of HFS in \altsci, and only incompletely considered the effects in \altscii.

\subsubsection{NLTE corrections}
\label{ScNLTE}

NLTE formation of solar Sc\,\textsc{i} and Sc\,\textsc{ii} lines has
been thoroughly investigated by \citet{Zhang08},
using the MAFAGS-ODF model.  As might be expected from the minority
status of neutral Sc in the Sun and its quite low ionisation potential
(6.56\,eV), \citet{Zhang08} found very large NLTE corrections to
Sc\,\textsc{i} abundances: about $+0.15$\,dex for flux profiles of the lines of interest in our analysis, when employing the standard \citet{Drawin69} recipe for treating collisions with hydrogen rescaled by a factor $S_\mathrm{H}=0.1$. Corrections to Sc\,\textsc{ii} abundances were less severe (about $-0.01$\,dex for lines of interest to us).  In the absence of any calculations for intensity profiles and/or in 3D, we simply adopt these results for disk centre in Table~\ref{table:lines}, noting that this way the NLTE corrections may be slightly overestimated.  For lines not studied by \citet{Zhang08}, given the size of corrections and the error likely induced by neglecting NLTE, we use the typical correction observed with similar lines.  Although we have NLTE corrections available in both flux and disk-centre intensity for most other iron-group elements, we choose not to rescale the NLTE flux corrections for Sc by the mean ratio of those corrections in order to estimate intensity corrections.  This is because the ratio of intensity to flux corrections, although sometimes substantially less than 1, is quite line specific; the line-to-line scatter in this ratio, across other elements, is actually comparable to the offset of the mean ratio from 1.  Dedicated calculations of Sc NLTE intensity abundance offsets, for the lines and model atmospheres that we employ here, would be most welcome.

\subsubsection{Line selection}
\label{Sclineselection}

We applied our line selection criteria (see the beginning of this Section) to all Sc\,\textsc{i} and Sc\,\textsc{ii} lines in the solar spectrum measured by \citet{Lawler89}.  We also compared with the previous work of \citet{Biemont74a}, \citet{Neuforge93}, \citet{Youssef89} and \citet{Reddy03}, retaining the five Sc\,\textsc{i} and nine Sc\,\textsc{ii} lines given in Table \ref{table:lines}.  We note that the $gf$-value of the very good Sc\,\textsc{ii} line at 660.5\,nm has a large uncertainty \citep[$>$$40$\%;][]{Lawler89}.  Rather than exclude this line, we reduced its weight (as indicated by the asterisk beside its weight in Table~\ref{table:lines}).

\subsection{Titanium}
\label{Tiatomicdata}

Our adopted wavelengths, transition designations and excitation potentials for Ti\,\textsc{i} come from \citet{Forsberg91}.
For Ti\,\textsc{ii}, we took wavelengths from \citet[][with erratum: \citealt{Pickering02}]{Pickering01} where
possible, based upon unpublished work of Zapadlik et al.~in Lund.  Otherwise, wavelengths came from \citet{Huldt82},
as did all excitation potentials and transition identifications.  Ti has five stable isotopes \citep{IUPAC98}:
$^{46}$Ti (8.2\% by number on Earth), $^{47}$Ti (7.4\%), $^{48}$Ti (73.7\%), $^{49}$Ti (5.4\%) and $^{50}$Ti (5.2\%).  $^{47}$Ti
has a nuclear spin of $I=\frac{5}{2}$ and $^{49}$Ti has $I=\frac{7}{2}$.

\subsubsection{Oscillator strengths}
\label{Tigfs}

\citet{Nitz98} and \citet{BW06} produced Ti\,\textsc{i} oscillator strengths by combining their own FTS BFs with
accurate TRLIF lifetimes from \citet{Salih90} and \citet{Lawler91}, respectively. 

\citet{Lawler13} have recently expanded and improved the work of \citet{Nitz98}, providing accurate oscillator strengths for nearly a thousand lines by combining their FTS and eschelle BFs with the lifetimes of \citet{Salih90} and \citet{Lawler91}.

\citet{Grevesse89} earlier produced accurate $gf$ values by renormalising the relative oscillator strengths of \citet{Blackwell2, Blackwell1, Blackwell3, Blackwell4}, which had been obtained by absorption spectroscopy in the Oxford furnace.  As opposed to the original Oxford works, in which relative oscillator strengths were set to an absolute scale using less accurate beam-foil lifetimes from \citet{Roberts73b} and the absolute data of \citet{Bell75}, \citet{Grevesse89} set their new values to an absolute scale using the accurate TRLIF lifetimes of \citet{Rudolph82}.  

We prefer the data of \citet{Lawler13} where possible, but we also performed some preliminary calculations of abundances arising from lines in common between \citet{Nitz98}, \citet{BW06} and the revised Oxford $gf$-values, in order to establish which set of data was the next most reliable.  Based upon the internal scatter and relative agreement between different lists, we concluded that the \citeauthor{Nitz98} values are to be preferred very slightly over the revised Oxford values, while the \citeauthor{BW06} $gf$-values are surprisingly discrepant. 

High-quality Ti\,\textsc{ii} oscillator strengths are available from the FTS BFs and TRLIF lifetimes of \citet{Bizzarri93}, and the extensive FTS and eschelle work by \citet{Wood13}.  The FTS study of \citet{Pickering01} also produced $gf$-values for many lines, where fractions for some branches were
completed using theoretical oscillator strengths of weak lines from \citet{Kuruczweb}.  \citeauthor{Pickering01} set different 
BFs to an absolute scale using either the \citeauthor{Bizzarri93} lifetimes or lifetimes derived from the theoretical
\citeauthor{Kuruczweb} $gf$-values.  Our preliminary investigations with lines common to the lists of \citet{Bizzarri93} and \citet{Pickering01}
revealed a much larger abundance scatter with the \citeauthor{Pickering01} data; we thus prefer the
\citet{Wood13} and \citeauthor{Bizzarri93} oscillator strengths to those from \citet{Pickering01}.

\subsubsection{Isotopic and hyperfine structure}
\label{Tihfs}

Much complimentary data exist on the isotopic splitting of Ti\,\textsc{i} lines, though unfortunately only for two of the
lines we use here.  The data we use come from laser fluorescence spectroscopy \citep[LFS, also known simply as laser-induced
fluorescence, LIF;][]{Gangrsky95}.  We prefer the data of \citeauthor{Gangrsky95} over the less accurate work of
\citet{Cruz94} and previous results from the same group \citep{Anastassov94}.  Isotopic separations can be estimated for many
of our our chosen Ti\,\textsc{ii} lines using the LFS measurements of \citet{Nouri10}. 

It has been consistently found that the hyperfine $A$ constants for $^{47}$Ti and $^{49}$Ti are
essentially equal, and $B(47) / B(49)\approx1.22$, for all levels \citep{Channappa65, Aydin90, Stachowska94, Gangrsky95}.
We therefore use the experimental values for the relevant isotope where available, but use rescaled experimental data from
the other where it does not exist for both isotopes.  Data on hyperfine structure for the Ti\,\textsc{i} lines for which
we have isotopic information are best obtained from \citet{Gangrsky95} and \citet{Aydin90}. In cases of overlap, the ABMR
data of \citeauthor{Aydin90} have smaller uncertainties than those of \citet{Johann81}, whereas the LFS data of
\citeauthor{Gangrsky95} is preferable to \citeauthor{Aydin90}'s LFS. LFS data from \citet{Jin09} is of similar quality to, and agrees well with, that 
of \citet{Gangrsky95}.  The only HFS data on Ti\,\textsc{ii} are the experimental ABMR and corresponding theoretical values produced by \citet{Berrah92}, and the LFS data of \citet{Nouri10}.

\subsubsection{NLTE corrections}
\label{TiNLTE}
The NLTE line formation of Ti lines has been extensively discussed by \citet{Bergemann11}. Our NLTE calculations rely on the same model atom, although we adopt a different scaling factor to the \citet{Drawin69} formula for inelastic \ion{H}{i} collision cross-sections ($S_\mathrm{H}=1$ rather than $S_\mathrm{H}=3$; cf Sect.\ \ref{model}). We computed \tii NLTE abundance corrections for disk-centre intensity with the \oneDAV, \marcs\ and \citetalias{HM} 1D model atmospheres; we adopt the \oneDAV\ results as an approximation to the real 3D NLTE corrections.  It is interesting that even with the relatively large value $S_\mathrm{H}=1$, the resulting NLTE corrections for the \tii lines are significant. For the \oneDAV\ model, they range from $0.04$ to $0.09$\,dex, whereas the use of the \citetalias{HM} model reduces them by a factor of two, mainly because its reduced temperature gradient makes over-ionisation less pronounced.  \citet{Bergemann11} found minimal NLTE effects on the relatively weak \tiii lines we consider, so we simply adopt the LTE results for this species.

\subsubsection{Line selection}
\label{Tilineselection}

We applied our selection criteria to numerous solar lines, including those used in previous works by \citet{Blackwell87}, \citet{Reddy03} and \citet{Bizzarri93}.  We ultimately retained 34 lines of Ti\,\textsc{i} and 14 of Ti\,\textsc{ii} (Table~\ref{table:lines}). Twenty-four of our Ti\,\textsc{i} lines we included by \citet{Lawler13}.

\subsection{Vanadium}
\label{Vatomicdata}

For V\,\textsc{i} we sourced wavelengths and excitation potentials from \citet{Davis78b}, calculating wavelengths of missing lines from the stated energy levels.  We adopted transition identifications from \citet{Whaling85}, with corrections to \mbox{V\,\textsc{i} 573.1\,nm} and \mbox{V\,\textsc{i} 609.0\,nm} following consultation with \citeauthor{Davis78b} and \citet{Martin88}.  V\,\textsc{ii} wavelengths and transition identifications came from \citet{Biemont89}, with excitation potentials from \citet{Sugar85}.

Vanadium has two stable isotopes: $^{51}$V ($I=7/2$) and $^{50}$V ($I=6$).  The isotopes are
present in the ratio $^{51}$V / $^{50}$V $\approx 400$ on Earth \citep{IUPAC98}; because of this large ratio, isotopic structure is of no
importance for vanadium lines.  V\,\textsc{i} and V\,\textsc{ii} lines are given in Table~\ref{table:lines} with corresponding atomic data.

\subsubsection{Oscillator strengths}
\label{Vgfs}

The best V\,\textsc{i} oscillator strengths available are those of \citet{Whaling85}, who measured both TRLIF lifetimes
and FTS BFs.   
In some cases we correct
this data for arithmetic errors in converting from BFs to transition probabilities, as per \citet{Martin88}.  There are
also a few accurate $gf$-values from \citet{Doerr85b}, who combined TRLIF lifetimes with BFs from hook absorption and hollow
cathode emission.  We prefer the data of \citet{Whaling85}, as their lifetime uncertainties are lower than
\citeauthor{Doerr85b}'s, and obtaining BFs by FTS is generally considered the most reliable method available.

For V\,\textsc{ii}, until very recently the most accurate $gf$-values come from the FTS BFs and TRLIF lifetimes of \citet{Biemont89}.  In addition
to their own, these authors drew on a large number of accurate TRLIF lifetimes measured by
\citet{Karamatskos86} to arrive at their
final oscillator strengths. 
 \citet{Karamatskos86} had also obtained FTS BFs, and also produced mostly accurate $gf$
values, but their results disagree with those of \citet{Biemont89} below around \mbox{350\,nm}.  
\citeauthor{Biemont89}
suggest that this is likely due to an FTS calibration error by \citet{Karamatskos86}, so we prefer \citeauthor{Biemont89}'s
results in general.  However, we do choose the $gf$ value of \citeauthor{Karamatskos86} over that of \citeauthor{Biemont89}
for the \mbox{V\,\textsc{ii} 395.2\,nm} line, as in this case the uncertainty of \citeauthor{Biemont89}'s measurement
is 50\%, whereas that of \citeauthor{Karamatskos86}'s is 8\%.  \citet{Schade87} also produced TRILF lifetimes, which
agree nearly perfectly with those of \citet{Karamatskos86}, and exhibit similar errors.  \citet{Biemont89} preferred
the lifetimes of \citet{Karamatskos86} because they were more extensive, but also because in the one case of disagreement,
the errors of \citeauthor{Karamatskos86} are smaller. 

During the final stages of refereeing of our article, we became aware of new experimental FTS+LIF measurements of V\,\textsc{ii} transition probabilities for a large number of UV/optical lines by the Wisconsin group \citep{Wood14V2}. Without a doubt, these should be the most accurate V\,\textsc{ii} data available now. Although it was too late to adopt these new $gf$-values, below we discuss how our results would have changed had we done so.

By comparing the claimed uncertainties of the vanadium abundances stated by \citet{Biemont89} with the internal
uncertainties of the sets of $gf$-values used to derive the abundances, we note that the uncertainty of their vanadium
abundance is almost certainly underestimated.

\subsubsection{Hyperfine structure}
\label{Vhfs}

Quite a lot of good data exists on the HFS of V\,\textsc{i}, with little overlap between the levels investigated by
different authors.  Based on the uncertainties assigned to levels common to different studies, we placed the data into a
preferential tier system.  In this system, no tier contained more than one value for any given level.  In the top tier
were the ABMR and LFS data of \citet{Childs79}, the ABMR results of \citet{ElKashef92}, \citet{Unkel89}, \citet{Johann81}
and \citet{Childs67b}, and the FTS data of \citet{Palmeri97}.  The second tier consisted of an earlier FTS study by
\citet{Palmeri95} and the crossed-beam results of \citet{Cochrane98}.  On the third tier were additional results from
\citet{Unkel89} using LFS, FTS data of \citet{Lefebvre02} and Doppler-free LFS results from \citet{Gough85}.
\citet{Whaling85} and \citet{Biemont89} included the effects of HFS in their analysis of V\,\textsc{i}, though we are now
able to draw upon better HFS data.

Until the recent fast-ion-beam LFS work of \citet{Armstrong11}, no HFS data existed in the literature for ionised vanadium.  \citet{Biemont89} estimated hyperfine broadening
of V\,\textsc{ii} lines empirically, adding multiple line components by eye to approximately reproduce line shapes and sufficiently desaturate modelled solar lines.  We have done something similar for the one line (399.7\,nm) where HFS data are not available from \citet{Armstrong11}, iteratively altering the hyperfine $A$ constants of the two levels involved until we achieved a synthetic spectral line that looked qualitatively similar to the observed line.  To account for the effects of convective velocities upon line shapes, it was necessary to use a 3D model for this exercise.  Due to the computational demands of recalculating the radiative transfer every time however, we performed these calculations on a single snapshot of the earlier 3D model \citep{AspI} only.\footnote{Interestingly, the resulting hyperfine constants for this line are in reasonable agreement with the experimental measurements of \citet{Wood14V2}, which appeared only at the end of the refereeing stage of this paper.}
The results of this estimation procedure, along with all other data pertaining to our chosen \vi and \vii lines, are given in Table~\ref{table:lines}.

\subsubsection{NLTE corrections}
\label{VNLTE}

NLTE formation of solar vanadium lines has not yet been investigated.  Like Sc, Ti and Cr, the rather low ionisation energy of V means that it is predominantly singly-ionised in the solar atmosphere.  As the minority species, \vi is expected to exhibit significant NLTE effects.  In the absence of any better guidance, we adopt a blanket NLTE correction of $+0.1$\,dex for all \vi lines; this is of a similar order as the mean NLTE offsets observed in \scani ($+0.15$\,dex), \tii ($+0.06$\,dex) and \cri ($+0.03$\,dex).  A dedicated NLTE study of V is sorely needed.

\subsubsection{Line selection}
\label{Vlineselection}

We retained 32 V\,\textsc{i} lines (Table~\ref{table:lines}) from previous analyses by \citet{Biemont78b}, \citet{Whaling85}, \citet{Reddy03} and \citet{McWilliam94}. The solar V\,\textsc{ii} lines are very poor quality, because of severe blending, and are ultimately only really useful as weak supporting indicators of the solar vanadium abundance.  After careful analysis of various lines used by \cite{Youssef89}, \citet{Biemont89} and \citet{McWilliam95}, we choose to keep only the five lines in Table~\ref{table:lines}.

\subsection{Chromium}
\label{Cratomicdata}

We sourced Cr\,\textsc{i} and Cr\,\textsc{ii} excitation potentials from \citet{Sugar85}, and used them to calculate
wavelengths.  
Where possible, we took transition identifications from \citet{Sobeck07} for Cr\,\textsc{i} and \citet{Nilsson06} for Cr\,\textsc{ii}.  Otherwise, we sourced transitions from VALD and checked them against the NIST database \citep{NIST}.  Chromium has four stable isotopes \citep{IUPAC98}: $^{50}$Cr (4.3\%), $^{52}$Cr (83.8\%), $^{53}$Cr (9.5\%) and $^{54}$Cr (2.4\%).  Only $^{53}$Cr has non-zero nuclear spin ($I=\frac{3}{2}$).

\subsubsection{Oscillator strengths}
\label{Crgfs}

Highly accurate Cr\,\textsc{i} oscillator strengths have been produced by \citet{Sobeck07}, who measured FTS BFs and normalised them with the extensive, very accurate TRLIF lifetimes of \citet{Cooper97}.  Other accurate lifetimes have been measured by TRLIF \citep{Hannaford81, Kwiatkowski81, Kwong80, Measures77} and level-crossing \citep{Becker77}; these data all agree well with \citeauthor{Cooper97}'s, and have comparable uncertainties. Other accurate $gf$-values were produced by \citet{Tozzi85}, also based upon FTS BFs but normalised to \citeauthor{Kwiatkowski81}'s lifetimes, and \citet{Blackwell84, Blackwell86}, who measured relative $gf$-values using absorption spectroscopy and set them to an absolute scale with the lifetimes of \citet{Hannaford81}, \citet{Kwiatkowski81} and \citet{Becker77}.  To complete their systems of lines, \citeauthor{Blackwell84} also drew upon some of the relative oscillator strengths carefully measured by \citet{Huber77} using the hook method.  Because they are all of high quality, we use $gf$-values from \citet{Sobeck07}, \citet{Blackwell84, Blackwell86} and \citet{Tozzi85} without any preference for data from one source or another; where data overlap, we take the mean of the $\log gf$-values available from each of these sources.

Oscillator strengths for Cr\,\textsc{ii} were recently produced by \citet{Gurell10} and \citet{Nilsson06}, who each combined their own FTS BFs with accurate TRLIF lifetimes; \citet{Gurell10} used their own lifetimes, whereas \citeauthor{Nilsson06} utilised a mixture of TRLIF lifetimes from \citet{Schade90}, \citet{Bergeson93b} and their own work.  Unfortunately, \citet{Gurell10} measured no useful solar lines, and \citeauthor{Nilsson06} only very few of them.  The small number of $gf$-values available from \citeauthor{Nilsson06} for good solar lines also return abundances that are highly inconsistent with each other.  In the absence of any good $gf$-values for the unblended Cr\,\textsc{ii} lines in the solar spectrum, we default to using the theoretical \citet{Kuruczweb} oscillator strengths.  Given that the \citeauthor{Kuruczweb} $gf$-values are known to often be inaccurate, especially for weak transitions, this is not a satisfactory situation; high-quality atomic data is urgently needed for Cr\,\textsc{ii}.

\subsubsection{Isotopic and hyperfine structure}
\label{Crhfs}

The only data on the isotopic splitting of Cr\,\textsc{i} lines come from the recent LFS work of \citet{Furmann05} and the much older Fabry-Perot spectroscopy of \citet{Heilig67}; we use the former.  HFS of $^{53}$Cr\,\textsc{i} has been measured very accurately with ABMR by \citet{Jarosz07}.  No data exist on the isotopic splitting of Cr\,\textsc{ii} lines, nor HFS of $^{53}$Cr\,\textsc{ii}.

\subsubsection{NLTE corrections}
\label{CrNLTE}
We computed \cri NLTE abundance corrections in intensity at disk-centre, for the \oneDAV , \marcs\ and \citetalias{HM} 1D model atmospheres, using the Cr model atom of \citet{Bergemann10}. For the majority of our \cri lines, the corrections are in the range $+0.02$ to $+0.04$\,dex for the \oneDAV\ model, and typically a factor of two lower for the \citetalias{HM} semi-empirical model. As for the other iron-peak elements except Sc (cf. Sect.\ \ref{model}), we used a scaling factor of $S_\mathrm{H}=1$ to the \citet{Drawin69} recipe for inelastic collisions with H. \citet{Bergemann10} excluded inelastic \ion{H}{i} collisions from their model atom ($S_\mathrm{H}=0$), in order to obtain ionisation balance with MAFAGS-ODF model atmospheres in a larger sample of late-type stars. This, together with the fact that they considered flux spectra, explains the rather large differences (of order $\sim$0.1\,dex) between our NLTE abundance corrections and theirs for solar \cri lines.  

For \altcrii, \citet{Bergemann10} found that LTE is an excellent approximation even without inelastic hydrogen collisions; we therefore do not apply any NLTE corrections for \altcrii.

\subsubsection{Line selection}
\label{Crlineselection}

Based on the solar analyses by \citet{Sobeck07} and \citet{Biemont78c}, we selected the 29 best Cr\,\textsc{i} and 10 best Cr\,\textsc{ii} lines in the solar spectrum.  These are given in Table~\ref{table:lines}.

\subsection{Manganese}
\label{Mnatomicdata}

We took Mn\,\textsc{i} wavelengths and transition designations from \citet{Adelman89}, and excitation potentials from \citet{Corliss77}.  Mn has just a single stable isotope \citep{IUPAC98}: $^{55}$Mn, with $I=\frac{5}{2}$.  Mn\,\textsc{i} lines and atomic data used in the current study are given in Table~\ref{table:lines}.  None of the Mn\,\textsc{ii} lines that we investigated were ultimately of sufficient quality for abundance determination.

\subsubsection{Oscillator strengths}
\label{Mngfs}

FTS Mn\,\textsc{i} BFs have been most recently measured by \citet{DenHartog11}, \citet{BW07} and \citet{BW05b}.  \citet{BW07} used highly accurate TRLIF lifetimes from \citet{Schnabel95}, along with one other lifetime from the laser-excited delayed coincidence of \citet{Marek75} to produce accurate oscillator strengths.  \citet{BW05b} used their own TRLIF lifetimes to also convert their BFs into accurate oscillator strengths, though for some levels slightly more accurate TRLIF lifetimes are also available from \citet{Schnabel95}.  \citet{DenHartog11} also measured TRLIF lifetimes with which to convert their BFs into very accurate $gf$-values, and averaged their data with previous accurate measurements in order to produce a set of recommended values.

Until these three recent studies, the most commonly used Mn oscillator strengths were those of \citet{Booth84c}.  These data were measured
as a set of relative $gf$-values in the Oxford furnace and set to an absolute scale using the laser-excited delayed coincidence
lifetimes of \citet{Becker80} and \citet{Marek75}, as well as the phase-shift results of \citet{Marek73}.
\citet{Booth84c} measured three different systems of lines. 
The first system consisted of eight lines with excitation potentials of around \mbox{0\,eV}, and was set to an
absolute scale using a single averaged lifetime from \citet{Marek75} and \citet{Marek73}.  The second system (24 lines
with excitation potential \mbox{$\sim$2\,eV}) was normalised using an average of the absolute scales implied by lifetimes of six
different levels, taken from \citet{Becker80}.  
The 27 lines of the third system (with excitations \mbox{$\sim$3\,eV})
were normalised using a pyrometry link to the second system, setting the two systems to the same absolute scale.

One concern with the $gf$-values of \citet{Booth84c} were some odd
discrepancies with the BFs derived earlier by
\citeauthor{Greenlee79} (\citeyear{Greenlee79}, \citetalias{Greenlee79}).  
There is no immediate
reason for the BFs by \citetalias{Greenlee79} to be unreliable. However, if one
compares $gf$-values given for the \mbox{3\,eV} lines by
\citeauthor{Booth84c} with $gf$-values for the same lines derived using
\citetalias{Greenlee79} BFs and either \citet{Becker80} or
\citet{Schnabel95} lifetimes, 
an odd dichotomy appears. Whilst we expect both sets to be reliable,
the $gf$-values of \citet{Booth84c} are consistently
\mbox{$\sim$0.15--0.20 dex} higher than the
\citetalias{Greenlee79}-\citeauthor{Becker80} or
\citetalias{Greenlee79}-\citeauthor{Schnabel95} values. This is
confirmed when the $gf$-values of \citet{BW07} are compared with the
data of \citet{Booth84c}: the values of \citeauthor{Booth84c} are
systematically larger, by $0.13$\,dex ($\pm0.02$). This is however
not the case for the \mbox{2\,eV} system, where the two sets agree
very well. These discrepancies have often been ignored in
the literature.

The obvious question is whether the pyrometry link utilised by \citeauthor{Booth84c} was indeed accurate, seeing as the discrepancy only exists for the \mbox{3\,eV} lines.  It seems that poor pyrometry is an unlikely explanation for a $\sim$40\% difference.  Clearly something is amiss, but we cannot explain the discrepancy with any confidence. The confusion in the \mbox{3\,eV}
oscillator strengths is our main reason for concluding that the stated uncertainty in the
solar manganese abundance of \citet[][$\log \epsilon_\mathrm{Mn}=5.39\pm0.03$]{Booth84b} probably substantially underestimated the
true error. In the end, two of our adopted \altmni\ lines are affected by the uncertainties in the \citeauthor{Booth84c} $gf$-values, as we explain below.

Wherever possible, we use the oscillator strengths of \citet{BW07} or those recommended by \citet{DenHartog11}.  
In cases of overlap, we use the recommended \citet{DenHartog11} values wherever the uncertainty of \citeauthor{DenHartog11}'s own measured value is smaller than the error given by \cite{BW07}; the differences are however tiny ($\sim$0.01\,dex or less).  For the 408.3\,nm line, where the value recommended by \citeauthor{DenHartog11} is the average of their own very accurate value and the slightly less accurate result of \citet{BW05b}, we adopt \citeauthor{DenHartog11}'s own raw result rather than the recommended value.

For lines without $gf$-values available from either \cite{DenHartog11} or \citet{BW07}, we derive new oscillator strengths from the BFs of \citetalias{Greenlee79} and the lifetimes of \citet{Schnabel95}.  For the two good \mbox{3\,eV} lines (426.59 and 445.70\,nm) measured only by \citeauthor{Booth84c}, we use the $gf$-values of \citeauthor{Booth84c} but renormalise them to the absolute scale of \citeauthor{BW07}, i.e. decrease them by 0.13\,dex.  Consummate with this rather approximate $gf$ derivation, we only give these lines a weighting of 1 in the final mean abundance.  These lines are marked with asterisks in Table~\ref{table:lines}.  For the remaining line in the \mbox{2\,eV} system (542.0\,nm), we continue to use the original oscillator strength of \citet{Booth84c}.

\subsubsection{Hyperfine structure}
\label{Mnhfs}

A wealth of data exists on HFS in Mn\,\textsc{i}, which we have classified into a similar tier system as for other elements.
The best original data come from the extremely accurate spin-exchange results of \citet{Davis71}, the ABMR of
\citet{Johann81}, ABMR by \citet{D79}, interference spectroscopy by \citet{B87} and laser-atomic-beam spectroscopy by
\citet{Kronfeldt85}.  The second and third tiers consist of FTS and OGS data obtained by \citet{BW05c} and \citet{Basar03}
respectively.  
We do not use the $B$ values of \citeauthor{Basar03} for the odd levels however, because in our opinion their accuracy is insufficient to clearly distinguish them from zero.
The next most accurate data come from  \citet{Lefebvre03}, followed by \citet{Luc72}, \citet{Handrich69} and \citet{Walther62}.
The solar abundance determinations of \citet{BW07} and \citet{Bergemann07} include extensive HFS data from many of the sources listed
above. 

\subsubsection{NLTE corrections}
\label{MnNLTE}
The NLTE formation of solar \mni lines was considered by \citet{Bergemann07}, using the 1D theoretical MAFAGS-ODF model atmosphere. Differences between the LTE and NLTE abundances determined using the solar flux spectrum were typically found to be around $+0.07$\,dex for the lines of interest to us.  We performed NLTE calculations with the same model atom, but adopted a scaling factor $S_\mathrm{H}=1$ to the \citet{Drawin69} formula (cf.\ Sect.\ \ref{model}), instead of \citeauthor{Bergemann07}'s default of $S_\mathrm{H}=0.05$. We calculated corrections in disk-centre intensity with the \oneDAV , \marcs\ and \citetalias{HM} 1D model atmospheres; as for other elements we adopt the \oneDAV\ results as proxies for the 3D case. The NLTE abundance corrections depend on the line properties, i.e. upper and lower excitation potentials, equivalent width and HFS. For example, the saturated $408.2$\,nm line ($E_\mathrm{exc}=2.2$\,eV) has an NLTE correction of only $+0.016$ dex. In contrast, the $542.0$\,nm line, with roughly the same equivalent width but
different upper level, has an NLTE correction of
$+0.07$ dex.  NLTE effects in the solar \mni lines are not very sensitive to the adopted efficiency of inelastic hydrogen collisions.
Reducing $S_\mathrm{H}$  to 0.05 increases the NLTE corrections for all investigated lines by a maximum of $\sim$0.02\,dex. Our adopted NLTE corrections are given in Table~\ref{table:lines}.

\subsubsection{Line selection}
\label{Mnlineselection}

The \mbox{3\,eV} system yields better lines for solar abundance determination than the \mbox{0\,eV} or \mbox{2\,eV} systems, as the \mbox{3\,eV} lines are formed lower in the photosphere, and are therefore less prone to uncertainties associated with the temperature structure of the model atmosphere.  Even amongst the \mbox{3\,eV} lines however, most usable Mn\,\textsc{i} lines are not particularly weak, so we were forced to consider mostly lines of intermediate strength.  The large HFS of many of these lines should at least mitigate the effects of line strength, by desaturating profiles and lowering formation heights.  Unfortunately, apart from \mni 408.3\,nm, all the lines with BFs available from \citet{BW05b} are too weak or blended to be useful in the Sun, so most of our chosen lines came from \citet{BW07}.  After considering previous solar abundance analyses \citep[e.g.][]{Blackwell72a, Biemont75a, Booth84b, BW07, Bergemann07}, we retained the 14 lines given in
Table~\ref{table:lines}.

\subsection{Iron}
\label{Featomicdata}

We used wavelengths from \citet{Nave94} for Fe\,\textsc{i}, and from \citet{2013ApJS..204....1N} for Fe\,\textsc{ii}.  Excitation potentials and transition designations for both species were taken from VALD.  Iron has four stable isotopes \citep{IUPAC98}: $^{54}$Fe (5.8\%), $^{56}$Fe (91.8\%), $^{57}$Fe (2.1\%) and $^{58}$Fe (0.3\%).  The only one of these
with non-zero nuclear spin is $^{57}$Fe, with $I=\frac{1}{2}$.  Iron lines therefore exhibit virtually no isotopic or
hyperfine structure.

\subsubsection{Oscillator strengths}
\label{Fegfs}

The best $gf$-values for Fe\,\textsc{i} have been obtained by quite different techniques. The Oxford dataset \citep[see][and references therein]{Oxford6} is based on absorption spectroscopy: very precise relative $gf$-values were measured in the Oxford furnace, and then normalised to an absolute scale using one line for which the absolute $gf$-value is known with high precision ($\pm$0.02 dex). Two other groups at Hannover \citep{Bard91,Bard94} and at Madison \citep{OBrian91} used emission spectroscopy, measuring lifetimes and BFs. These three sources provide our adopted $gf$-values. When $gf$-values were available from more than one of these sets for any given line, we adopted an unweighted mean of the values from the different sets.  The exception to this rule was a group of three lines where we gave less weight to the \citet{OBrian91} data, because of their larger uncertainties for these specific lines. For one line (\fei 829.4\,nm) where the error on the $gf$-value remains large, we degrade the weight of the line in our analysis by one unit, as indicated by the asterisk in Table~\ref{table:lines}; the uncertainty of the other \fei $gf$-values given in Table~\ref{table:lines} is probably of order 5--10\%.  Newer oscillator strengths are also available from \citet{Ruffoni13}, but for the only line in our list to have been remeasured (\fei 578.4\,nm), the newer oscillator strength results in a clearly discrepant abundance (by $\approx$$0.1$\,dex).

Fe\,\textsc{ii} oscillator strengths increased in accuracy over the past 20 years as progressively more accurate TRLIF
lifetimes were measured by \citet{Hannaford92}, \citet{Schnabel99} and \citet{Schnabel04}, and used to normalise earlier
FTS and grating spectrometer emission BFs from \citet{Heise90} and \citet{Kroll87}.  Probably the most accurate $gf$-values now
come from the compilation of \citet{Melendez09}, who used these and other experimental lifetimes to recalibrate
and average a raft of theoretical and experimental BFs; we adopt these data for all our Fe\,\textsc{ii}
lines. All of our \altfeii\ lines have laboratory-based rather than astrophysical $gf$-values from \citet{Melendez09}. 

\subsubsection{NLTE corrections}
\label{FeNLTE}
We computed NLTE corrections for \fei lines using the Fe model atom of \citet{MB_fe}, which was constructed from the most up-to-date theoretical and experimental atomic data available for \fei and \altfeii. We computed the disk-centre intensity spectrum using a scaling factor $S_\mathrm{H}=1$ to the \citet{Drawin69} recipe for inelastic H collisions, as preferred by the analysis of \citeauthor{MB_fe} (cf.\ Sect.\ \ref{model}). We did the calculations with the \oneDAV , \marcs\ and \citetalias{HM} 1D model atmospheres, resulting in mean NLTE corrections of $+0.01$\,dex.  Larger \fei NLTE effects were advocated by \citet{2011A&A...528A..87M}, who also used an extended Fe model atom, but a lower efficiency for hydrogen collisions ($S_\mathrm{H}=0.1$).  This choice, together with the fact that \citet{2011A&A...528A..87M} considered flux spectra, mostly explains the difference with our results. \citeauthor{2011A&A...528A..87M}'s NLTE corrections were $+0.04$\,dex for lines with excitation energies up to 1\,eV, and about $+0.03$\,dex for higher-excitation lines.  NTLE corrections are negligible for \feii \citep{MB_fe}, so we adopt LTE results for the ionised lines.

\subsubsection{Line selection}
\label{Felineselection}

We selected the 22 best \fei and 9 best \feii lines, i.e.~those lines for which equivalent widths are easily measured, that do not show any intractable trace of blending, and are not too strong. Our selected lines and atomic data are given in Table~\ref{table:lines}. For Fe\,\textsc{i}, we made sure to have a sample of lines that covers a large range of excitation potentials (0--4.6\,eV) to probe the performance of the different atmospheric models over a range of heights.

\subsection{Cobalt}
\label{Coatomicdata}

We took excitation potentials of Co\,\textsc{i} from \citet{Pickering96b}, as well as wavelengths and transition designations where available; otherwise, we sourced wavelengths and transitions from \citet{Cardon82}.  The only stable isotope of cobalt is $^{59}$Co \citep{IUPAC98}, which has nuclear spin $I=\frac{7}{2}$.  Our chosen Co\,\textsc{i} lines and atomic data are given in Table~\ref{table:lines}.  None of the Co\,\textsc{ii} lines in the solar spectrum are suitable for abundance analyses.

\subsubsection{Oscillator strengths}
\label{Cogfs}

The most reliable Co\,\textsc{i} oscillator strengths currently available come from \citet{Nitz99}, who measured
FTS BFs and set them to an absolute scale using their own TRLIF lifetimes \citep{Nitz95}. The next most accurate
data are those of \citet{Cardon82}, who measured BFs that they set to an absolute
scale using the TRLIF lifetimes of \citet{Marek77} and \citet{Figger75}.  
For some lines, the $gf$-values of \citet{Cardon82} are accurate to better than 10\%, which
is comparable to the accuracy obtained by \citet{Nitz99}; for other lines the uncertainties are much larger, of 
order 20--30\%. BFs contemporary with those of \citet{Cardon82} are also available from 
\citet{Guern82}, but we prefer the data of \citeauthor{Cardon82} as they are based upon FTS recordings and include
a more complete set of branches.

\subsubsection{Hyperfine structure}
\label{Cohfs}

We used stated uncertainties to classify the wealth of data available on Co\,\textsc{i} HFS into a similar tier system as
for other elements.  Our first choice of HFS data were the ABMR results of \citet{Childs68}, and the combined Doppler-free
and Doppler-limited LFS / OGS results of \citet{Guthlorien90}.  The next most accurate data come from the FTS of
\citet{Pickering96a}.  Also available are unpublished data obtained by J. Ibrahim-R\"ud and R. Wenzel, reproduced in 
the paper of \citet{Guthlorien90}.  
We fit these data into the hierarchy on a level-by-level
basis around \citet{Guthlorien90}, \citet{Pickering96a} and \citet{Childs68}.
\citet{Bergemann10Co} included extensive HFS data in their calculation of the solar Co abundance, showing that neglect or inaccurate treatment of HFS can lead to severe
errors in derived abundances.

\subsubsection{NLTE corrections}
\label{CoNLTE}

Non-LTE formation of solar Co\,\textsc{i} and Co\,\textsc{ii} lines has been investigated by \citet{Bergemann10Co} using the MAFAGS-ODF models. The results indicate large departures from LTE in Co\,\textsc{i}, leading to NLTE abundance corrections of $+0.1$--$0.2$\,dex at $S_\mathrm{H}=0.05$ \citep[][cf.~their Table 4]{Bergemann10Co}, resembling the situation with \scani lines \citep{Zhang08}. We use the same Co model atom, but adopt $S_\mathrm{H}=1$ (cf.\ Sect.\ \ref{model}) and disk-centre intensity spectra for the \oneDAV , \marcs\ and \citetalias{HM} 1D model atmospheres.  This leads to somewhat smaller NLTE abundance corrections, of order $+0.09$\,dex for \oneDAV\ and $+0.07$ dex for the \citetalias{HM} model.

\subsubsection{Line selection}
\label{Colineselection}

Unfortunately, there are rather few good lines in the solar spectrum with oscillator strengths available from
\citet{Nitz99}, so the bulk of our lines have $gf$ values drawn from \citet{Cardon82}.  For some of the cleanest weak lines, the \citeauthor{Cardon82} $gf$-values have rather large uncertainties (over 20\% in some cases).  We include such lines because of their excellent profiles, but downgrade their weightings; affected lines are marked with an asterisk in Table~\ref{table:lines}.  From the lines considered in the abundance analyses of \citet{Cardon82}, \citet{Biemont78b}, \citet{Kerola76} and \citet{Holweger71}, we retained the 13 transitions given in Table~\ref{table:lines}.

\subsection{Nickel}
\label{Niatomicdata}

We obtained \nii wavelengths and excitation potentials from \citet{Litzen93}.  Transition identities came from
\citet{Wickliffe97}, except for \mbox{\nii 481.2\,nm}, where the transition designation is from VALD \citep{VALD}.
Our selected \nii lines are given in 
Table~\ref{table:lines}.  We also considered \altniii, but it ultimately played a very minimal role in our
analysis; we omit it from Table~\ref{table:lines}, give a truncated discussion of its atomic data
and line selection in this section, and discuss only briefly the mean implied Ni abundance in Sect.~\ref{Niresults}.

Nickel has five stable isotopes, so exhibits significant isotopic structure
\citep[as seen by e.g.][]{Brault81, Melendez99a}.  These are $^{58}$Ni, $^{60}$Ni, $^{61}$Ni, $^{62}$Ni and $^{64}$Ni,
present in the approximate ratios 68:26:1:4:1 \citep{IUPAC98}.  In practice, the isotopic structure of nickel lines is dominated by
$^{58}$Ni and $^{60}$Ni due to their much greater natural abundances.  As an even-$Z$ element, all the even-$A$ nuclei of nickel
have $I=0$, so nickel lines do not exhibit any HFS apart from $^{61}$Ni.  The contribution of $^{61}$Ni to the HFS of nickel is minimal,
given its very low abundance relative to $^{58}$Ni and $^{60}$Ni.  
For spectroscopic purposes, one can thus effectively regard nickel as consisting of four isotopes, and devoid of HFS.

\subsubsection{Oscillator strengths}
\label{Nigfs}

High-quality \nii oscillator strengths are available from the FTS BFs of \citet{Wickliffe97}, which the authors placed
on an absolute scale using the TRLIF lifetimes of \citet{Bergeson93a}.  These have recently been updated and greatly extended by \citet{Wood14}. 
A small number of high-quality $gf$-values are also available from \citet{Johansson03}, based upon FTS BFs and a single 
TRLIF lifetime.  \citet{Bergeson93a} used their
new lifetimes to produce other accurate $gf$-values from the BFs of \citet{Blackwell89}, but these lines are all in the UV,
so of little use to us because the ultraviolet solar spectrum is so crowded.  Accurate oscillator strengths for optical 
lines of \niii do not exist, so we turned to the extensive theoretical transition probabilities of \citet{Fritzsche00}.

\subsubsection{Isotopic structure}
\label{Nihfs}

Wherever available, we employ isotopic separations from \citet{Wood14}, who fitted the isotopic shifts of a large number of \nii energy levels to earlier spectroscopic data.  Much of the power of that analysis can be attributed to the accurate FTS wavelengths of $^{58}$Ni and $^{60}$Ni line components recorded by \citet{Litzen93}.  As in \citeauthor{Wood14}'s analysis, we model $^{58}$Ni and $^{60}$Ni components explicitly, and estimate the contribution of the remaining isotopes by placing them in a single line component, which we offset from $^{60}$Ni by the same amount as $^{60}$Ni is offset from $^{58}$Ni.  These data are included in Table~\ref{table:hfs}.
 
\subsubsection{NLTE corrections}
\label{NiNLTE}

The only explicit investigation of non-LTE effects on solar nickel line formation so far has been that of \citet{Bruls93}, who looked at the \nii 676.8\,nm line often used for helioseismology.  Although \citeauthor{Bruls93} did not give any explicit NLTE abundance correction for this line, his Fig.~4 would imply a correction of about $+$0.06\,dex.  This line corresponds to a transition between low-lying atomic levels and is thus formed higher than those we employ here.  It may therefore be expected to show stronger NLTE effects than our weaker high-excitation lines. Because \citet{Bruls93} completely neglected inelastic H collisions, his results can probably be taken as an upper limit for possible NLTE effects. We therefore do not expect significant departures from NLTE for our own weak, high-excitation lines, and simply adopt LTE results for \altnii.  Further investigation of NLTE Ni line formation \citep[e.g.][]{Vieytes13} would be welcome however, as this expectation bears additional verification.

\subsubsection{Line selection}
\label{Nilineselection}

From the most accurate $gf$-values available for \altnii, we have retained the 16 weak, unblended lines of Table~\ref{table:lines}. We also included the slightly stronger \mbox{\nii 617.7\,nm} line, because of its pristine appearance in the solar spectrum and the quality of its atomic data. 

Although the situation with \niii is better than for \mnii or \altcoii, most of the \niii lines in the IR are too weak to be useful for abundance purposes, and those in the optical are generally at very short wavelengths and severely blended.  We attempted to use the lines at 340.2, 342.1, 345.4 and \mbox{376.9\,nm}; all are perturbed to some degree, so the scatter in resultant abundances probably reflects both large intrinsic errors in the theoretical $gf$-values and the crowding in this spectral region.

\begin{figure*}[p]
\centering
\begin{minipage}[t]{0.37\textwidth}
\centering
\includegraphics[width=\linewidth]{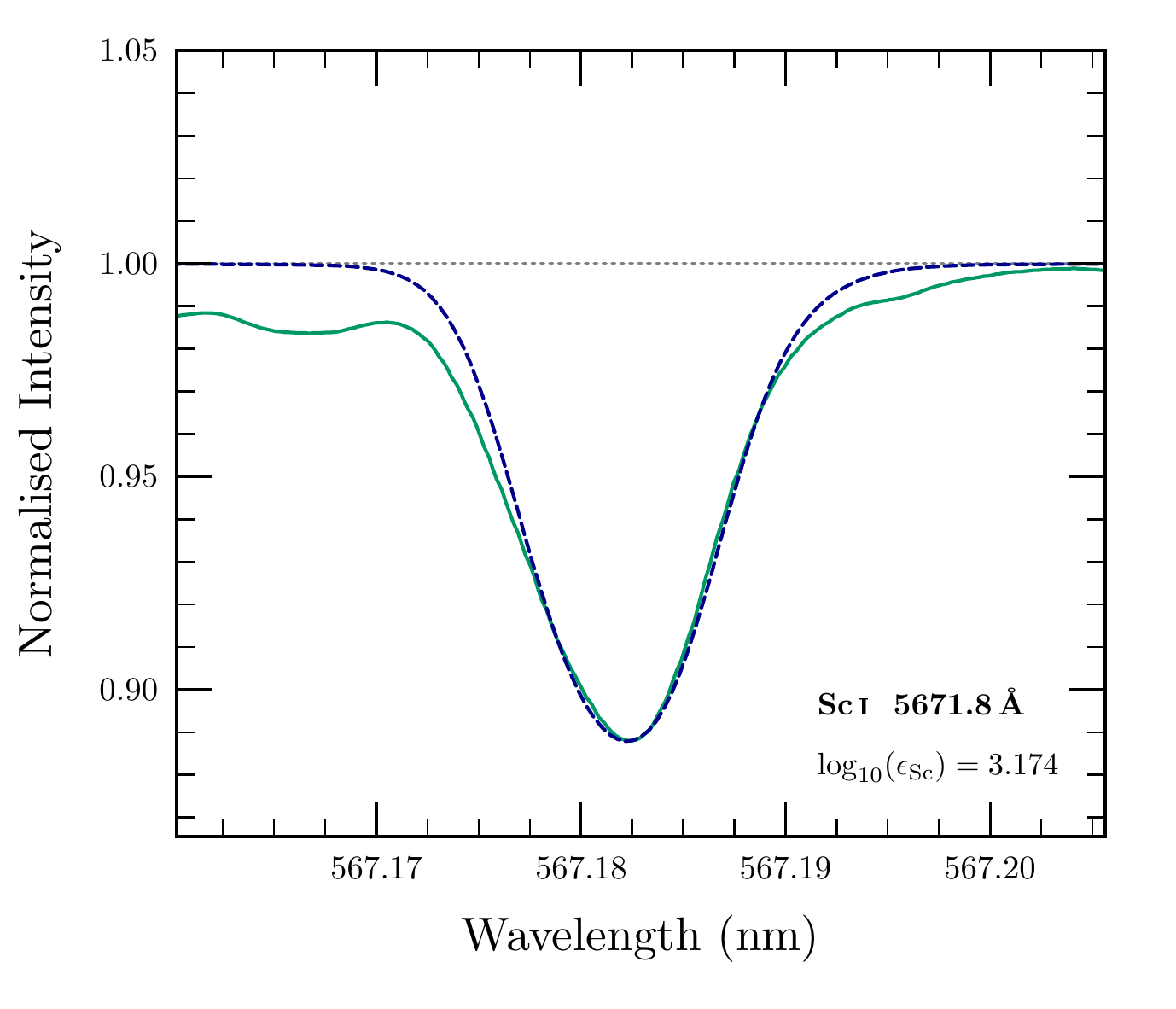}
\includegraphics[width=\linewidth]{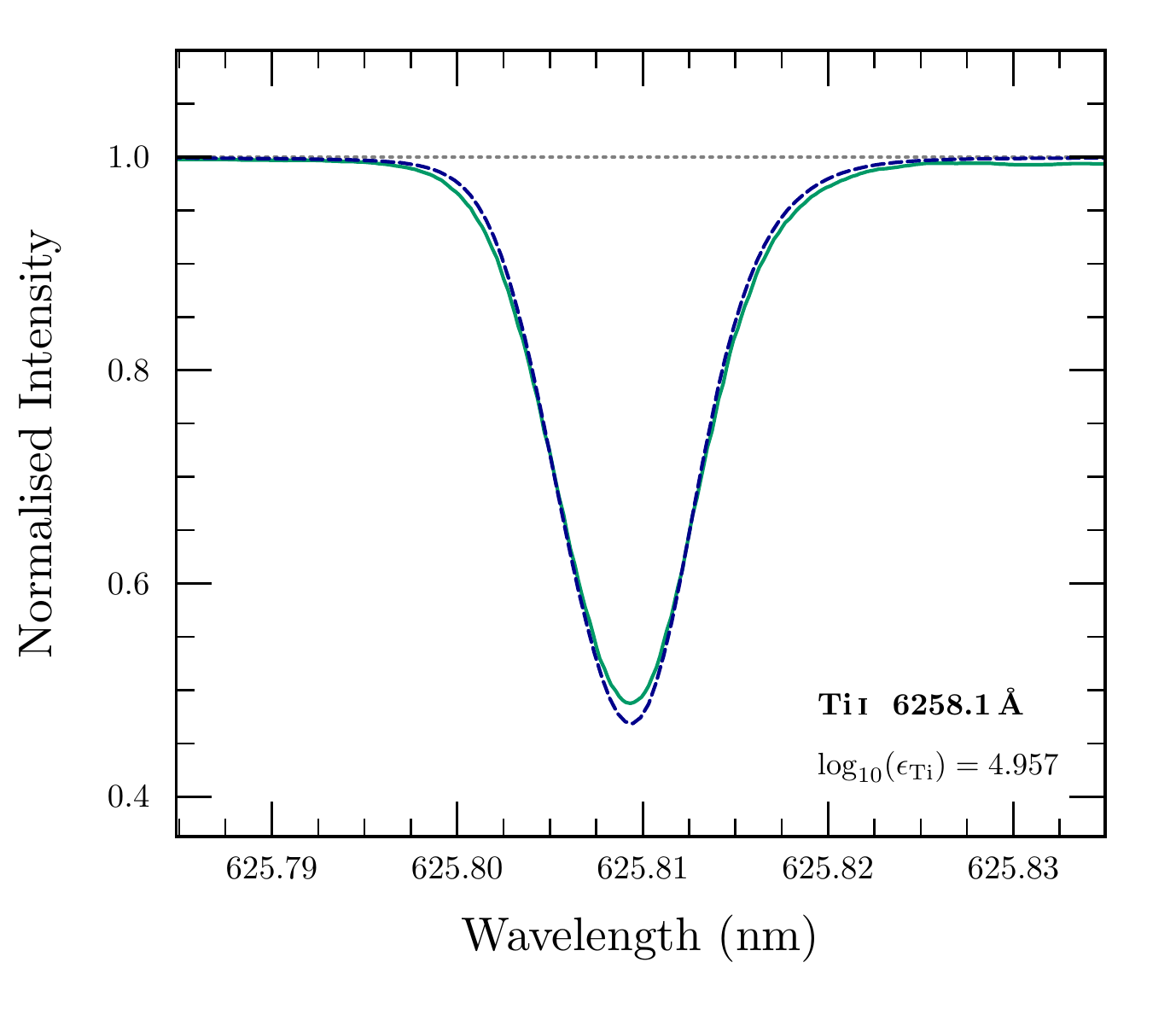}
\includegraphics[width=\linewidth]{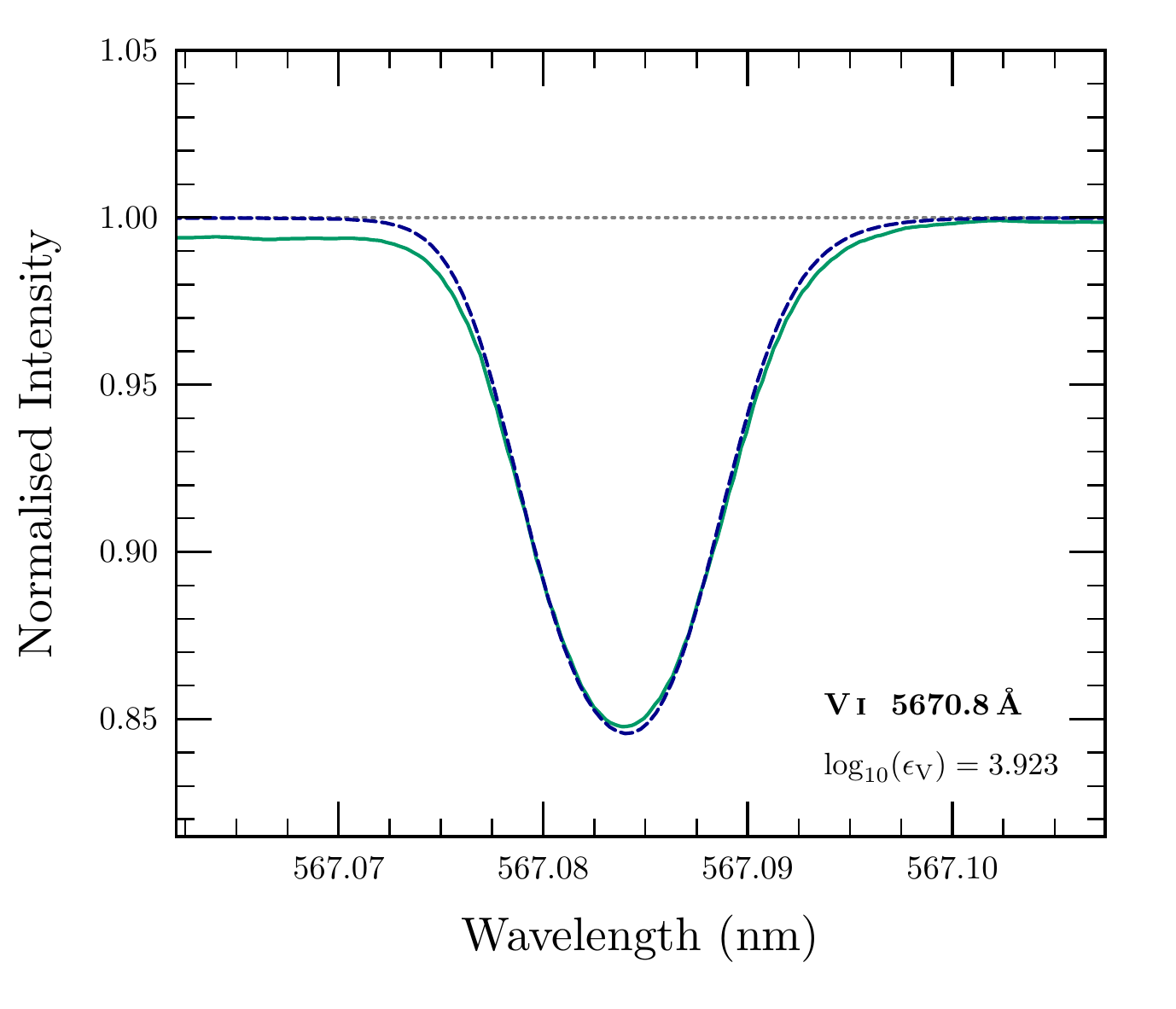}
\includegraphics[width=\linewidth]{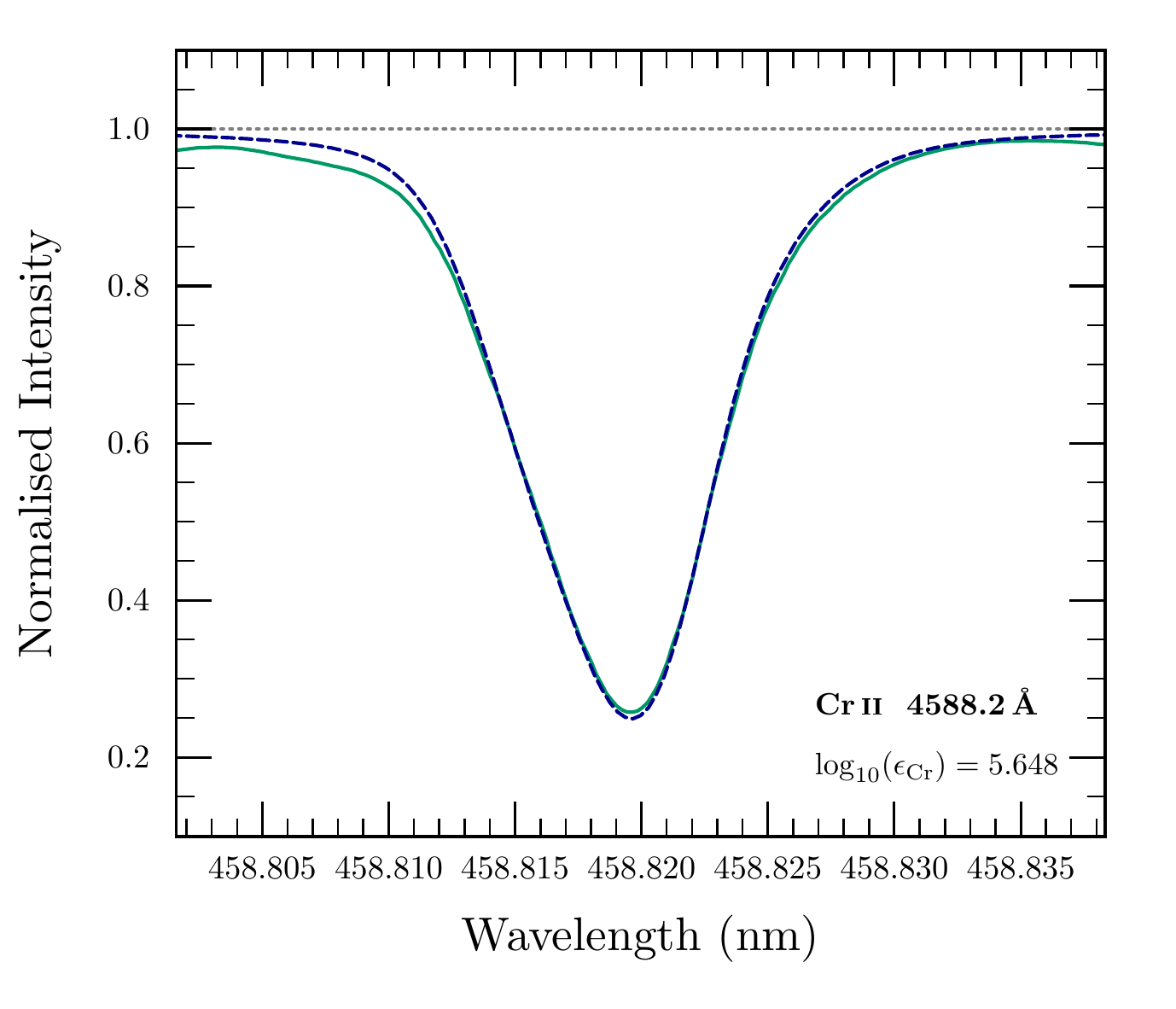}
\end{minipage}
\hspace{0.05\textwidth}
\begin{minipage}[t]{0.37\textwidth}
\centering
\includegraphics[width=\linewidth]{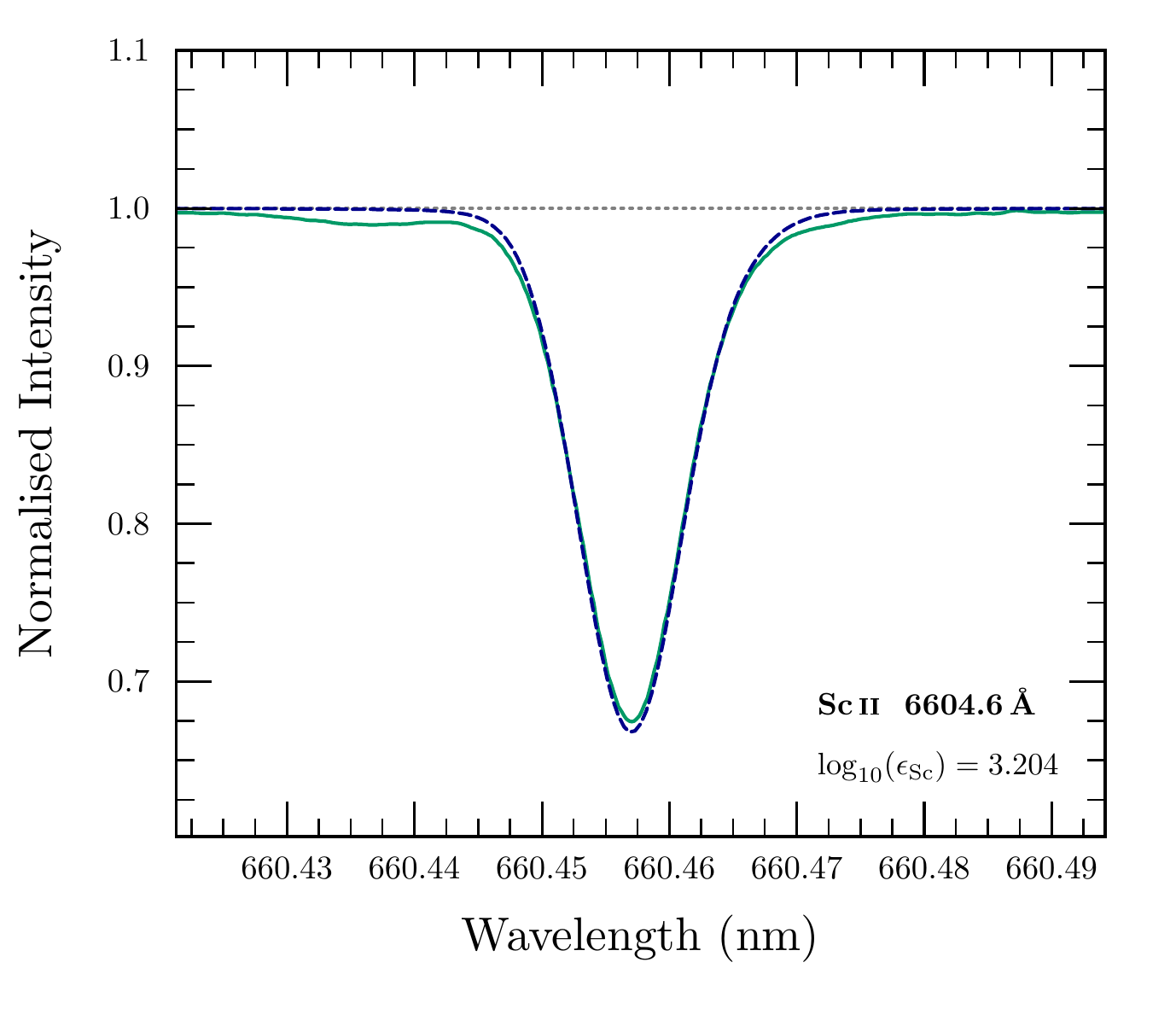}
\includegraphics[width=\linewidth]{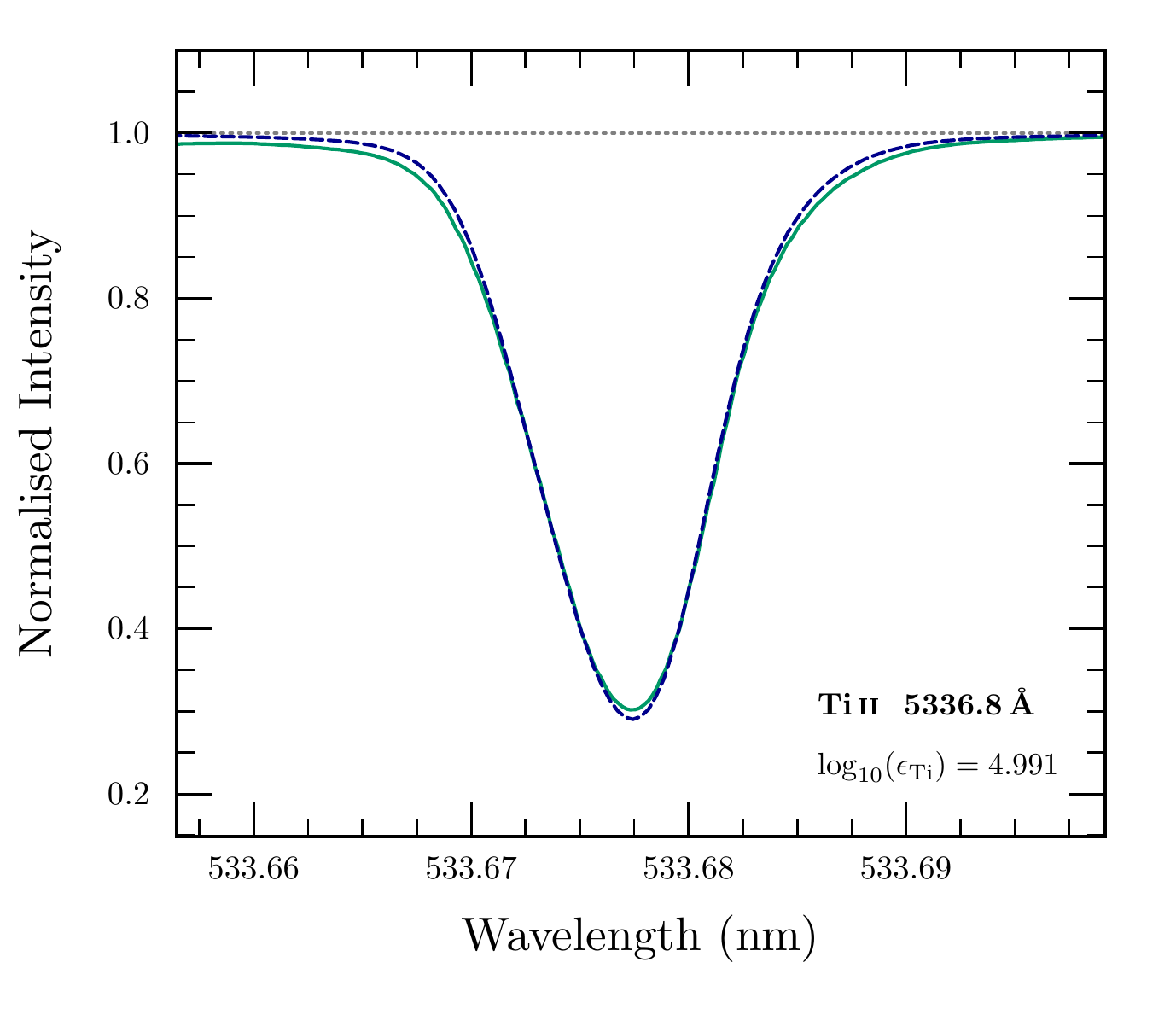}
\includegraphics[width=\linewidth]{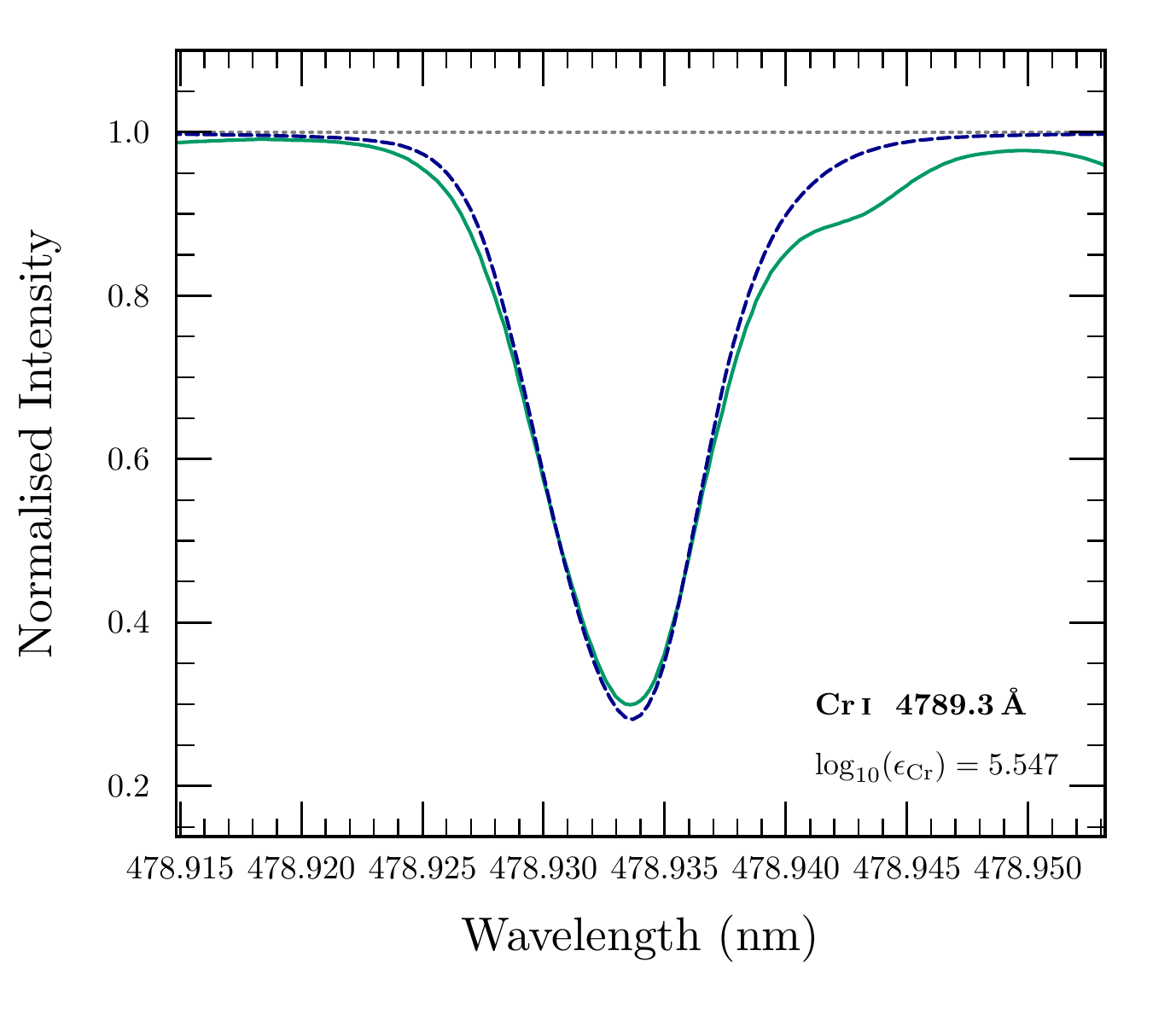}
\includegraphics[width=\linewidth]{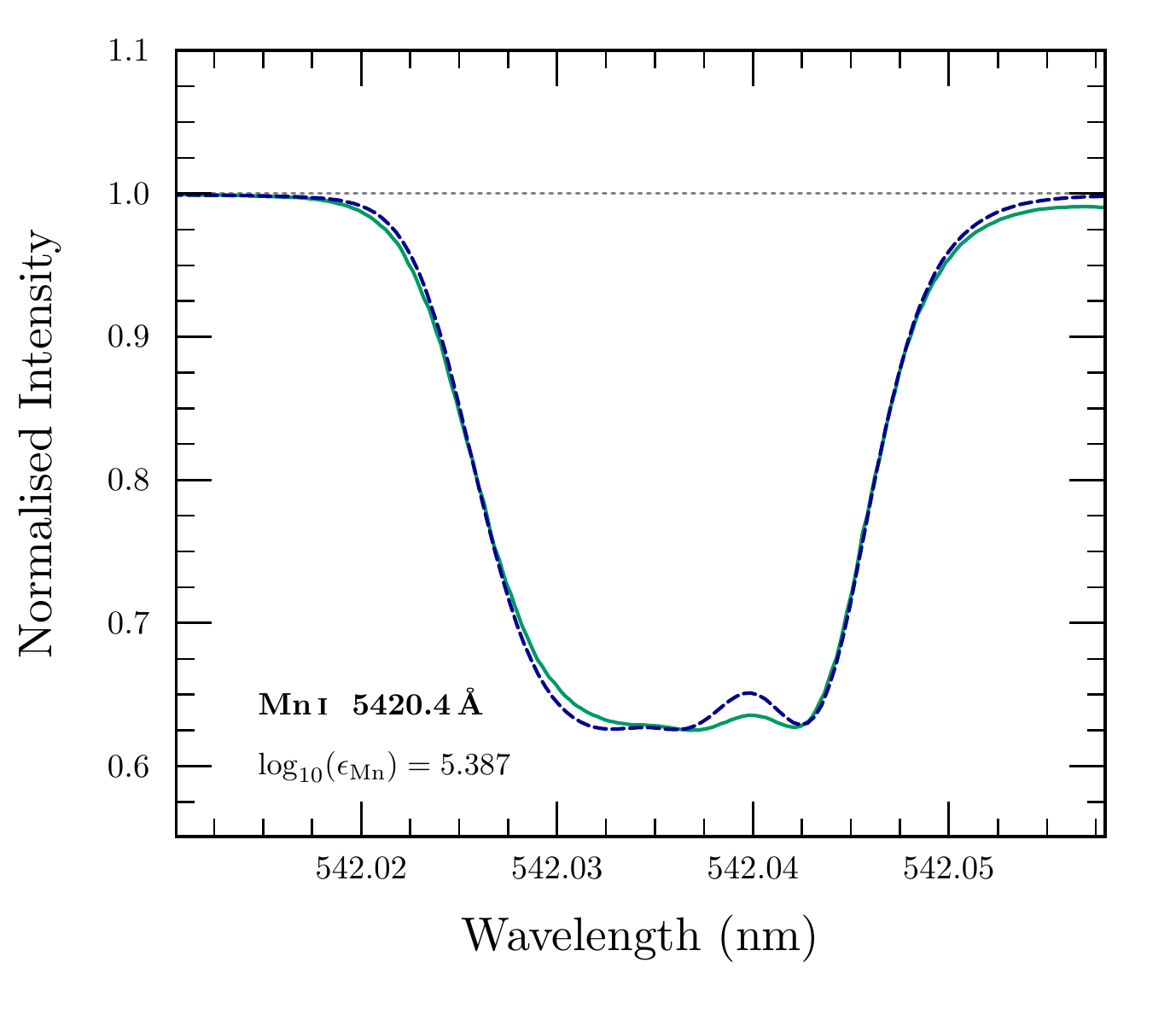}
\end{minipage}
\caption{Example spatially and temporally averaged, disk-centre synthesised \altsci, \altscii, \alttii, \alttiii, \altvi, \altcri, \crii and \mni line profiles (blue dashed), shown in comparison to the observed FTS profile (solid green).  We removed the solar gravitational redshift from the FTS spectrum, convolved the synthesised profile with an instrumental sinc function and fitted it in abundance.  Wavelengths and continuum placements have been adjusted for display purposes.  Plotted profiles are computed in LTE, but quoted abundances in each panel include NLTE corrections computed in 1D (wherever available).}
\label{fig:profiles}
\end{figure*}

\begin{figure*}
\centering
\begin{minipage}[t]{0.39\textwidth}
\centering
\includegraphics[width=\linewidth]{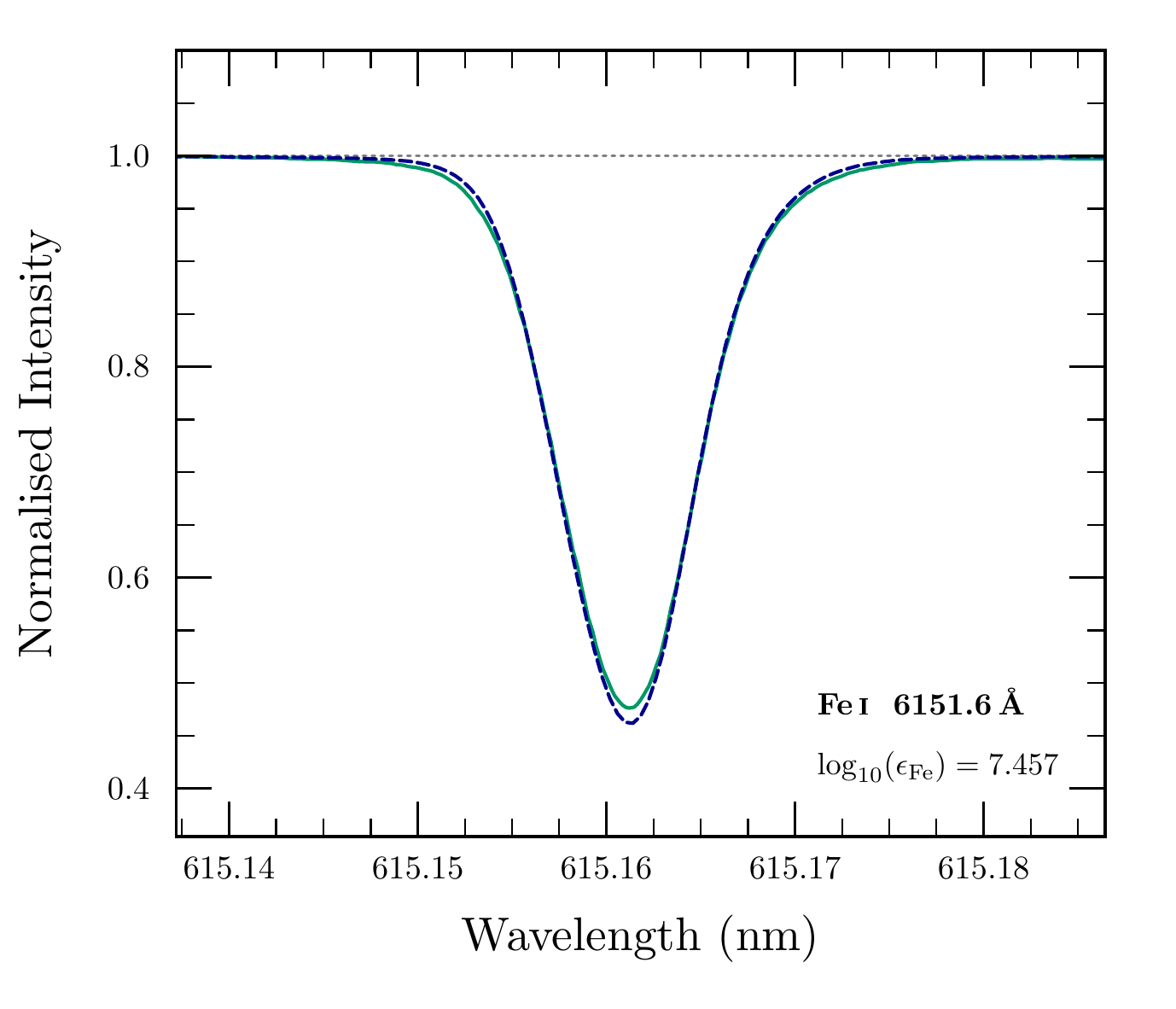}
\includegraphics[width=\linewidth]{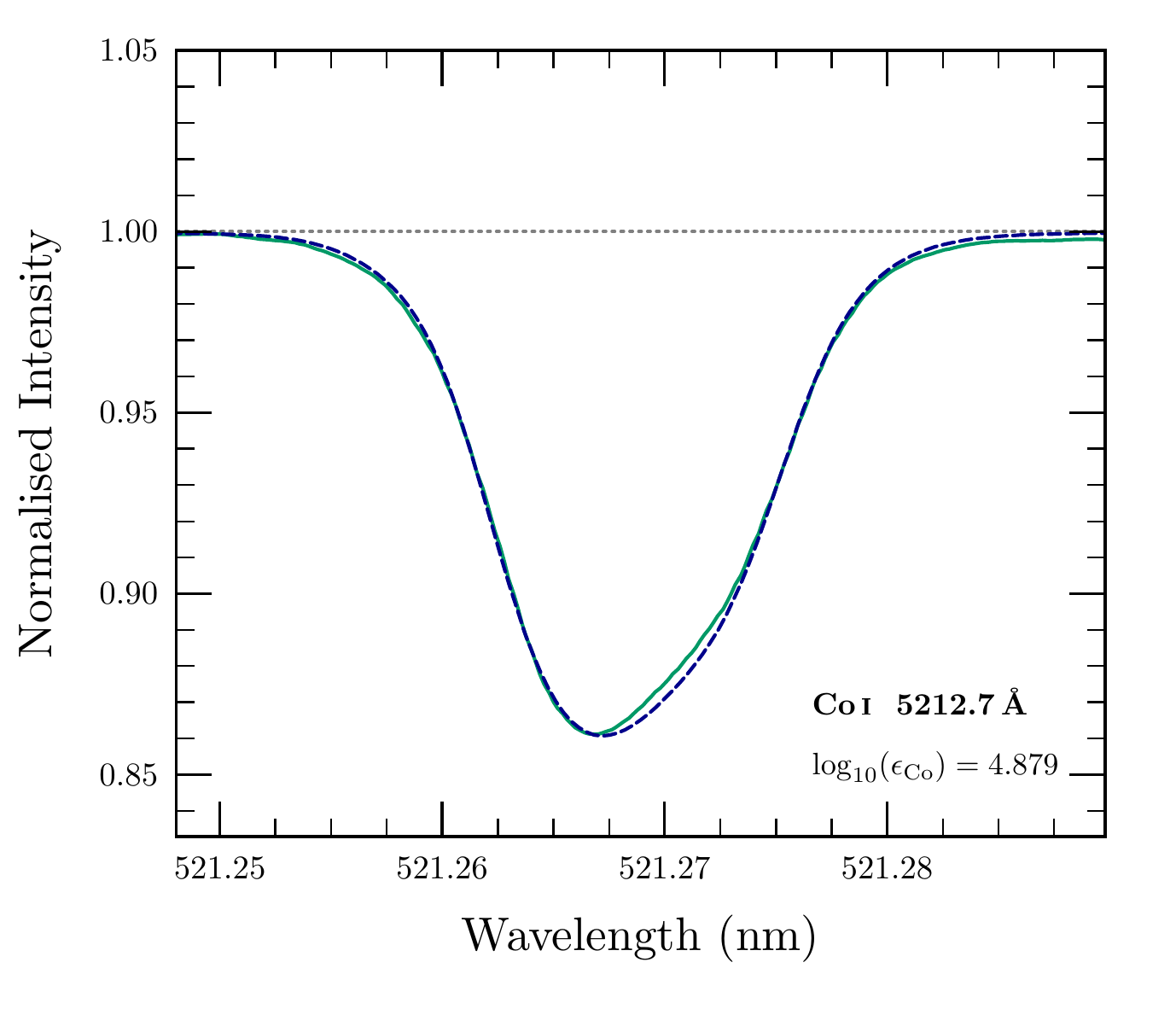}
\end{minipage}
\hspace{0.05\textwidth}
\begin{minipage}[t]{0.39\textwidth}
\centering
\includegraphics[width=\linewidth]{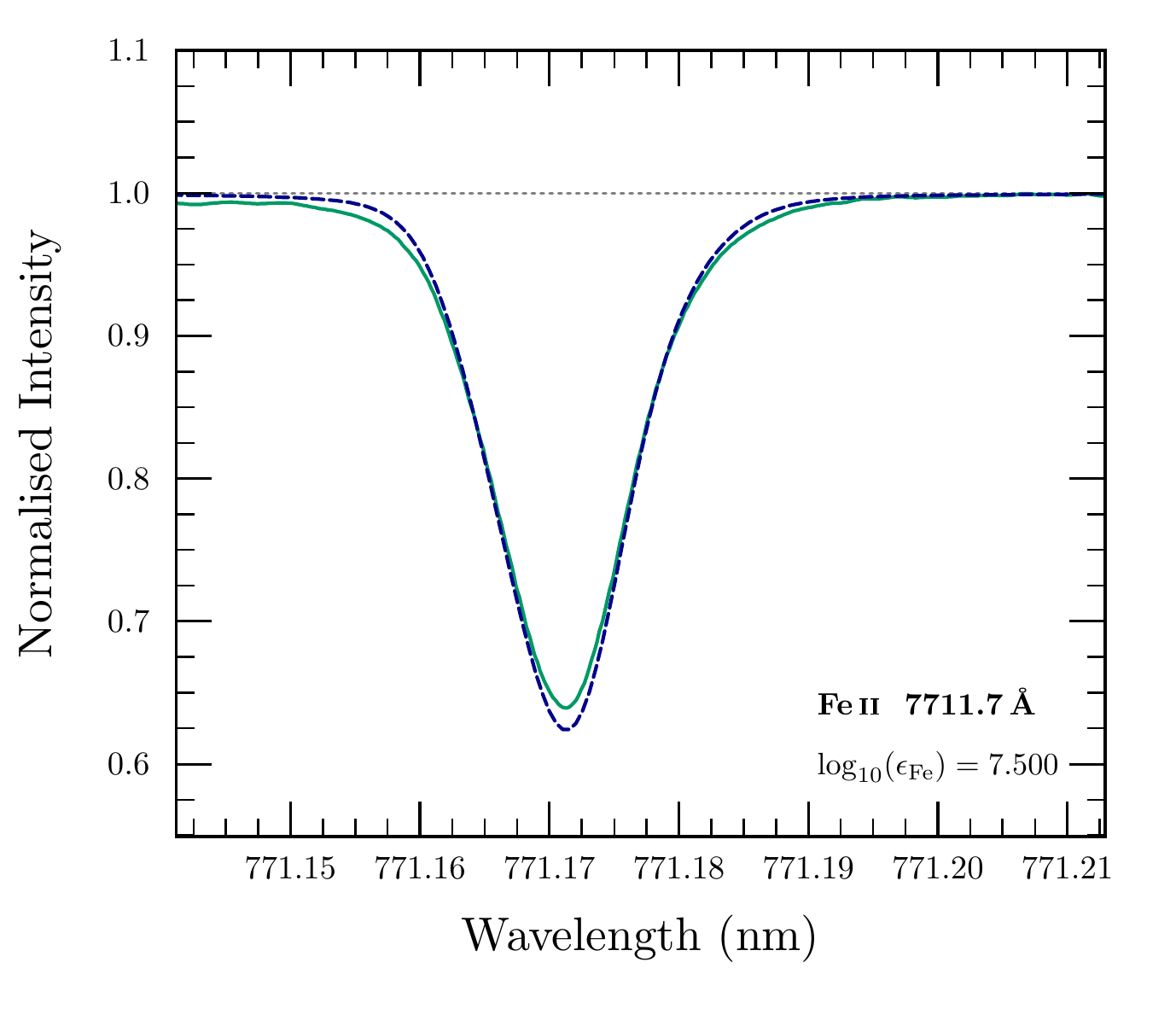}
\includegraphics[width=\linewidth]{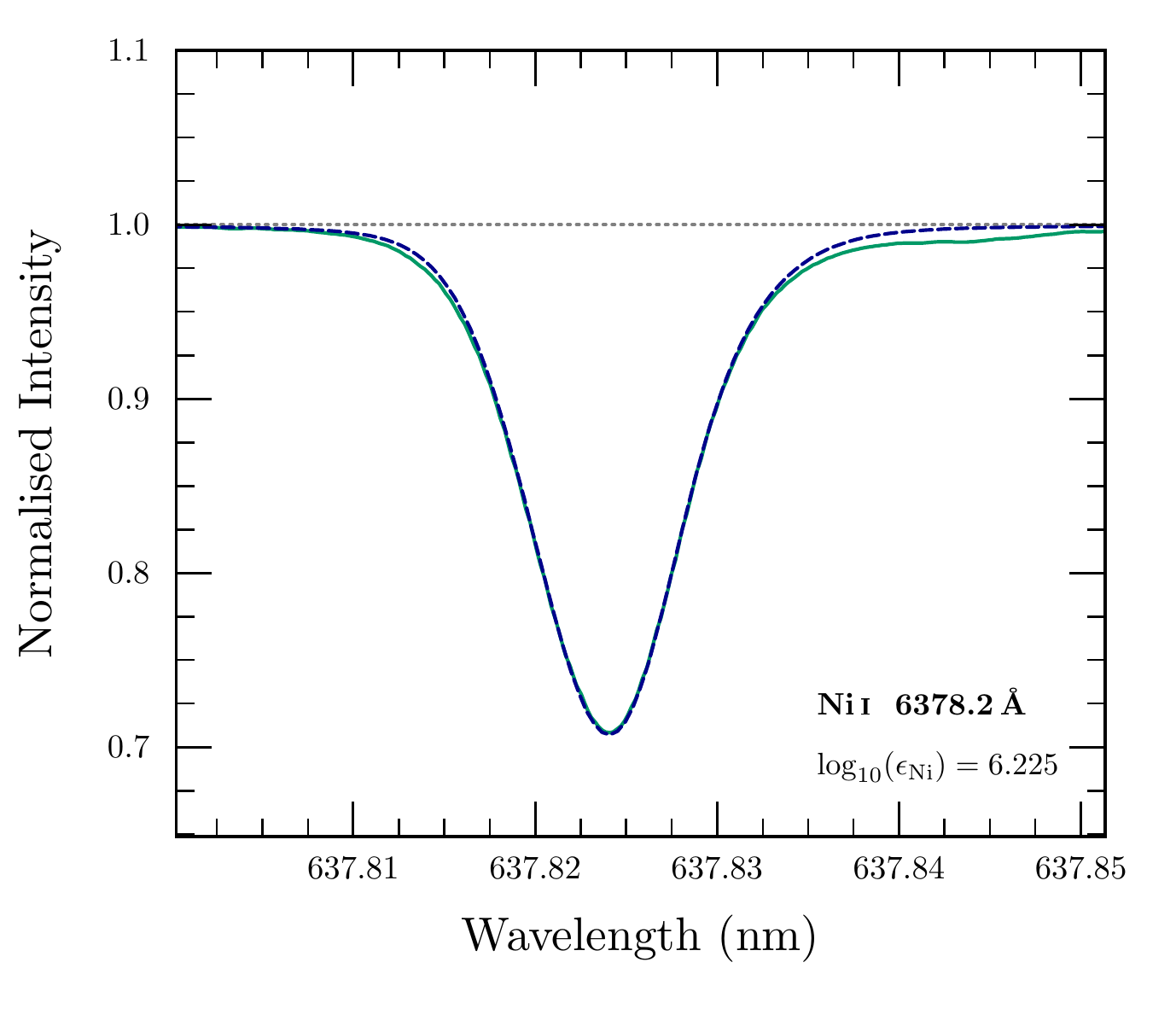}
\end{minipage}
\caption{As per Fig.~\protect\ref{fig:profiles}, but for \altfei, \altfeii, \coi and \altnii.}
\label{fig:profiles2}
\end{figure*}

\begin{figure*}
\centering
\begin{minipage}[t]{0.4\textwidth}
\centering
\includegraphics[width=\linewidth]{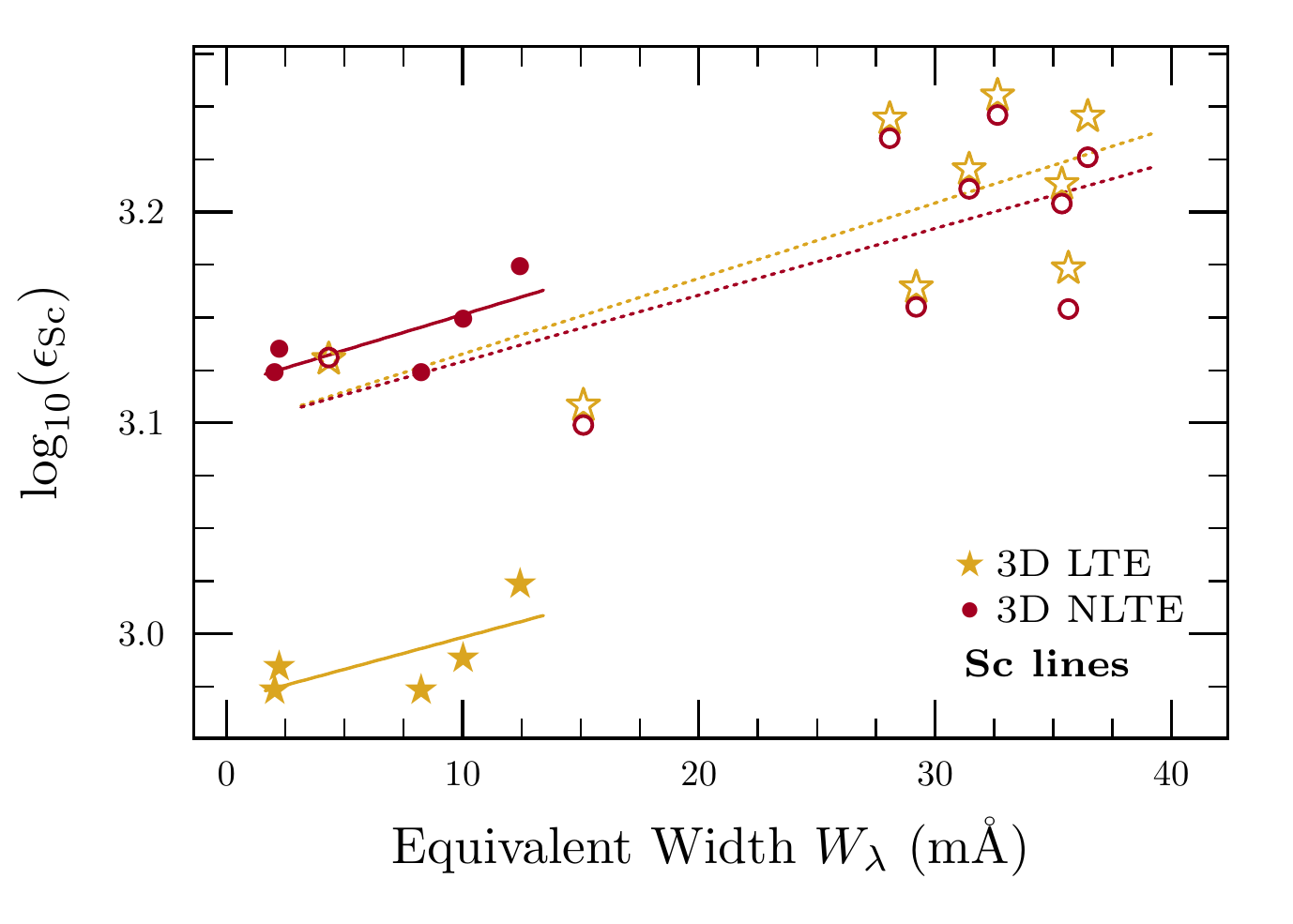}
\includegraphics[width=\linewidth]{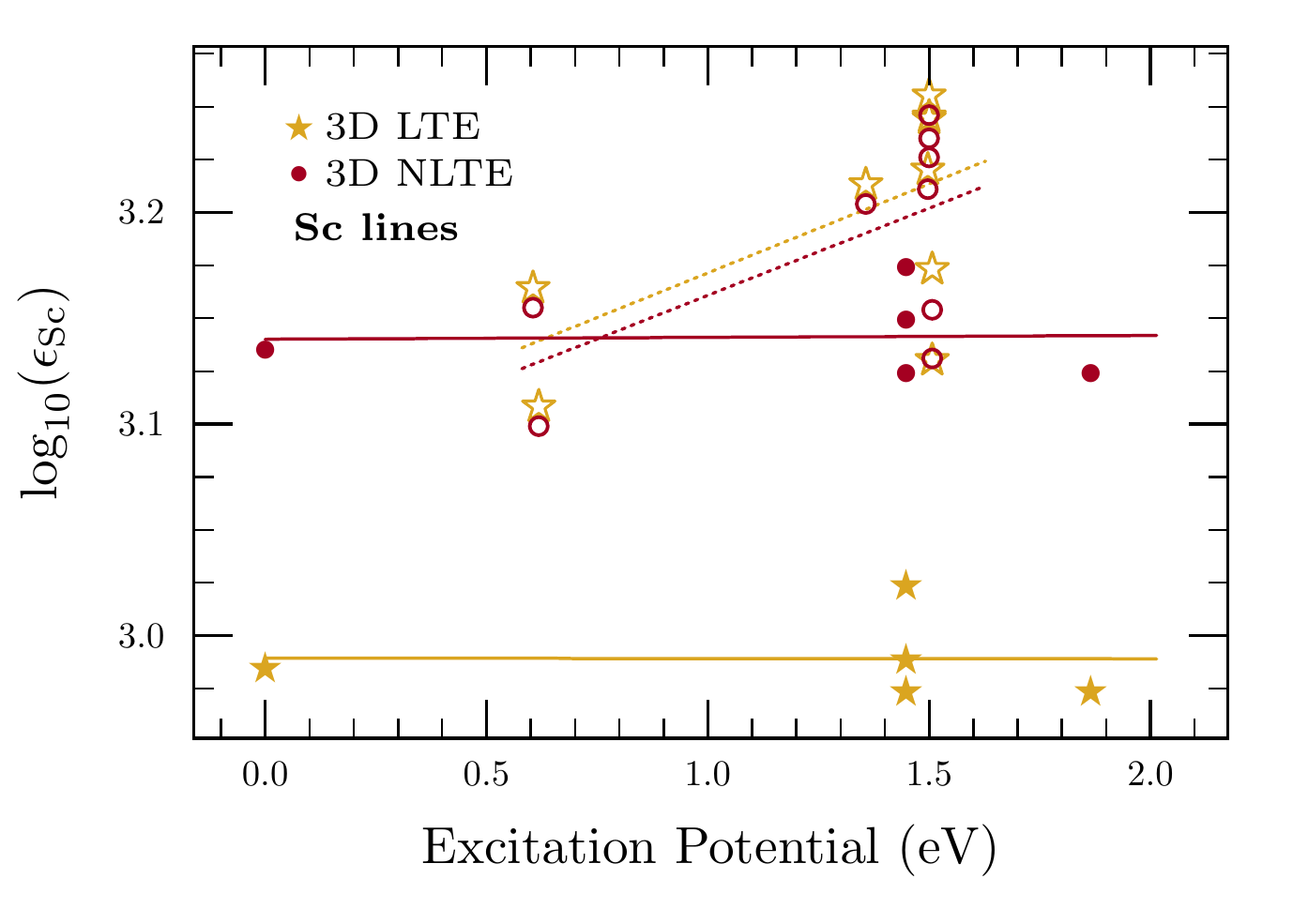}
\end{minipage}
\hspace{0.05\textwidth}
\begin{minipage}[t]{0.4\textwidth}
\centering
\includegraphics[width=\linewidth]{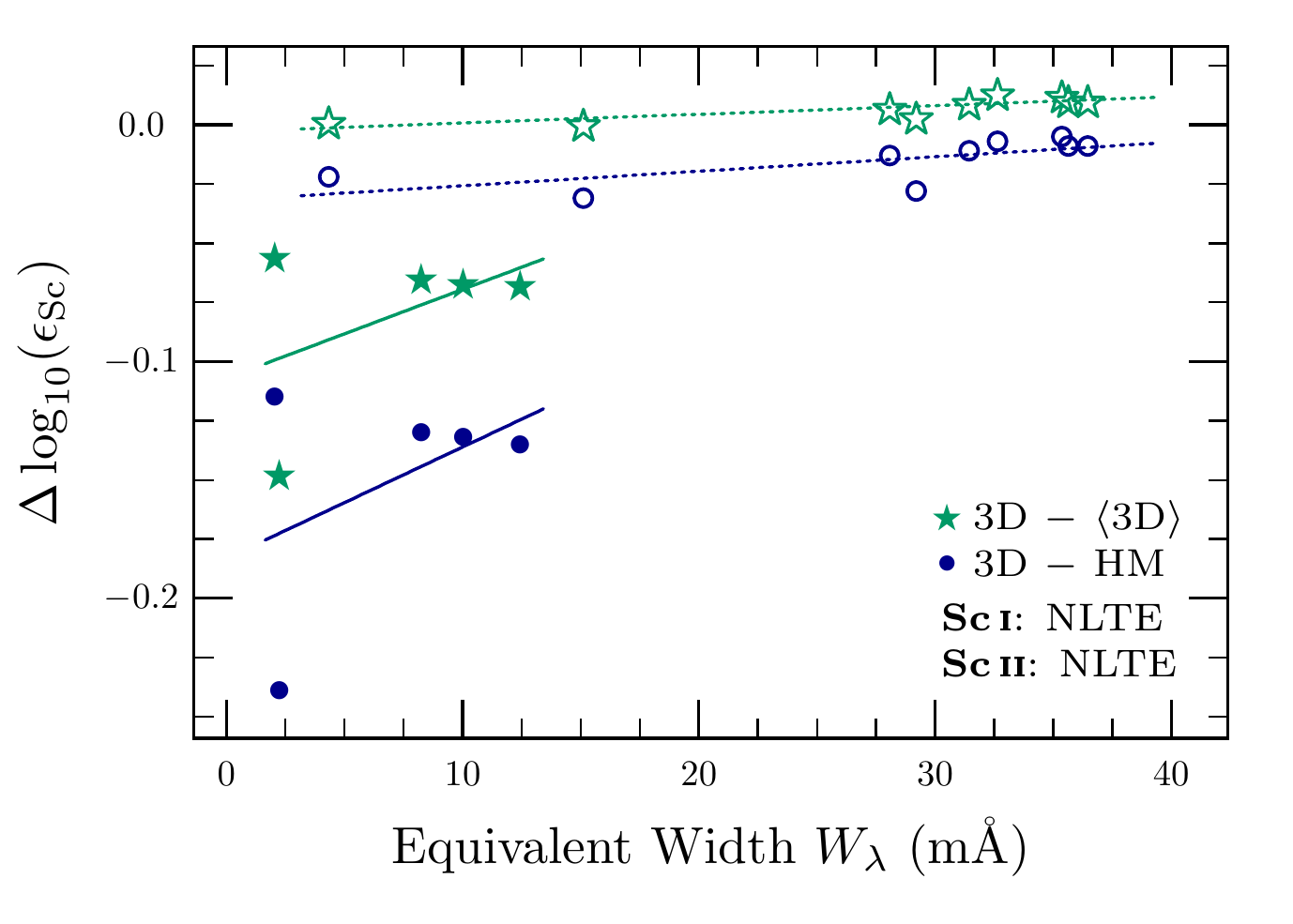}
\includegraphics[width=\linewidth]{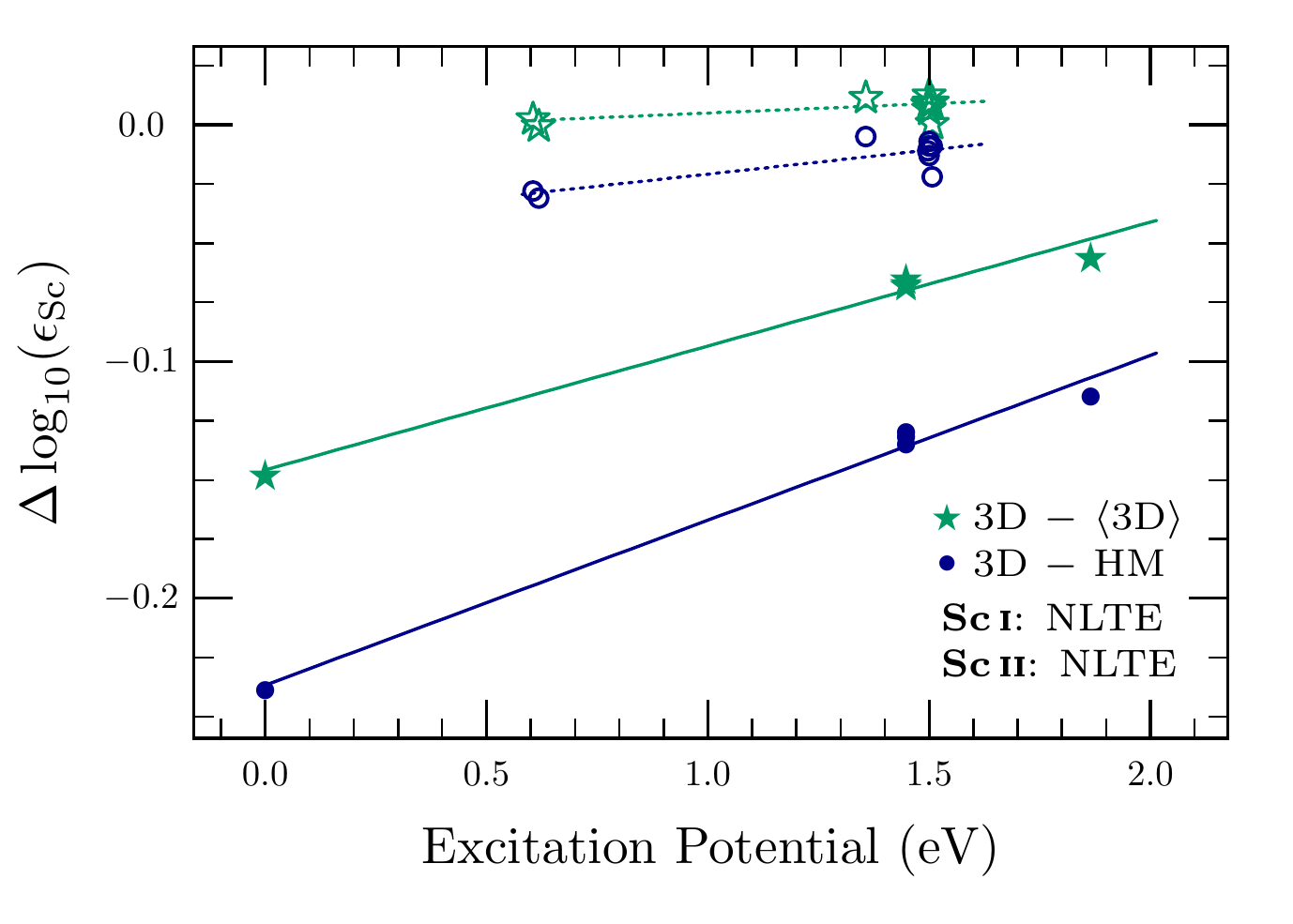}
\end{minipage}
\caption{\textit{Left}: Sc abundances derived from \scani and \scanii lines with the 3D model, shown as a function of equivalent width and lower excitation potential.  \textit{Right}: Line-by-line differences between Sc abundances obtained with the 3D and \oneDAV\ models, and between those obtained with the 3D and \citetalias{HM} models.  Filled symbols and solid trendlines indicate lines of the neutral species (\altsci), whereas open symbols and dotted lines indicate singly-ionised (\altscii) lines.  Trendlines give equal weight to each line (unlike our mean abundances, where we give larger weights to better lines).}
\label{fig:sc}
\end{figure*}

\begin{figure*}
\centering
\begin{minipage}[t]{0.4\textwidth}
\centering
\includegraphics[width=\linewidth]{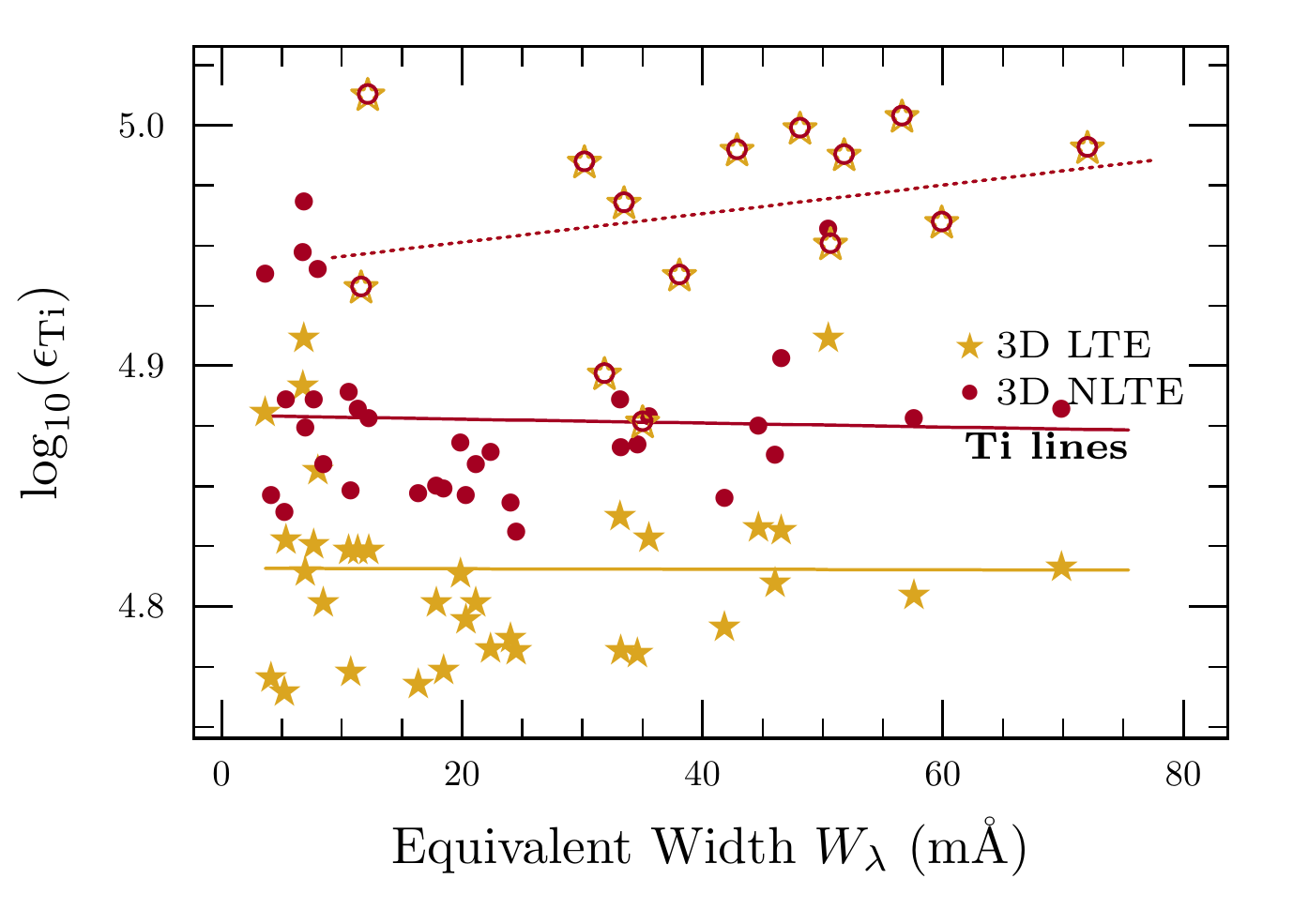}
\includegraphics[width=\linewidth]{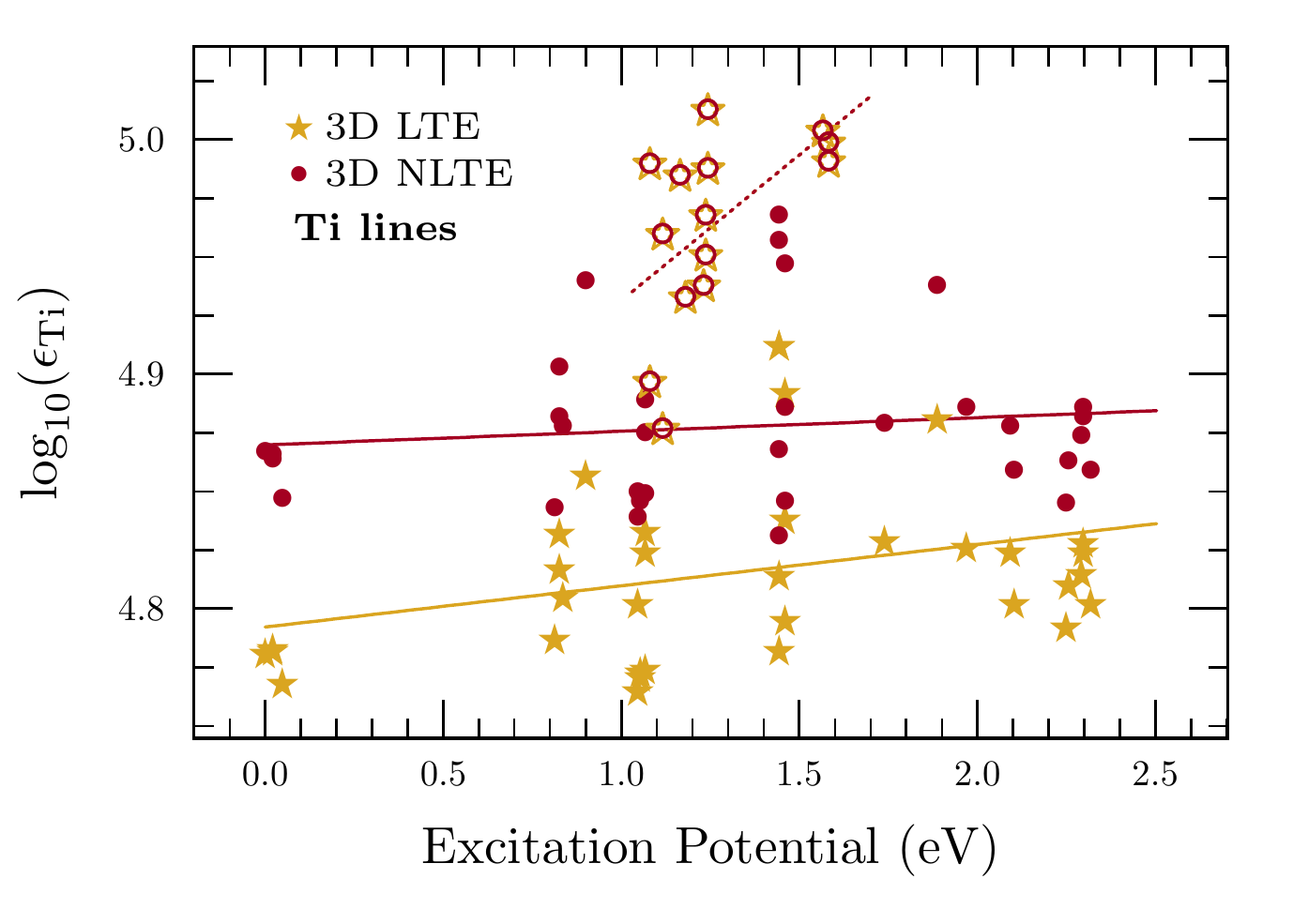}
\end{minipage}
\hspace{0.05\textwidth}
\begin{minipage}[t]{0.4\textwidth}
\centering
\includegraphics[width=\linewidth]{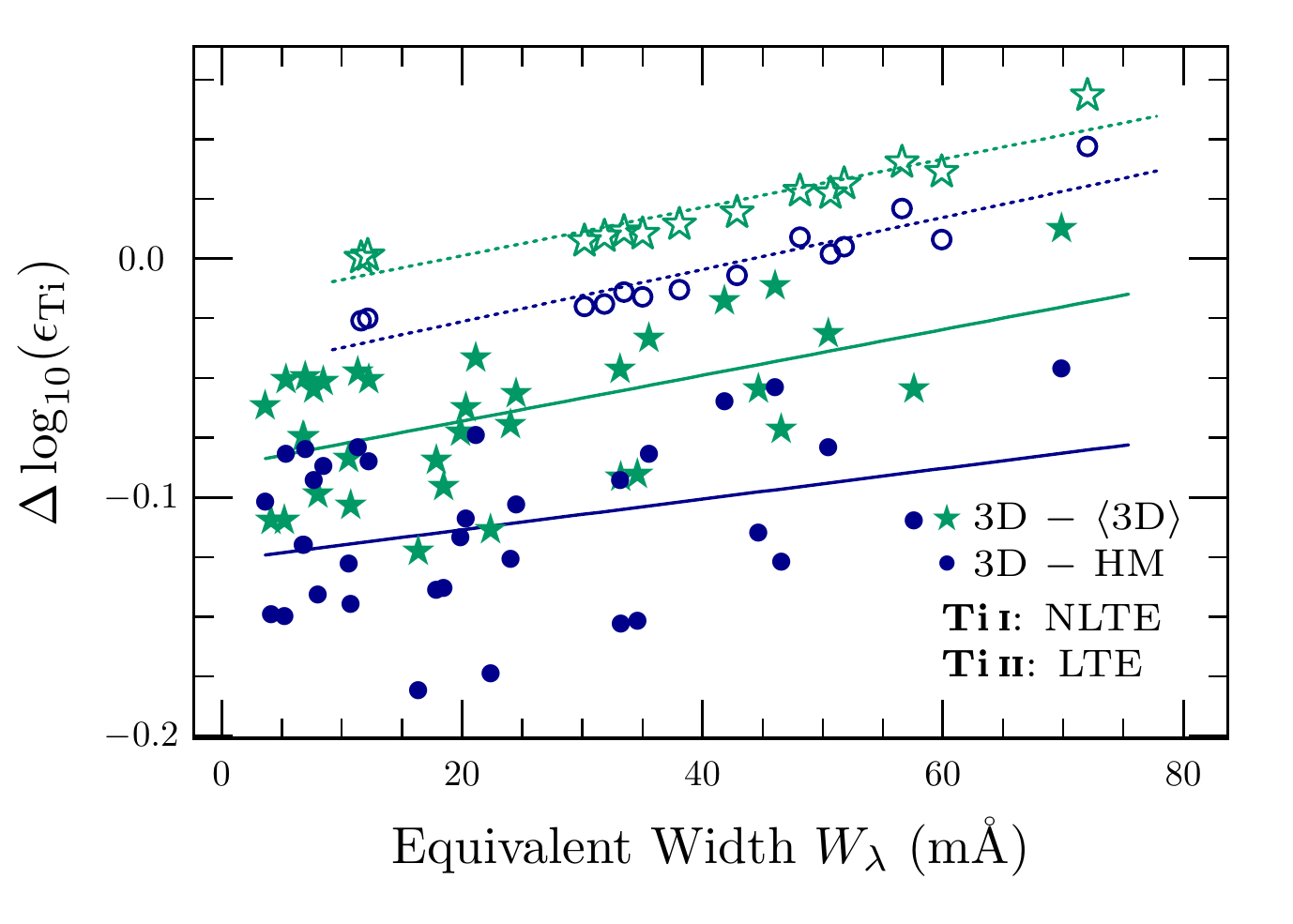}
\includegraphics[width=\linewidth]{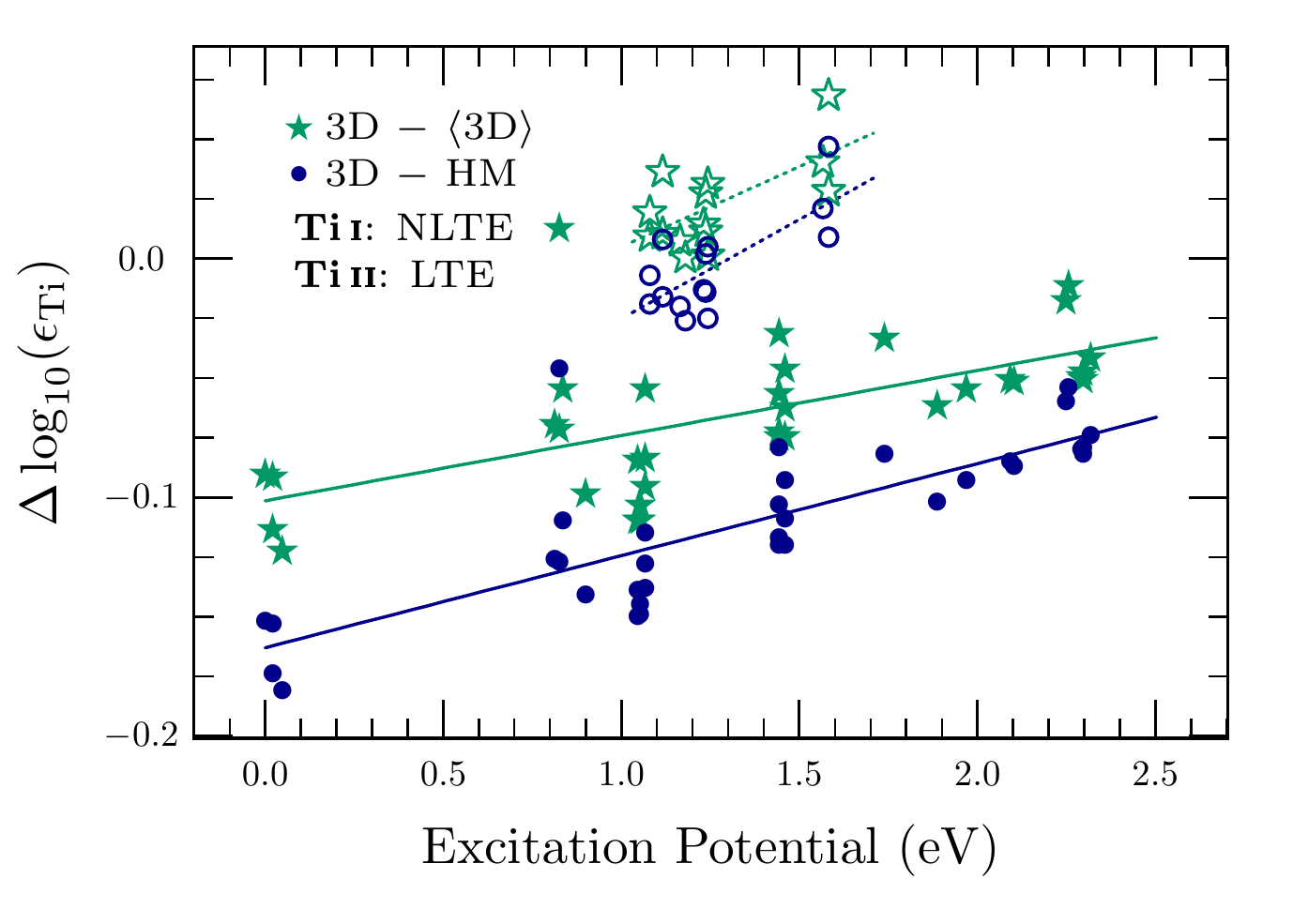}
\end{minipage}
\caption{\textit{Left}: 3D Ti abundances from \tii and \tiii lines, as a function of equivalent width and lower excitation potential.  \textit{Right}: Line-by-line differences between abundances obtained with the 3D and \oneDAV\ models, and between those obtained with the 3D and \citetalias{HM} models.  Filled symbols and solid trendlines indicate neutral lines, open symbols and dotted lines indicate singly-ionised lines.}
\label{fig:ti}
\end{figure*}

\begin{figure*}
\centering
\begin{minipage}[t]{0.4\textwidth}
\centering
\includegraphics[width=\linewidth]{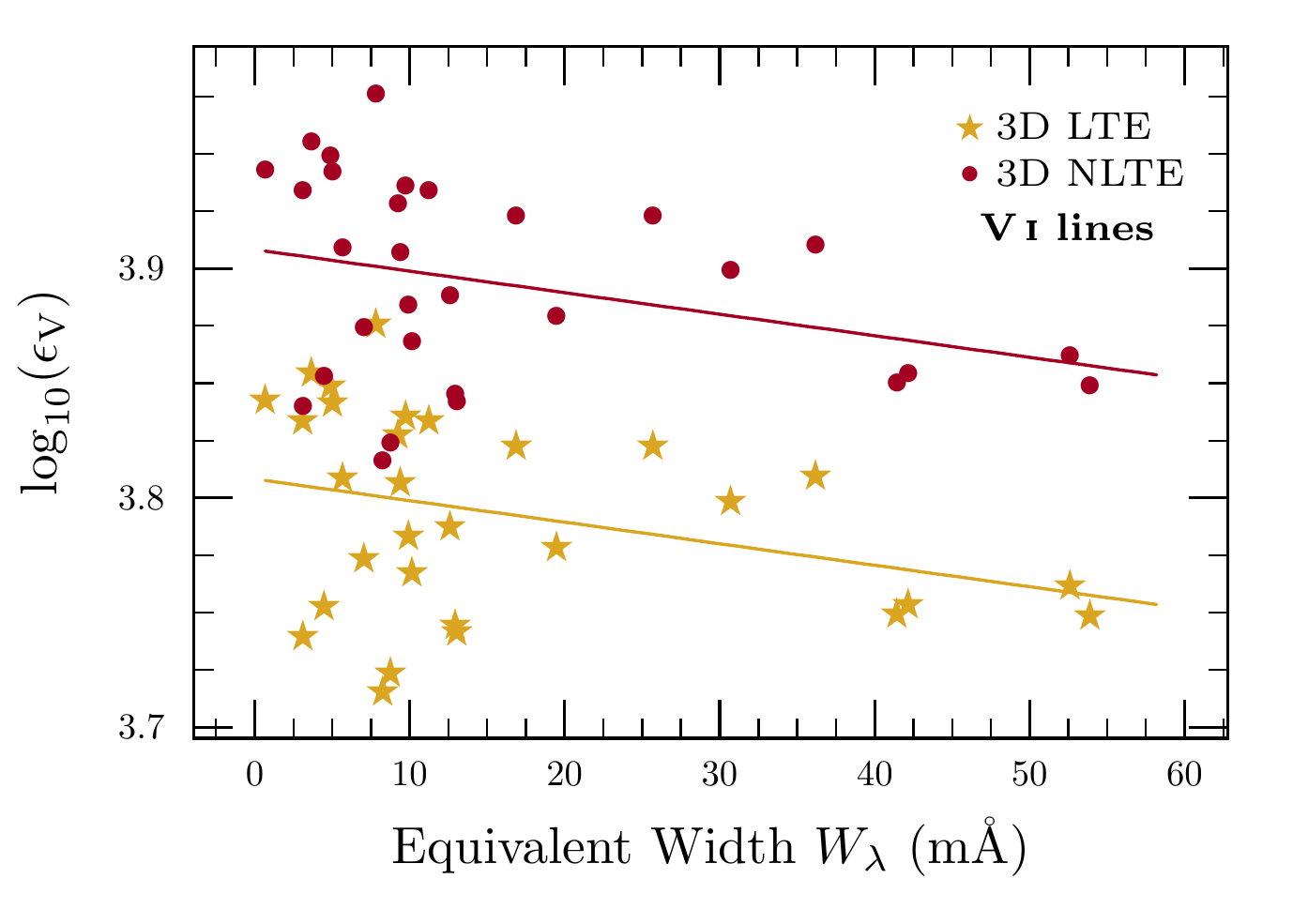}
\includegraphics[width=\linewidth]{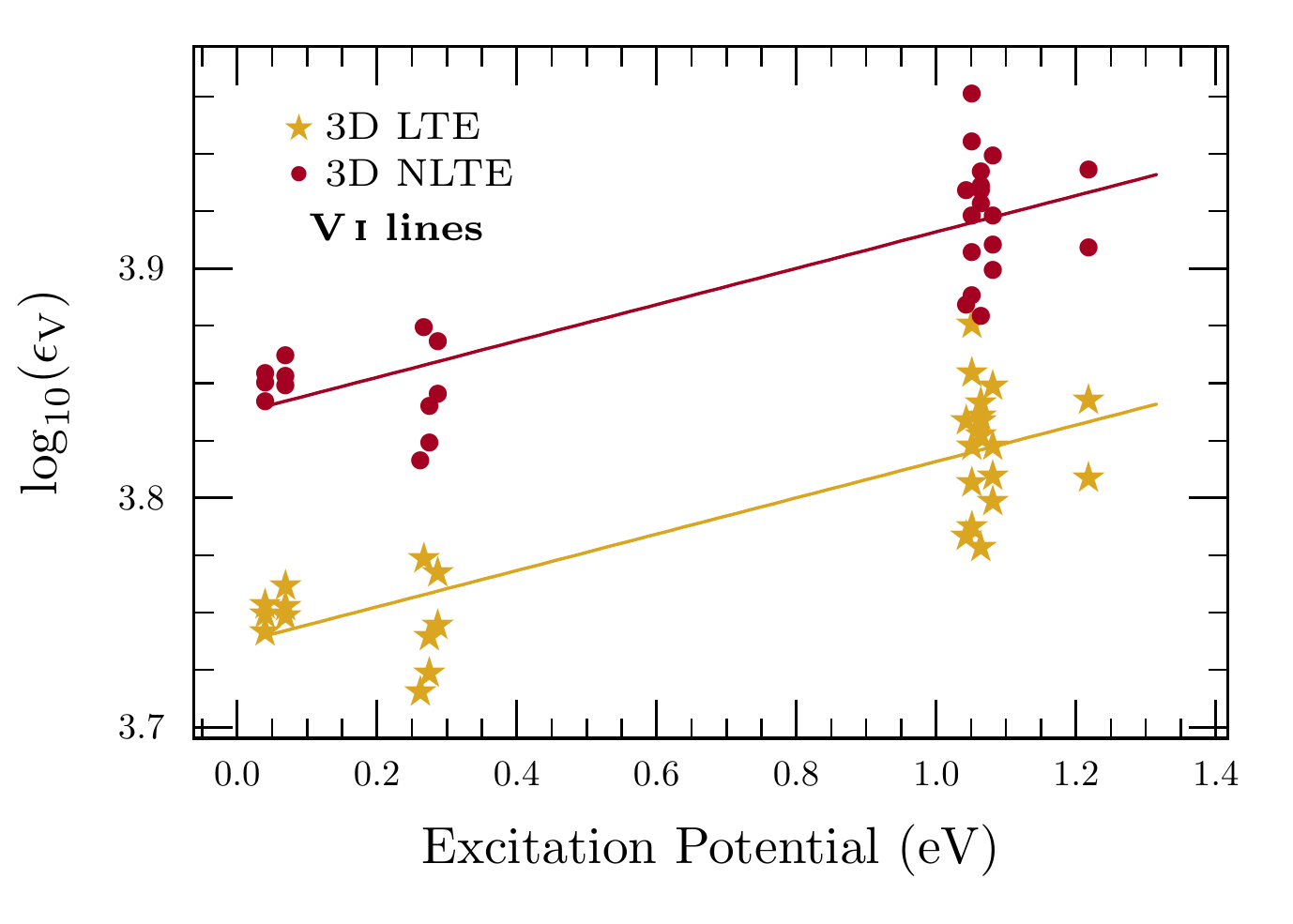}
\end{minipage}
\hspace{0.05\textwidth}
\begin{minipage}[t]{0.4\textwidth}
\centering
\includegraphics[width=\linewidth]{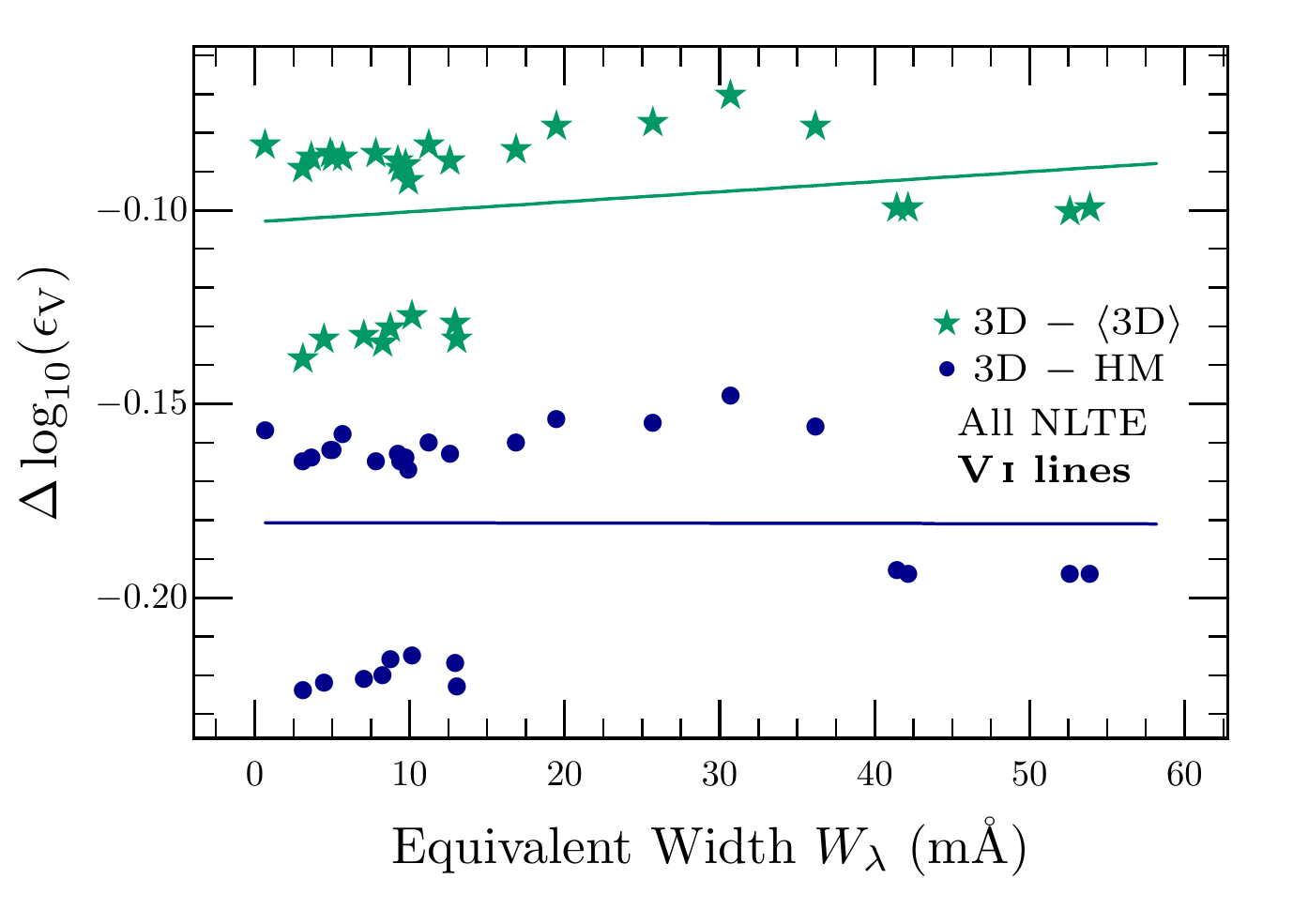}
\includegraphics[width=\linewidth]{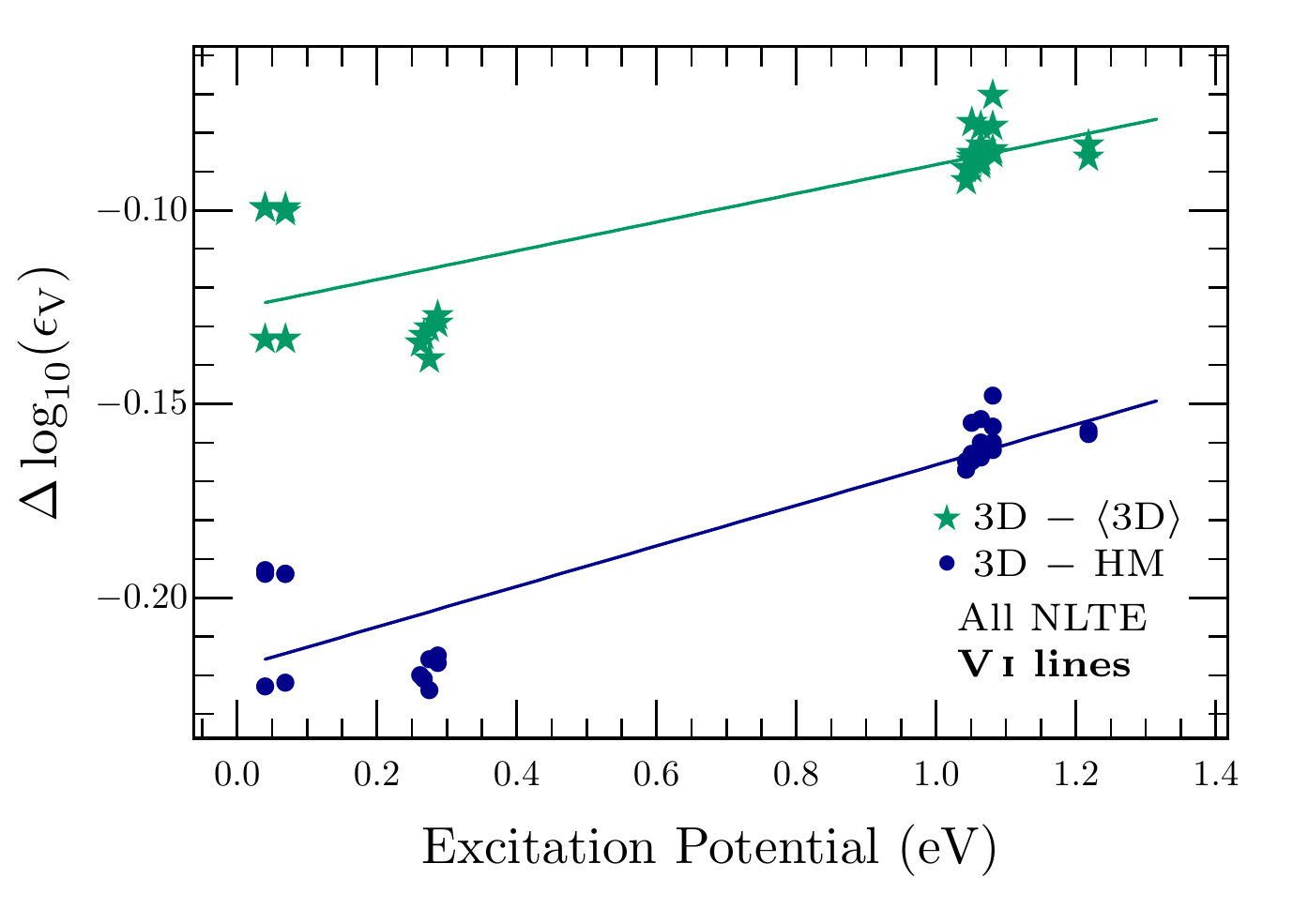}
\end{minipage}
\caption{\textit{Left}: 3D V abundances from \vi lines, as a function of equivalent width and lower excitation potential.  \textit{Right}: Line-by-line differences between abundances obtained with the 3D and \oneDAV\ models, and between those obtained with the 3D and \citetalias{HM} models.}
\label{fig:v}
\end{figure*}

\begin{figure*}[t]
\centering
\begin{minipage}[t]{0.4\textwidth}
\centering
\includegraphics[width=\linewidth]{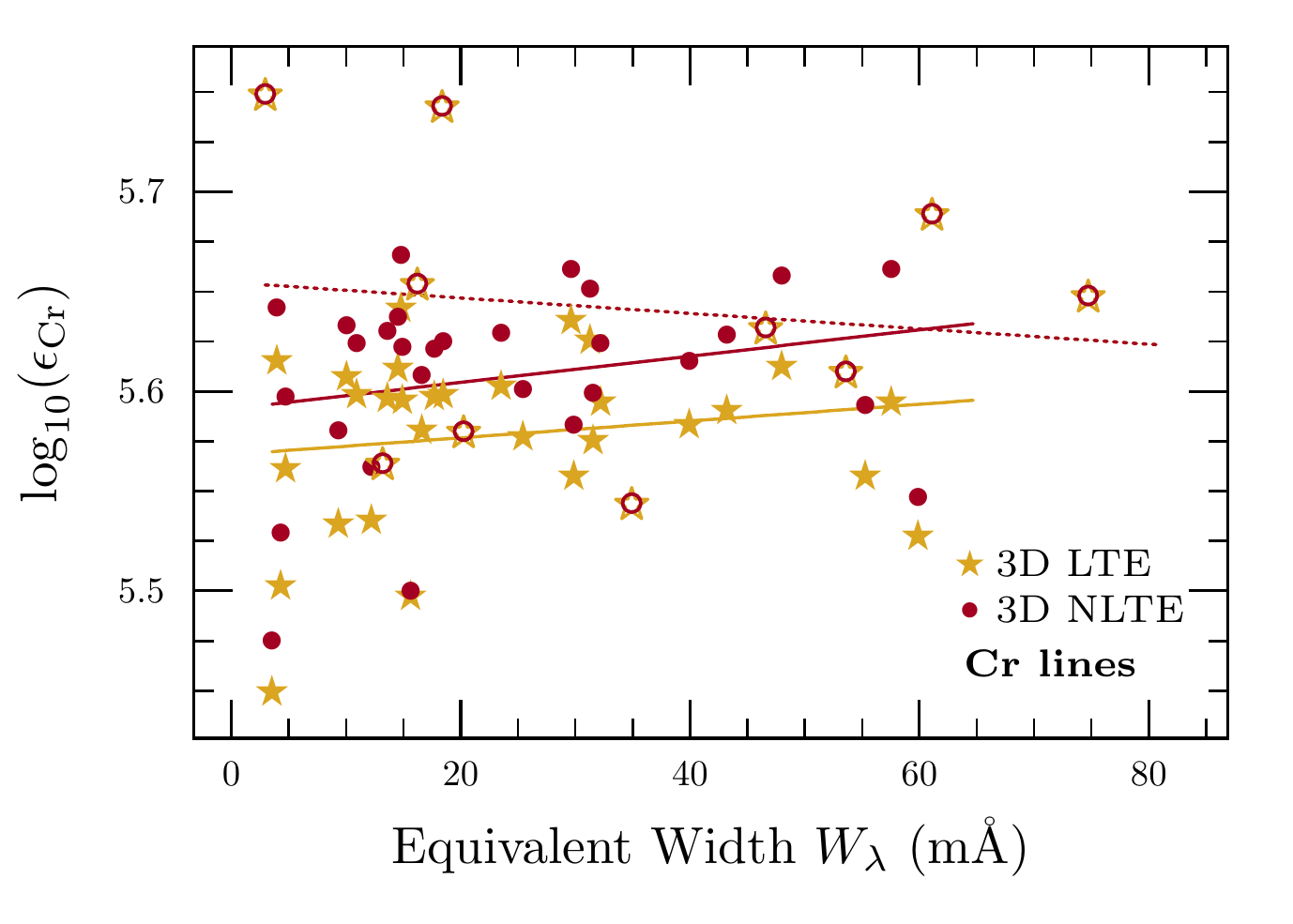}
\includegraphics[width=\linewidth]{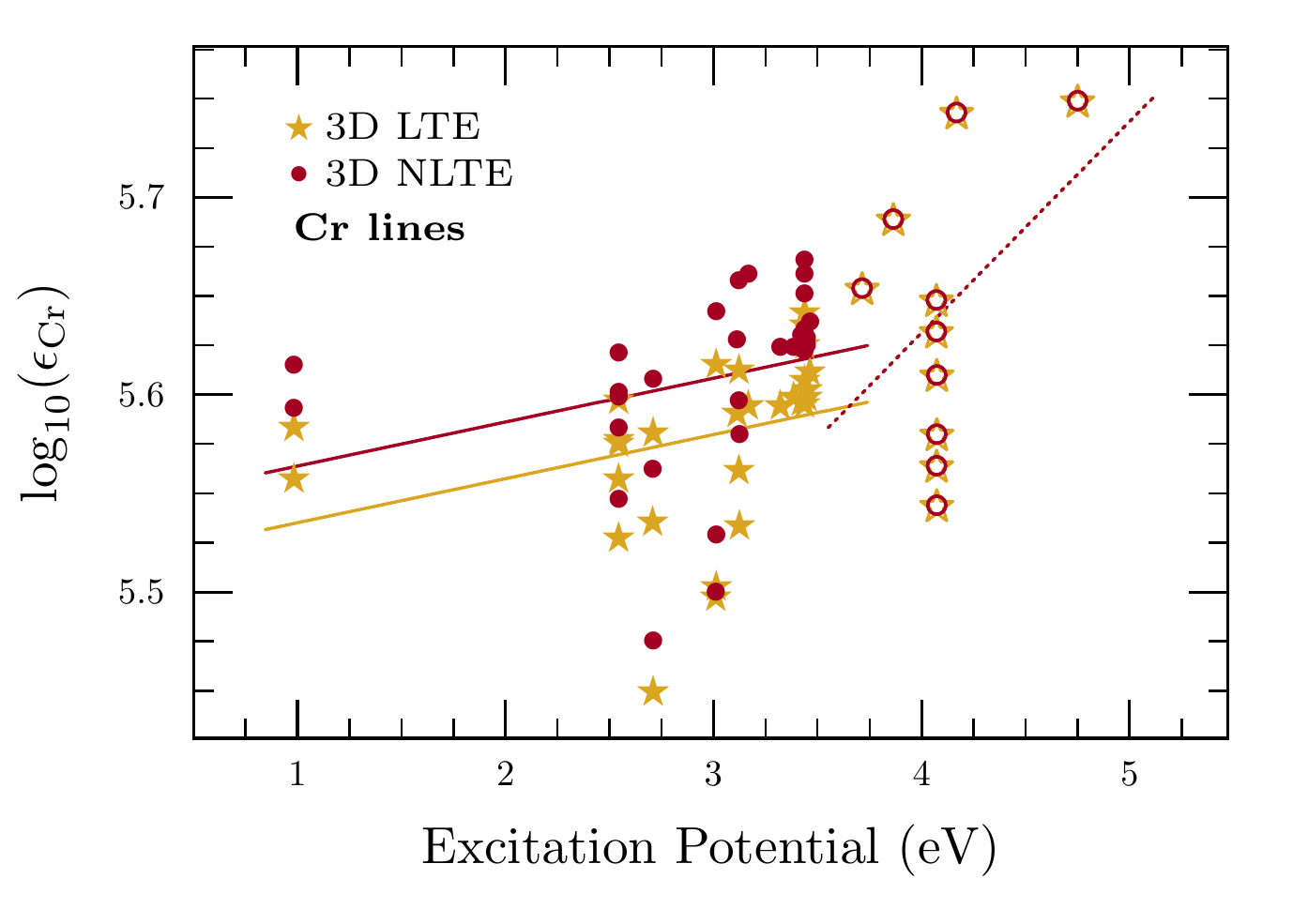}
\end{minipage}
\hspace{0.05\textwidth}
\begin{minipage}[t]{0.4\textwidth}
\centering
\includegraphics[width=\linewidth]{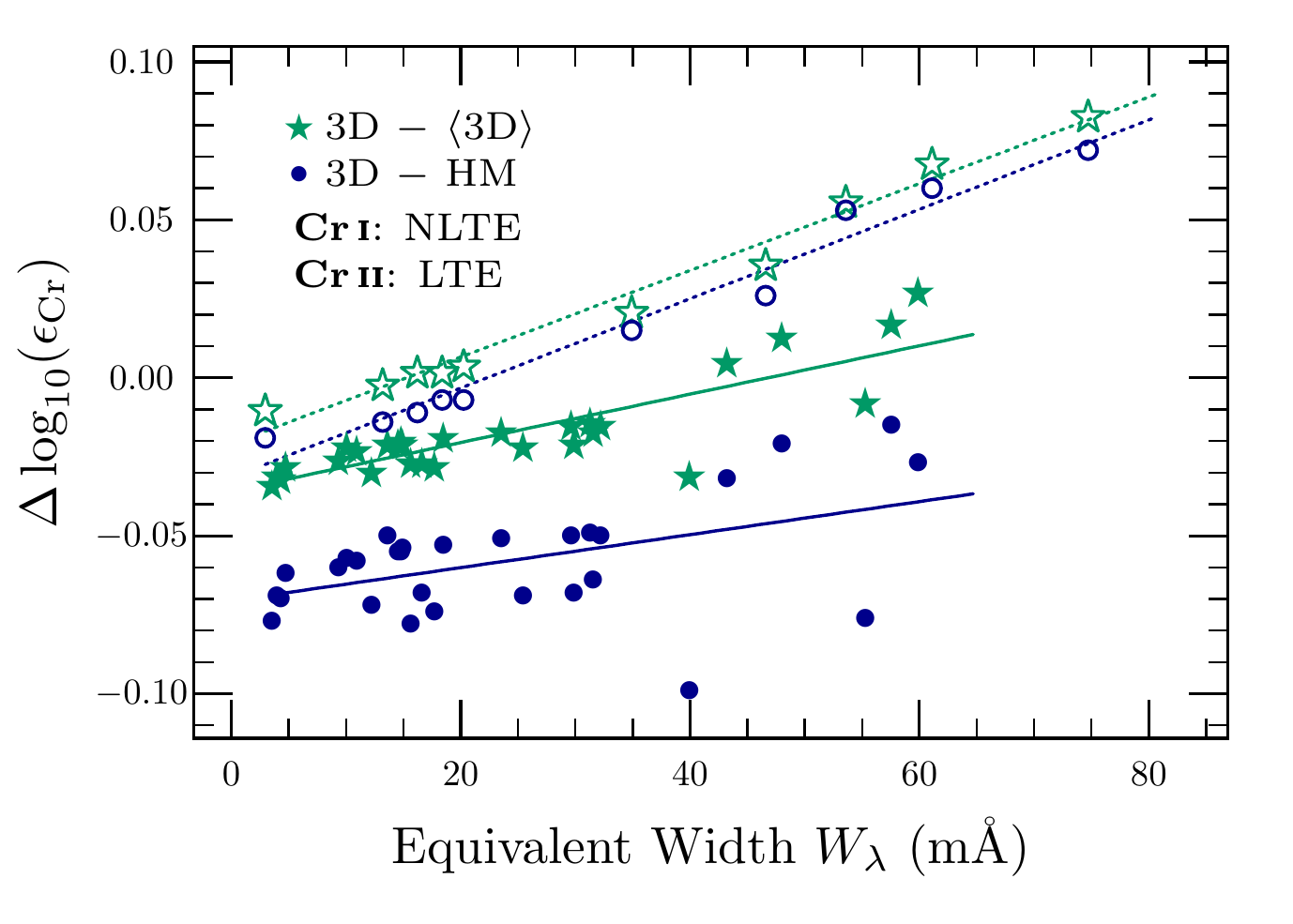}
\includegraphics[width=\linewidth]{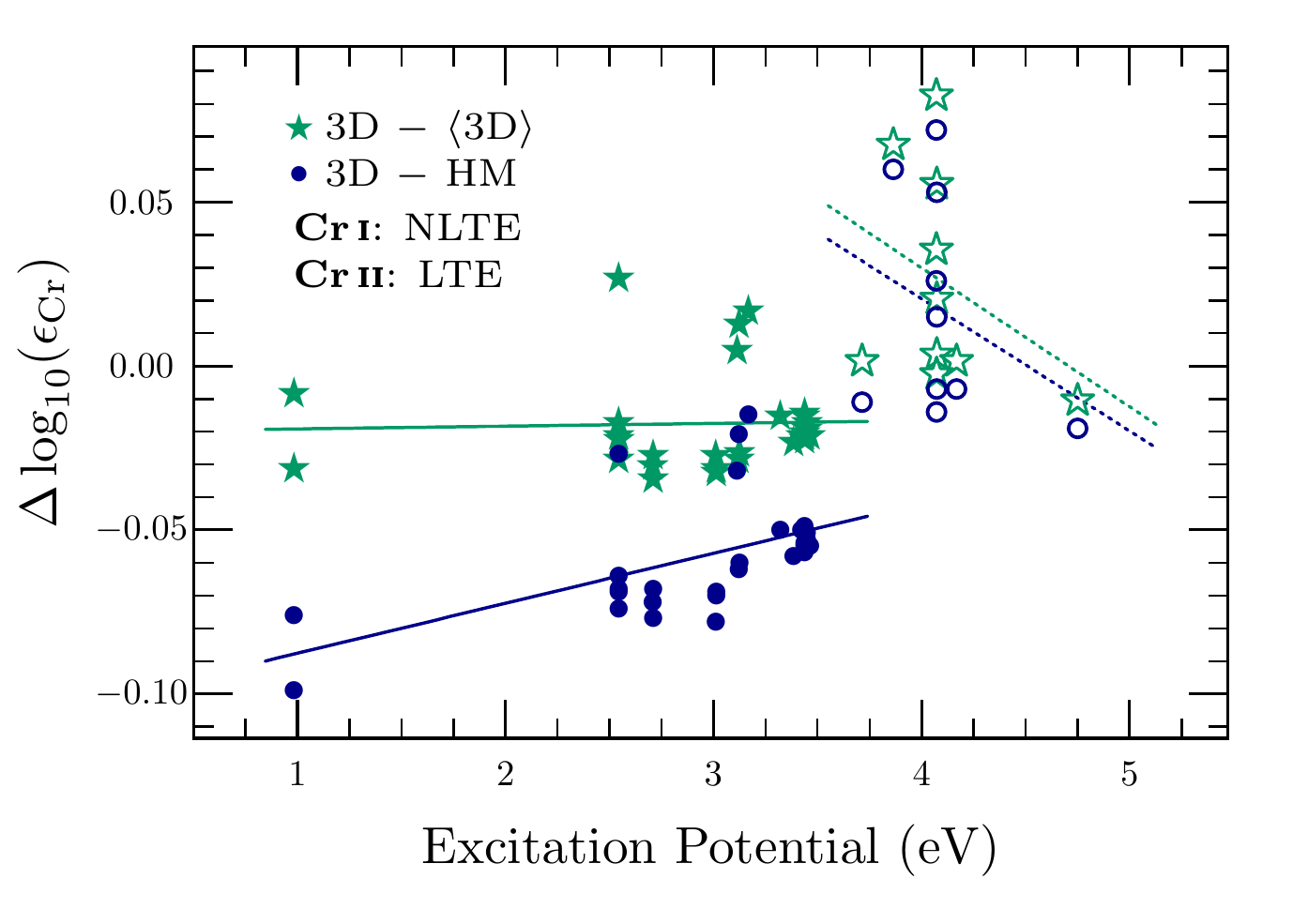}
\end{minipage}
\caption{\textit{Left}: 3D Cr abundances from \cri and \crii lines, as a function of equivalent width and lower excitation potential.  \textit{Right}: Line-by-line differences between abundances obtained with the 3D and \oneDAV\ models, and between those obtained with the 3D and \citetalias{HM} models.  Filled symbols and solid trendlines indicate neutral lines, open symbols and dotted lines indicate singly-ionised lines.}
\label{fig:cr}
\end{figure*}

\begin{figure*}[t]
\centering
\begin{minipage}[t]{0.4\textwidth}
\centering
\includegraphics[width=\linewidth]{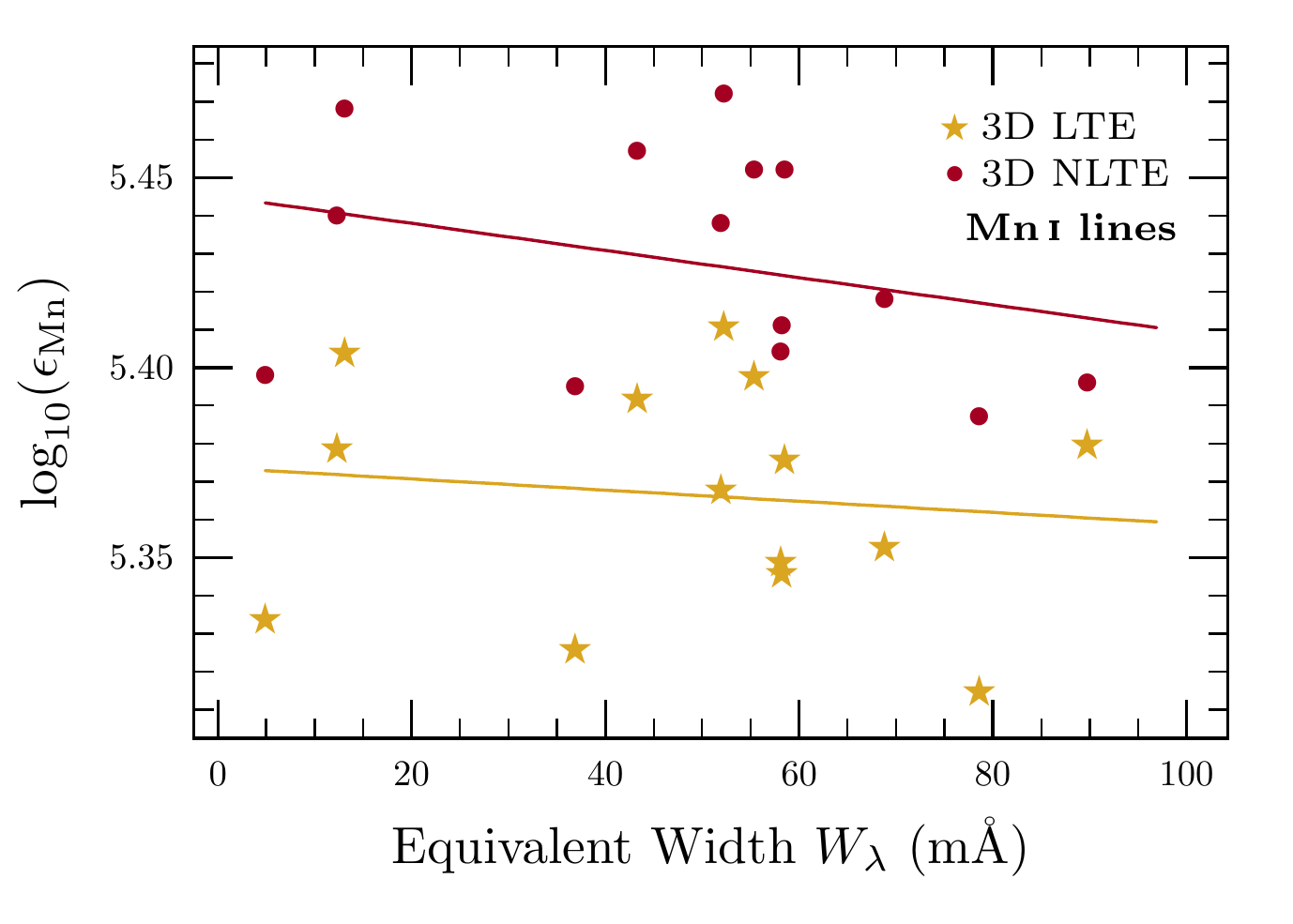}
\includegraphics[width=\linewidth]{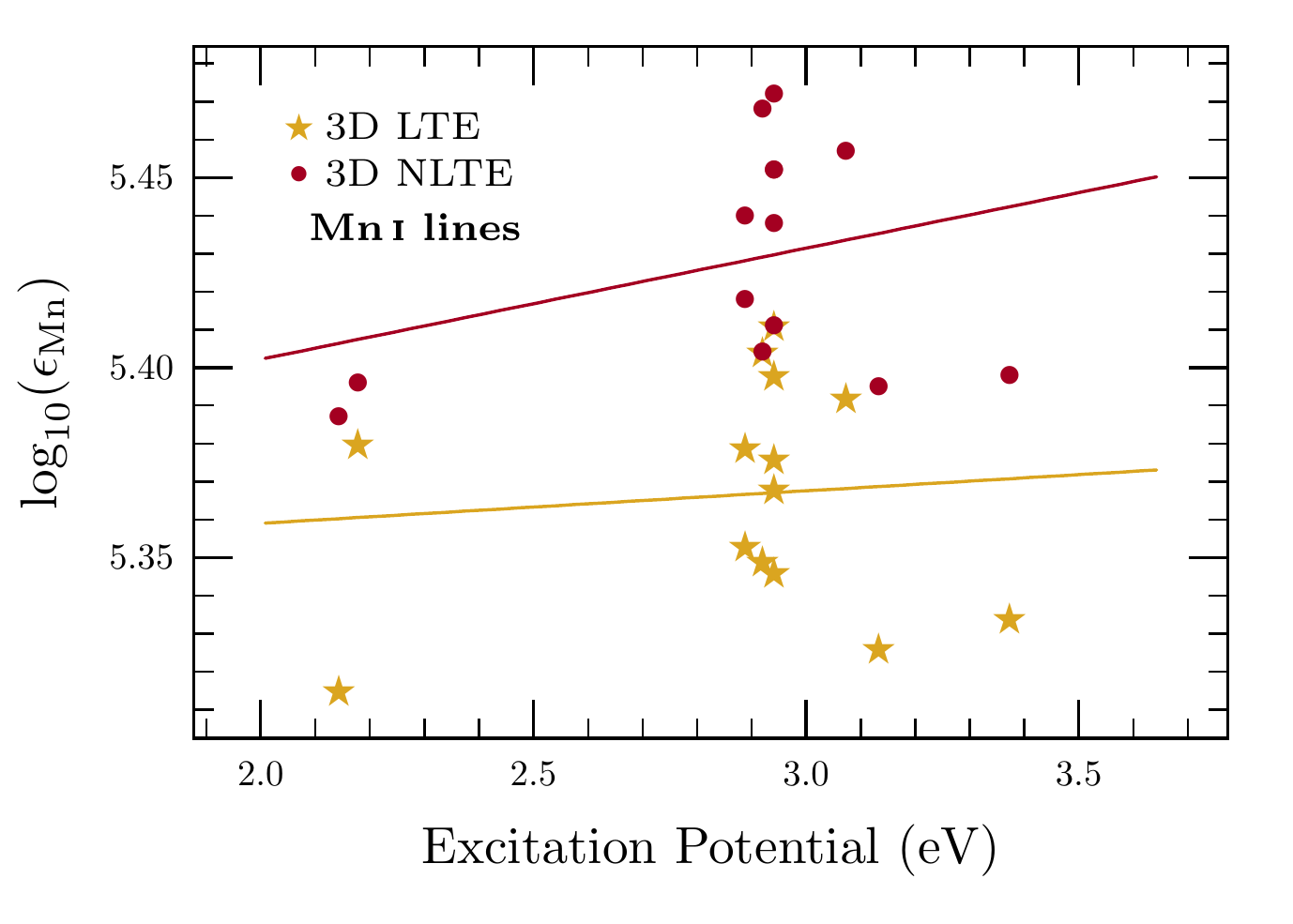}
\end{minipage}
\hspace{0.05\textwidth}
\begin{minipage}[t]{0.4\textwidth}
\centering
\includegraphics[width=\linewidth]{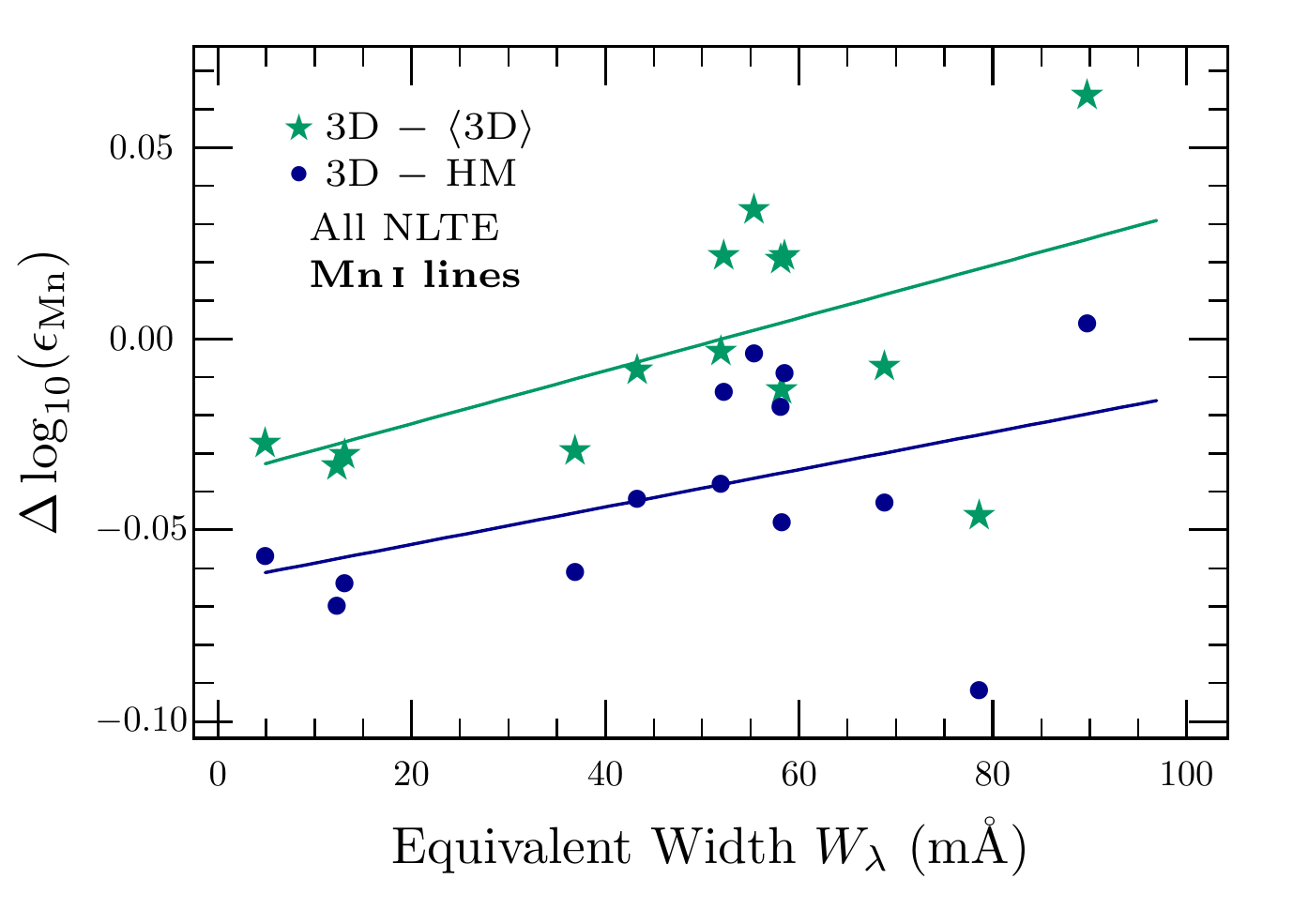}
\includegraphics[width=\linewidth]{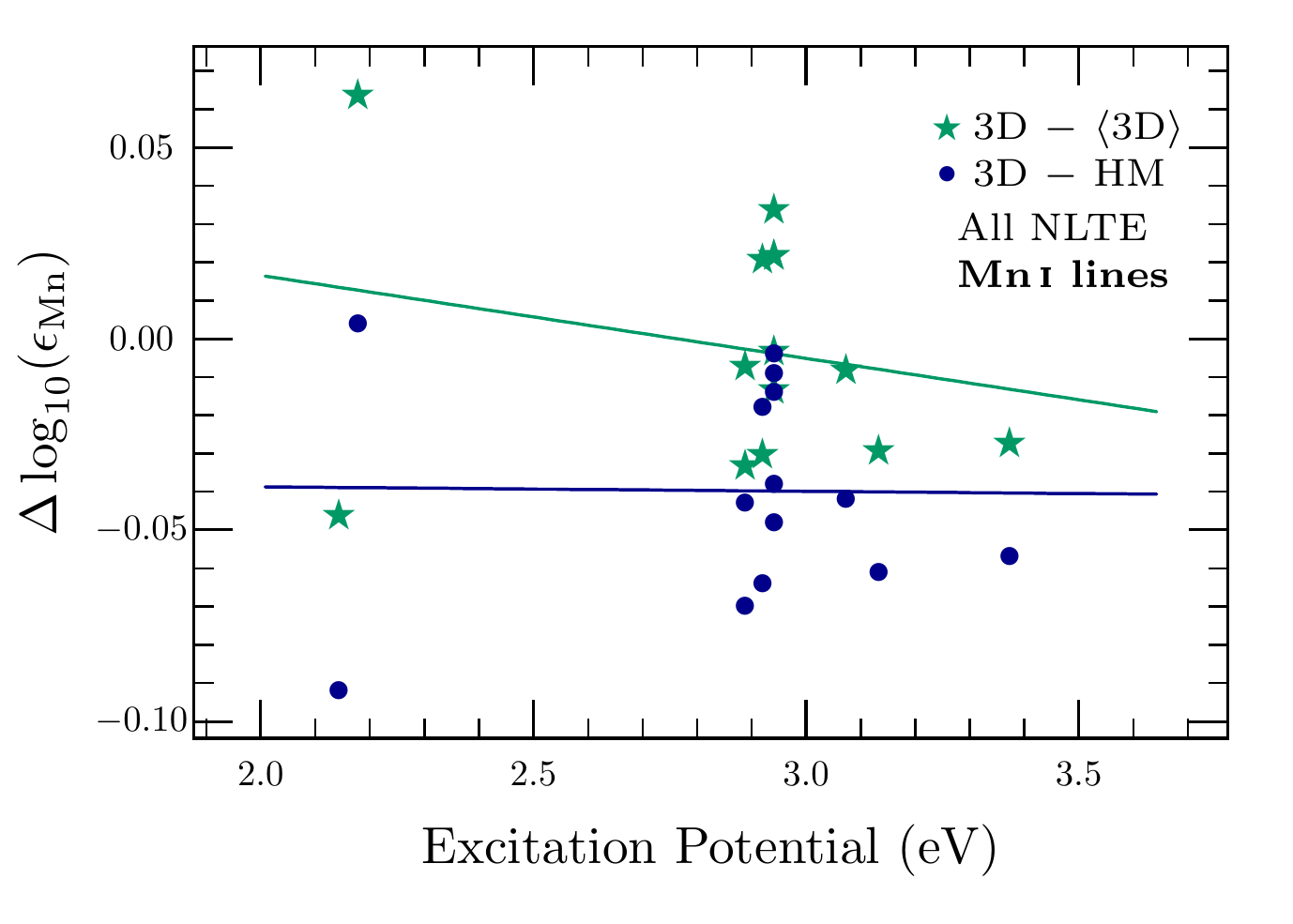}
\end{minipage}
\caption{\textit{Left}: 3D Mn abundances from \mni lines, as a function of equivalent width and lower excitation potential.  \textit{Right}: Line-by-line differences between abundances obtained with the 3D and \oneDAV\ models, and between those obtained with the 3D and \citetalias{HM} models.}
\label{fig:mn}
\end{figure*}

\begin{figure*}[t]
\centering
\begin{minipage}[t]{0.4\textwidth}
\centering
\includegraphics[width=\linewidth]{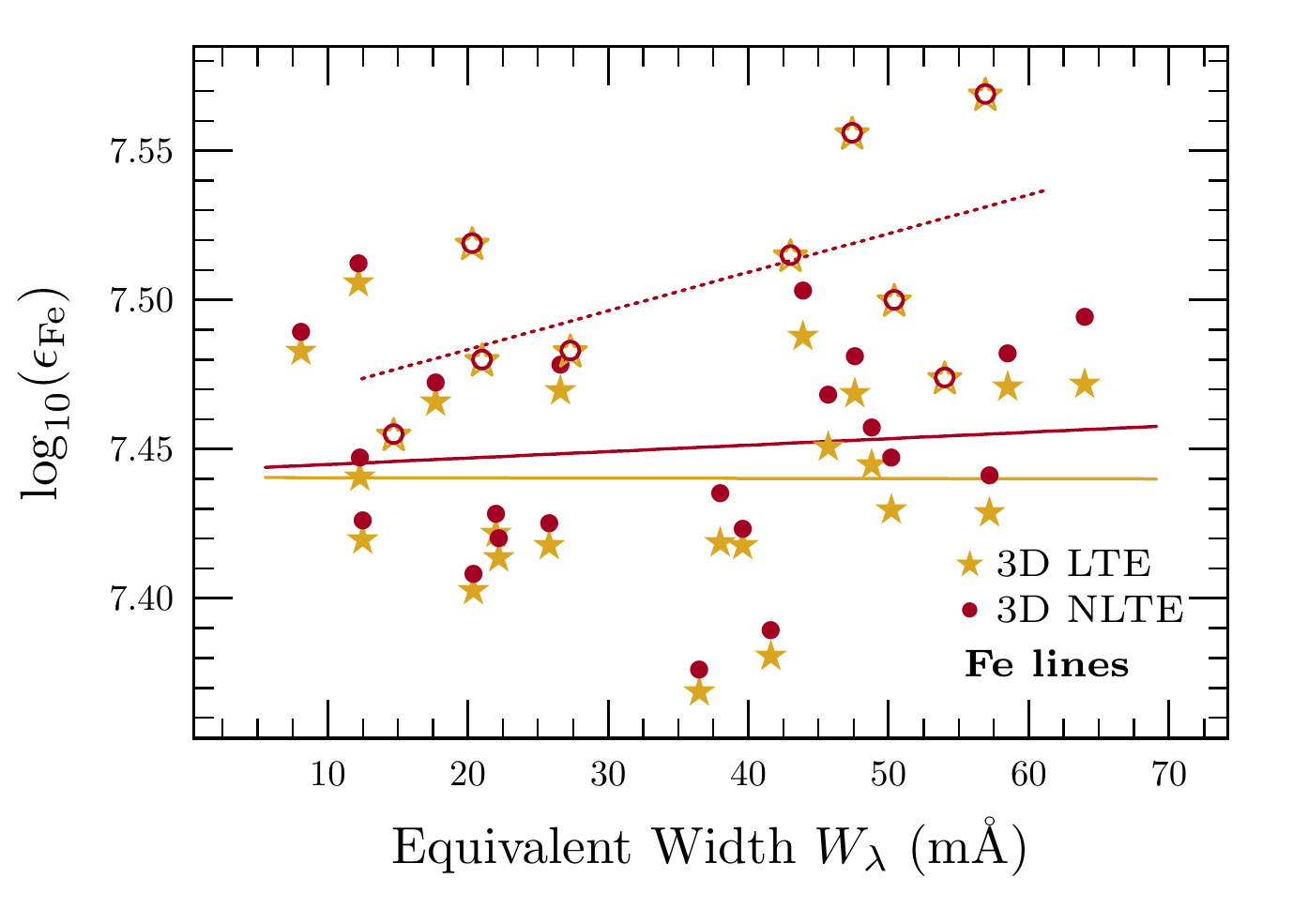}
\includegraphics[width=\linewidth]{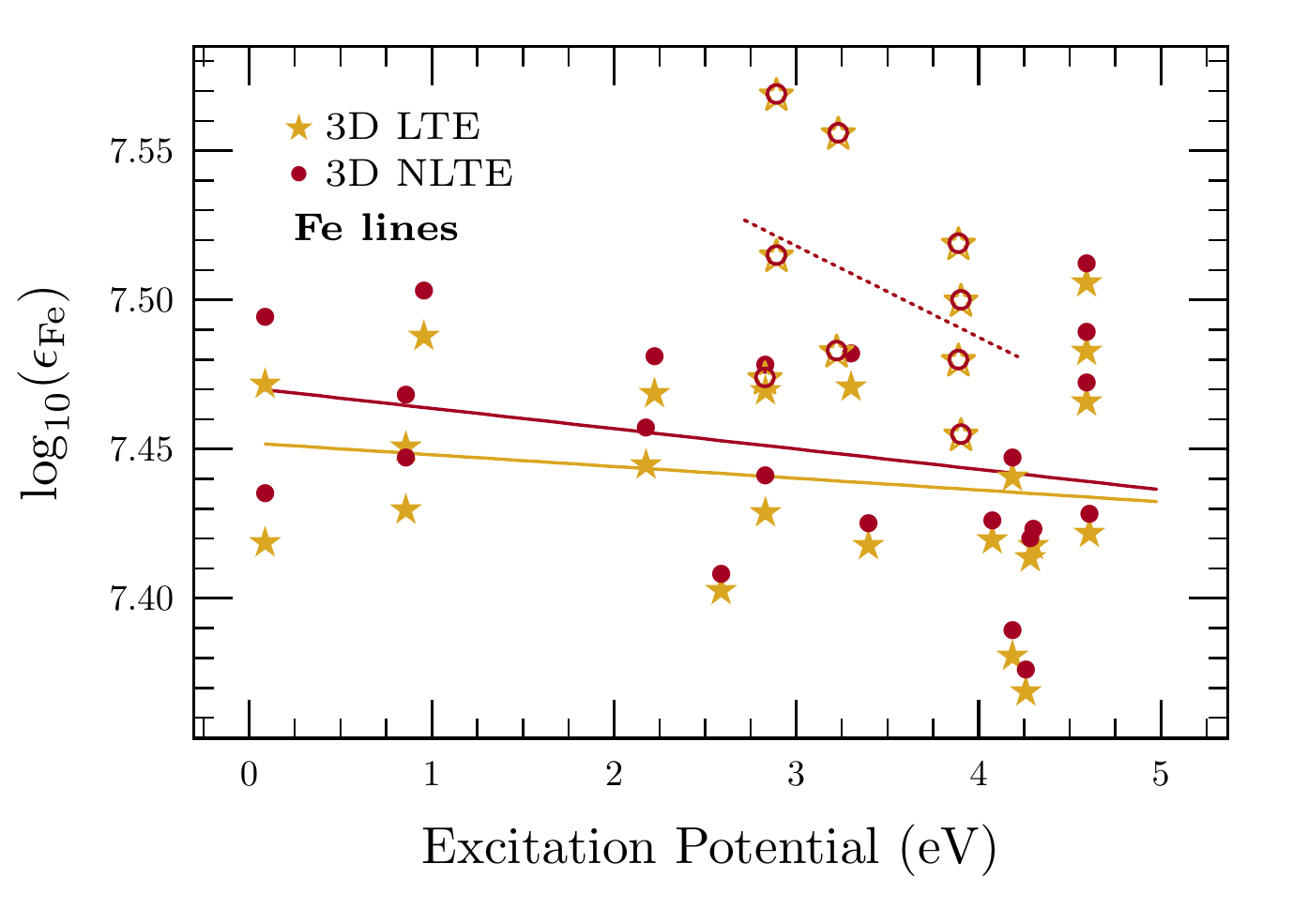}
\end{minipage}
\hspace{0.05\textwidth}
\begin{minipage}[t]{0.4\textwidth}
\centering
\includegraphics[width=\linewidth]{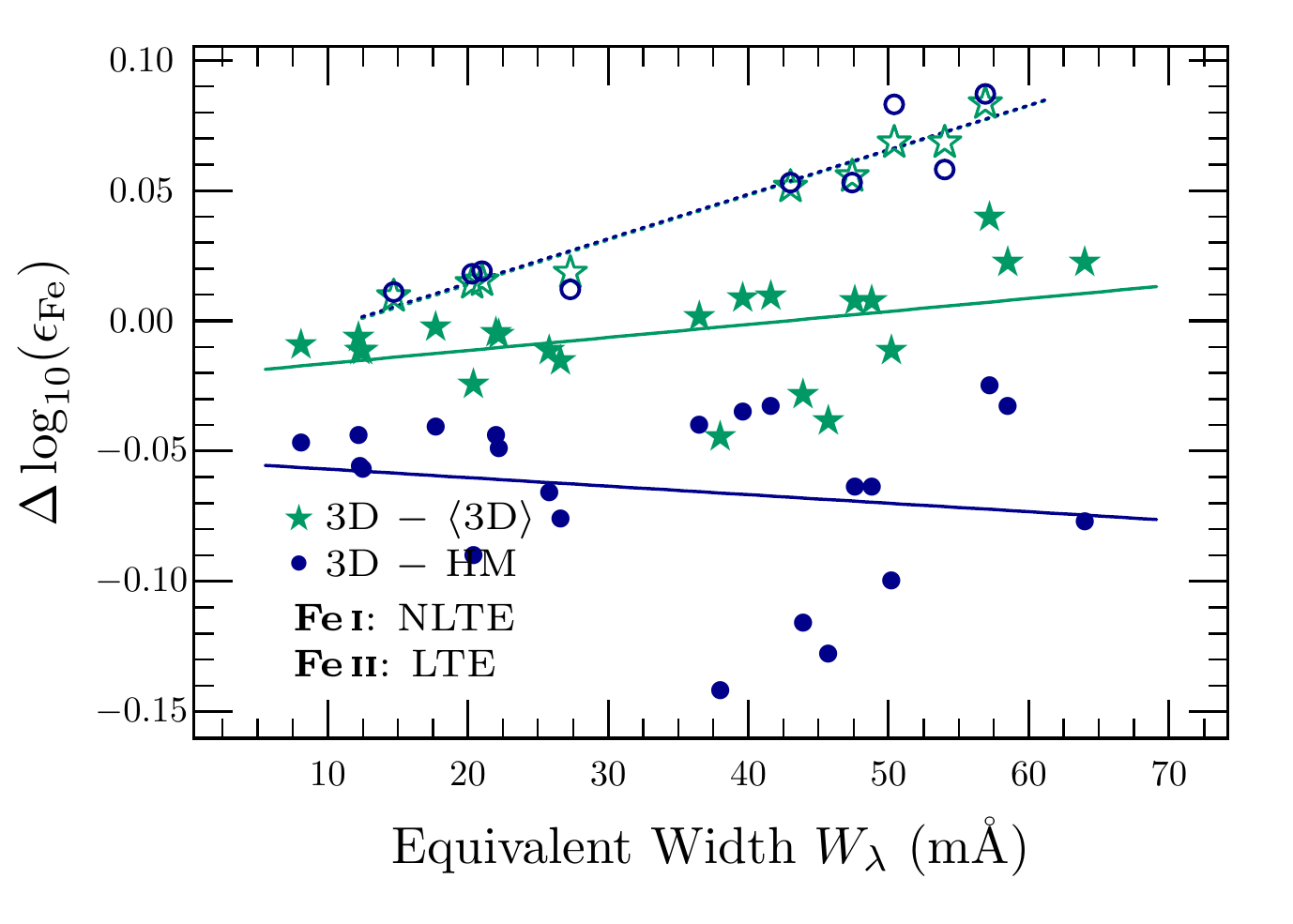}
\includegraphics[width=\linewidth]{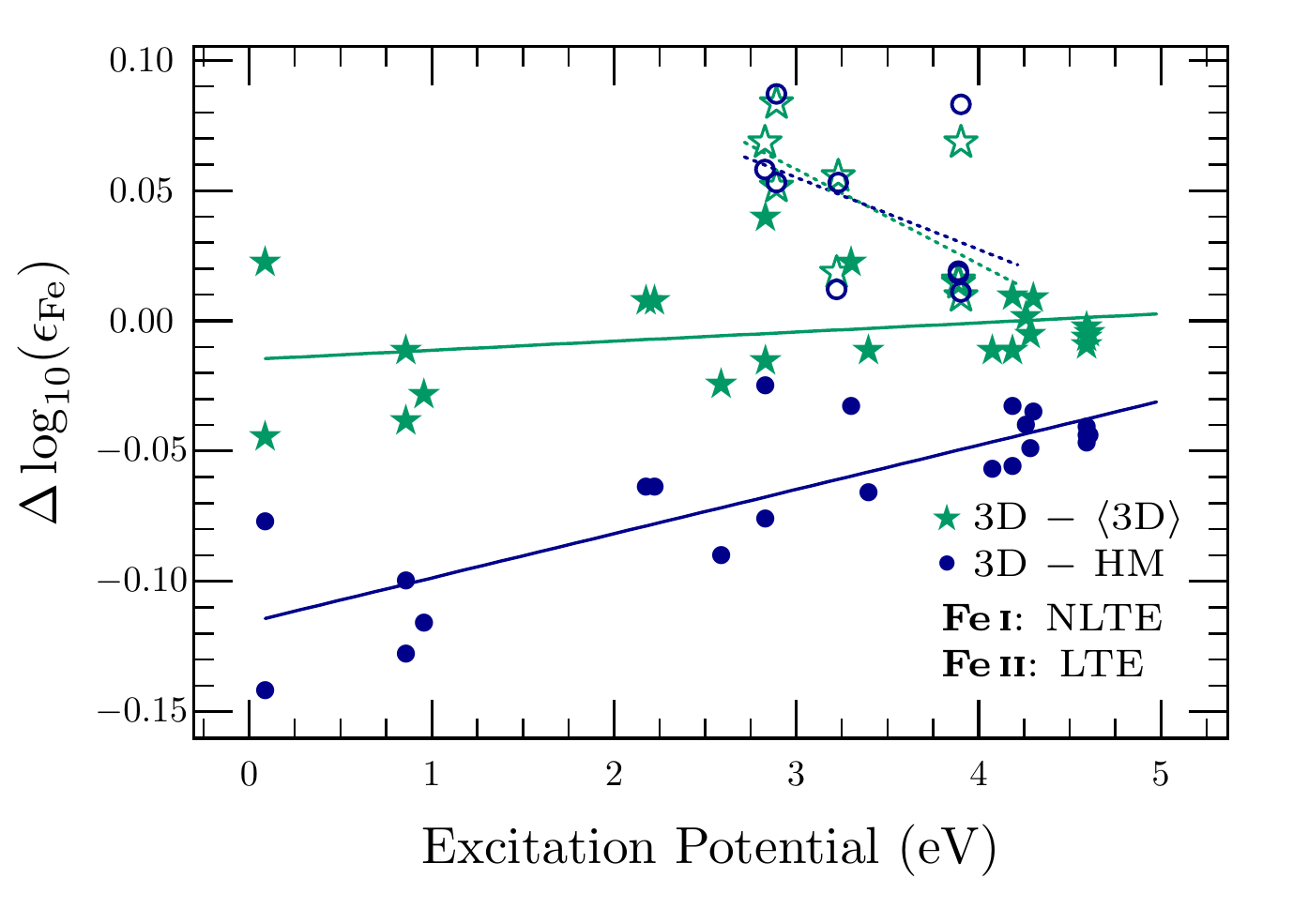}
\end{minipage}
\caption{\textit{Left}: 3D Fe abundances from \fei and \feii lines, as a function of equivalent width and lower excitation potential.  \textit{Right}: Line-by-line differences between abundances obtained with the 3D and \oneDAV\ models, and between those obtained with the 3D and \citetalias{HM} models.  Filled symbols and solid trendlines indicate neutral lines, open symbols and dotted lines indicate singly-ionised lines.}
\label{fig:fe}
\end{figure*}

\begin{figure*}
\centering
\begin{minipage}[t]{0.4\textwidth}
\centering
\includegraphics[width=\linewidth]{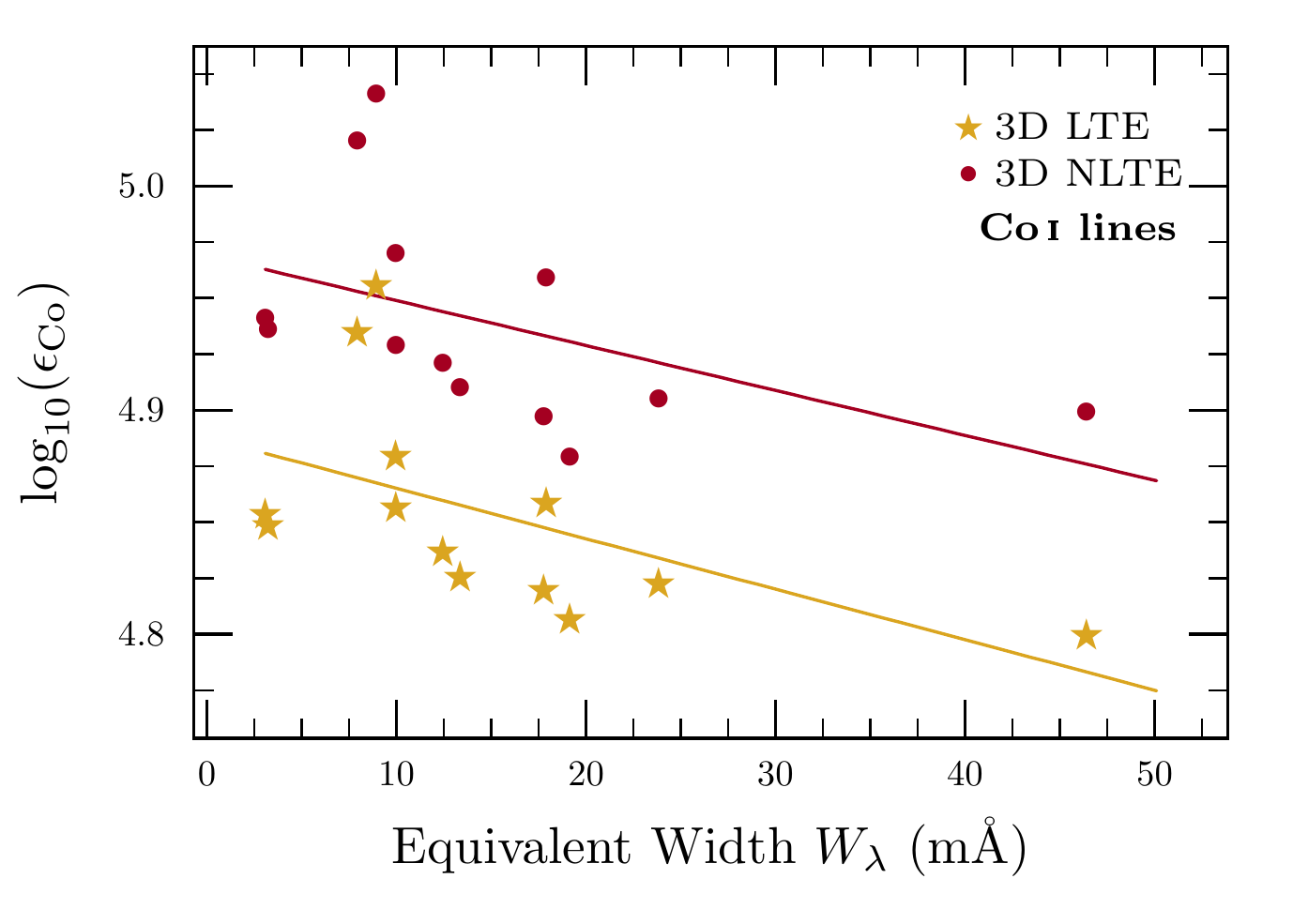}
\includegraphics[width=\linewidth]{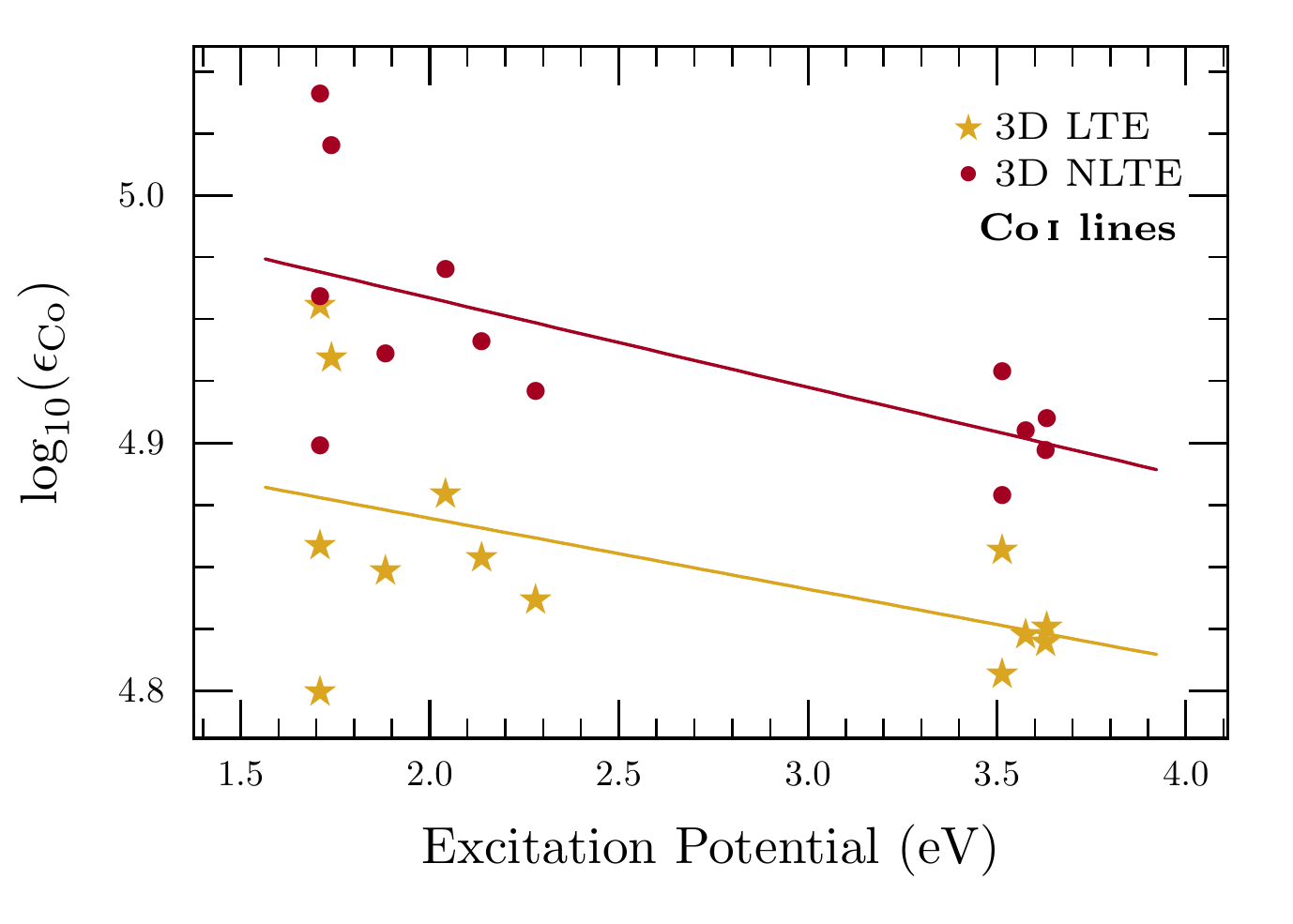}
\end{minipage}
\hspace{0.05\textwidth}
\begin{minipage}[t]{0.4\textwidth}
\centering
\includegraphics[width=\linewidth]{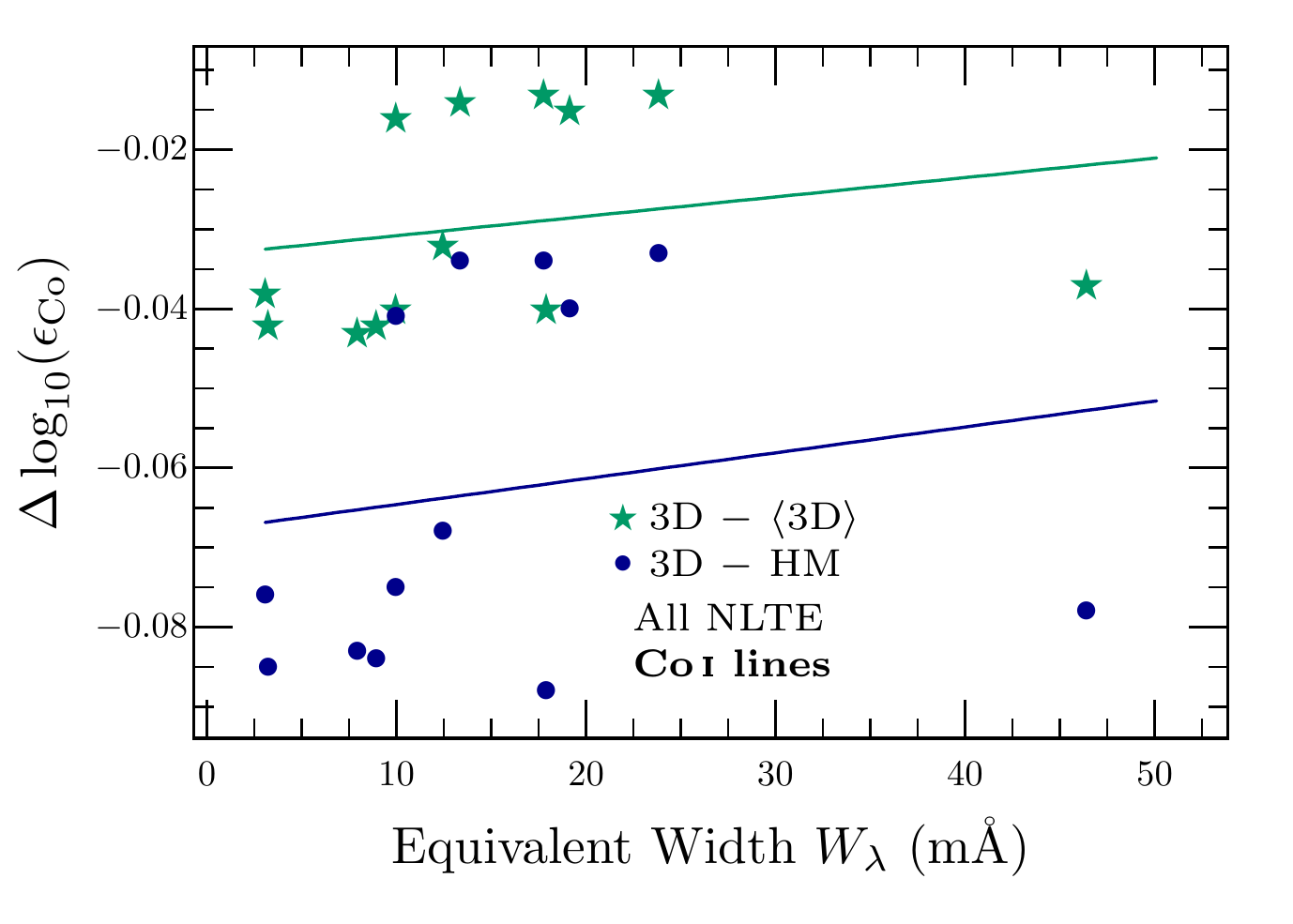}
\includegraphics[width=\linewidth]{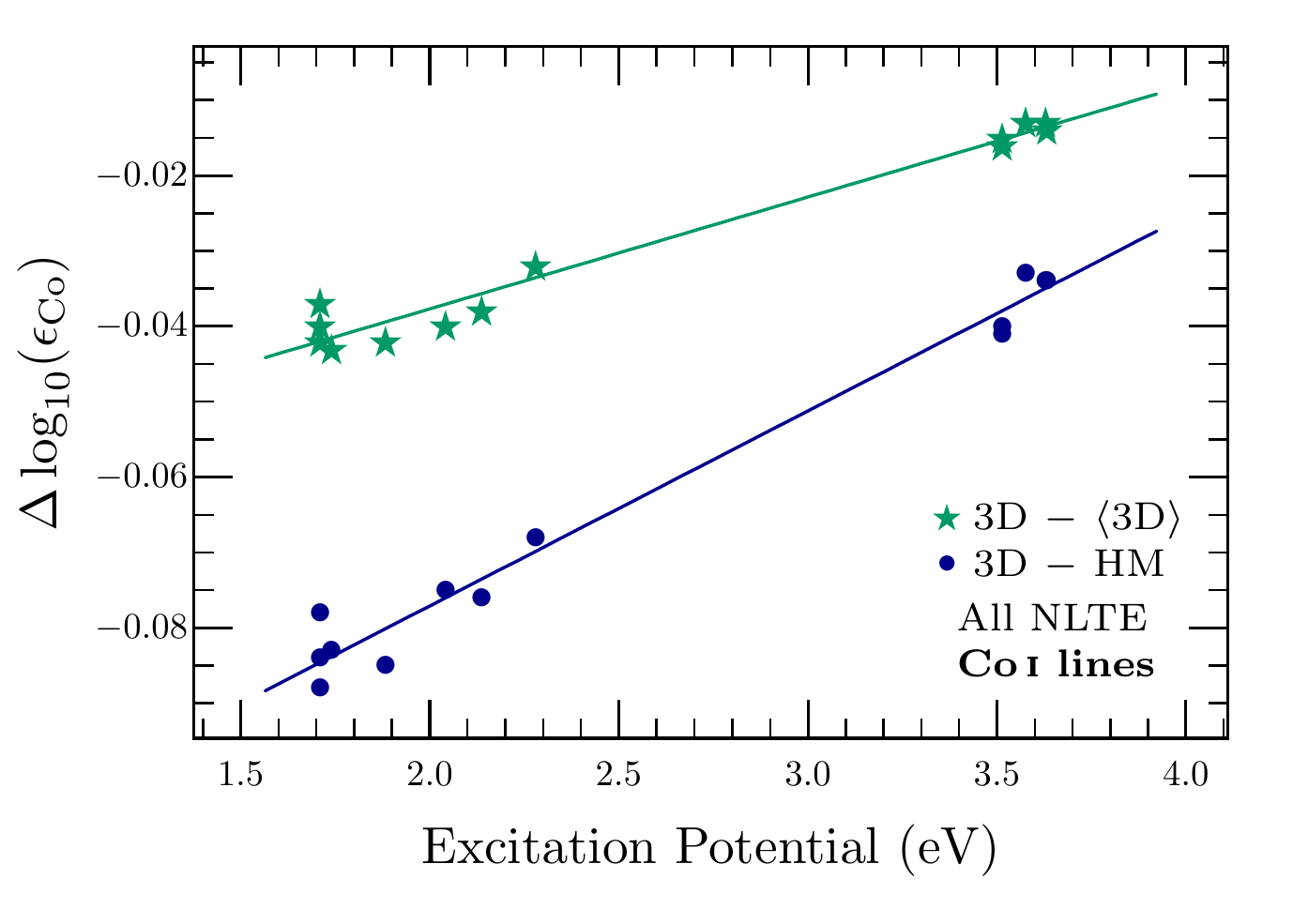}
\end{minipage}
\caption{\textit{Left}: 3D Co abundances from \coi lines, as a function of equivalent width and lower excitation potential.  \textit{Right}: Line-by-line differences between abundances obtained with the 3D and \oneDAV\ models, and between those obtained with the 3D and \citetalias{HM} models.}
\label{fig:co}
\end{figure*}

\begin{figure*}
\centering
\begin{minipage}[t]{0.4\textwidth}
\centering
\includegraphics[width=\linewidth]{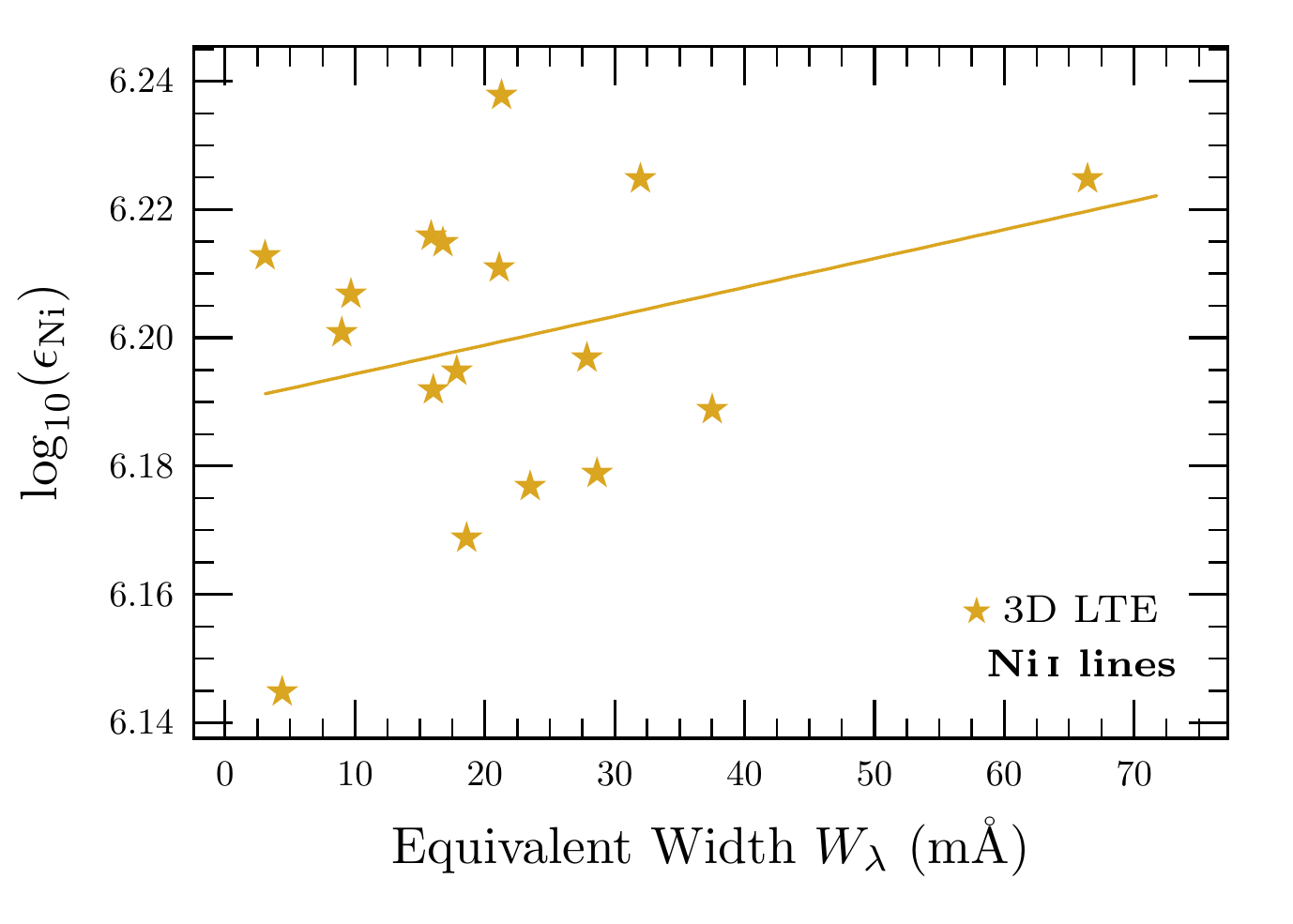}
\includegraphics[width=\linewidth]{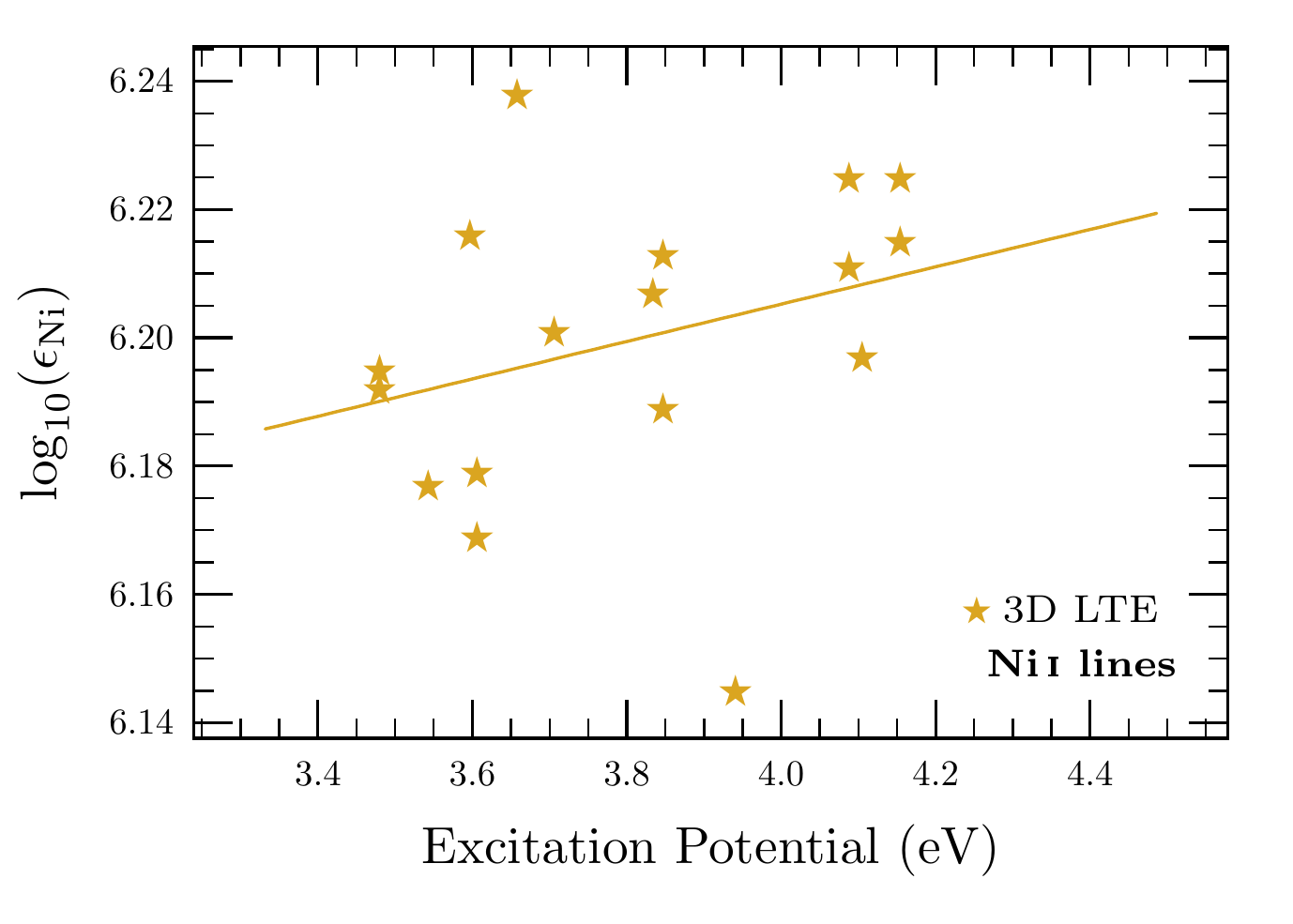}
\end{minipage}
\hspace{0.05\textwidth}
\begin{minipage}[t]{0.4\textwidth}
\centering
\includegraphics[width=\linewidth]{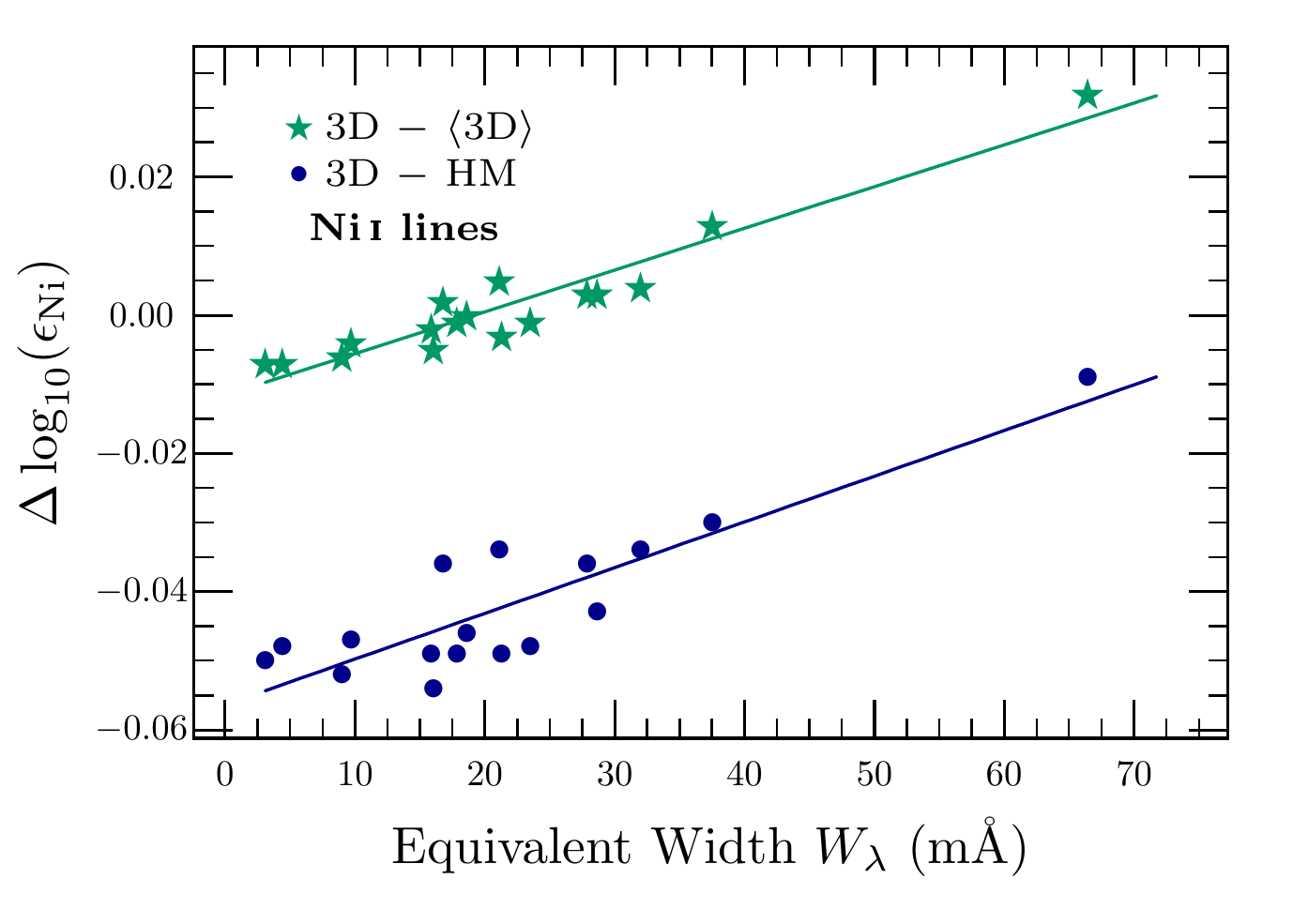}
\includegraphics[width=\linewidth]{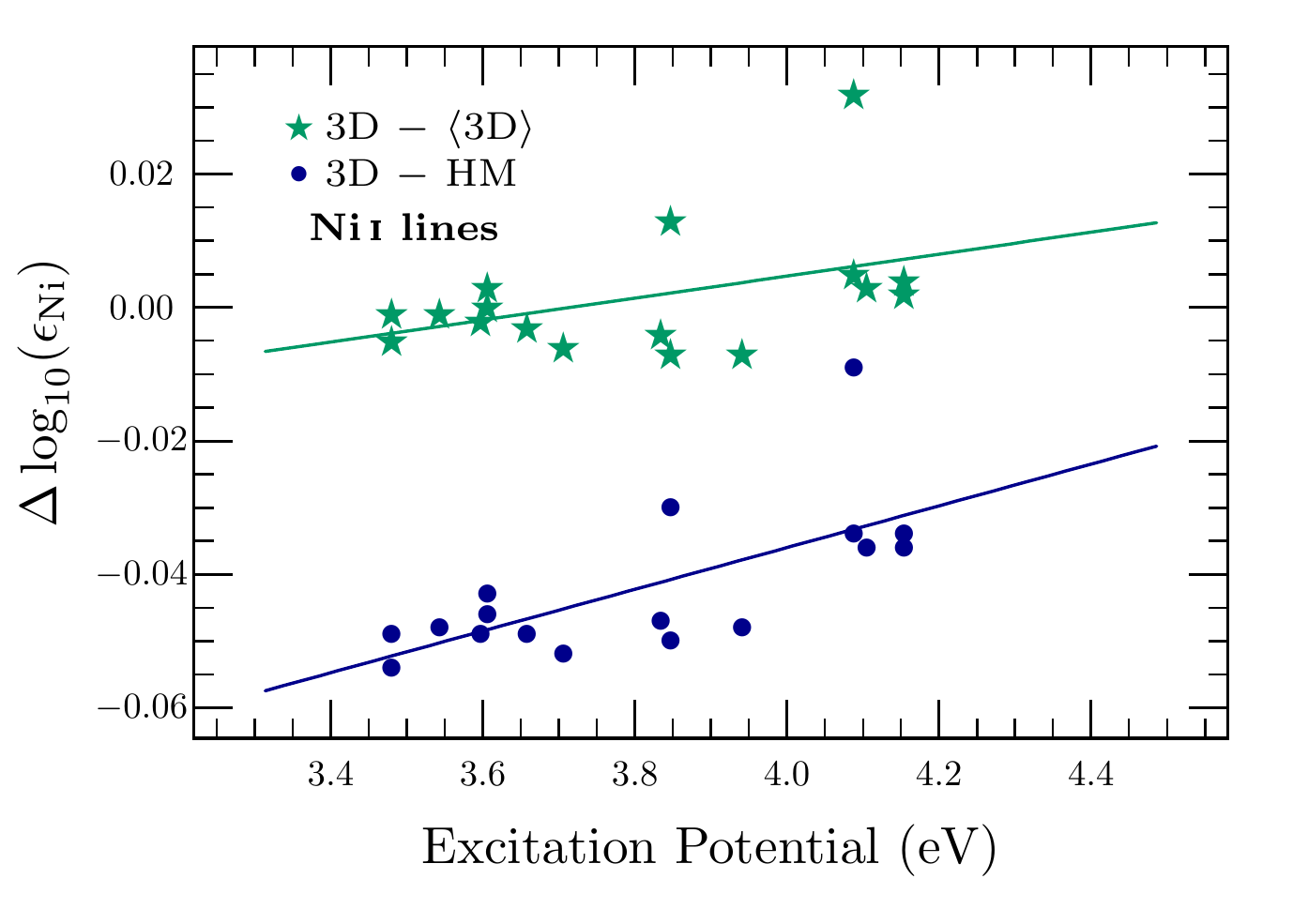}
\end{minipage}
\caption{\textit{Left}: 3D Ni abundances from \nii lines, as a function of equivalent width and lower excitation potential.  \textit{Right}: Line-by-line differences between abundances obtained with the 3D and \oneDAV\ models, and between those obtained with the 3D and \citetalias{HM} models.}
\label{fig:ni}
\end{figure*}

\section{Derived solar elemental abundances}
\label{results}

We have derived the solar abundances of Sc, Ti, V, Cr, Mn, Fe, Co and Ni from each of the lines given in Table~\ref{table:lines}.  The interpolated theoretical 3D line profiles show good agreement with the observed solar spectrum, as can be seen in the sample of lines shown in Figs.~\ref{fig:profiles} and \ref{fig:profiles2}.  This agreement is a result of the inhomogeneous, three-dimensional temperature and velocity structure of the 3D model atmosphere, and the inclusion of HFS and isotopic structure wherever necessary.  A small systematic deviation of the theoretical profiles from the observed spectrum can be seen in the cores and wings of some lines: the lines are slightly too deep in the core and too shallow in the wings, which may signal NLTE effects (note that full 3D line formation calculations have not been attempted). These small discrepancies may also indicate that the photospheric velocity structure of the 3D model, whilst certainly highly realistic, is not quite perfect.

The 3D abundance results are given for each line in Table~\ref{table:lines}, including NLTE corrections where possible.  In Table~\ref{table:lines} we also give the LTE abundances derived from each line with the \oneDAV, HM, \textsc{marcs} and {\sc miss} 1D models.  In Figs.~\ref{fig:sc}--\ref{fig:ni} we plot the 3D abundances as a function of line strength and excitation potential.  We also show in these figures the difference between the abundances derived using the 3D and HM models, and between those derived from the 3D and \oneDAV\ models, as a function of line strength and excitation potential.  

In the following sections we discuss the results for each element in detail, comparing with previous determinations of the solar abundance and with the meteoritic values.  The latter we take from the recent careful compilation and analysis of \citet{Lodders09}, renormalised to the photospheric abundance of silicon determined in \citetalias{AGSS_NaCa} ($\log\epsilon_\mathrm{Si}=7.51$, as already done in \citetalias{AGSS}).  We also describe the updates we have made for specific elements since \citetalias{AGSS}.  In addition to those updates, before reiterating on all abundance calculations a final time, we updated all partition functions and ionisation potentials (Table~\ref{table:partition}), and updated our equation-of-state tables and base atmospheric composition to the published \citetalias{AGSS} mixture.  We summarise our full results in Table~\ref{table:abuns}, including our final recommended abundances.  We compare these results to previous solar abundance compilations in Table~\ref{table:compilations}.  We remind the reader that the error treatment used in the following sections is summarised in Sect.\ \ref{calculations}.

\subsection{Scandium}
\label{Scresults}

For \scani lines, the 3D+NLTE result is $\log \epsilon_\mathrm{Sc}=3.14\pm0.09$ ($\pm$0.01 stat, $\pm$0.09 sys).  The corresponding result from \scanii lines is $\log \epsilon_\mathrm{Sc}=3.17\pm0.04$ ($\pm$0.02 stat, $\pm$0.04 sys), in good agreement with the \scani result.  Taking all \scani and \scanii lines together, our final recommended Sc abundance becomes
\begin{displaymath}
\log \epsilon_\mathrm{Sc}=3.16\pm0.04\ (\pm0.01\ \mathrm{stat},\ \pm0.04\ \mathrm{sys}).
\end{displaymath}
Intriguingly, this value is 0.11\,dex, or more than two standard deviations, larger than the meteoritic value \citep[$3.05\pm0.02$;][]{Lodders09}.  We leave speculation as to the importance of this and other photospheric-meteoritic differences for future work.

We see from Table~\ref{table:abuns} and the right-hand panels of Fig.~\ref{fig:sc} that the model atmosphere plays only a minor role for \scanii lines, as is generally true for the dominant ionisation stage.  On the other hand, abundances derived from \scani are extremely sensitive to the choice of model, as expected for a low ionisation neutral species.  We also see in Fig.~\ref{fig:sc} that 3D Sc abundances do not show any significant trend with excitation potential or line strength.  Abundance scatter is generally higher with Sc\,\textsc{ii} than Sc\,\textsc{i} lines, reflecting the greater uncertainty in the \scanii $gf$-values.

Our results are in good agreement with the  Sc abundance of \citet{Zhang08}. Their value of $\log \epsilon_\mathrm{Sc}=3.13\pm0.05$ was essentially based on \scanii lines, and derived using the same experimental $gf$-values as we do here, along with a theoretical 1D photospheric model.  Compared to the result we presented in \citetalias{AGSS} ($\log \epsilon_\mathrm{Sc}=3.16\pm0.04$), in this analysis we have discarded the three lines discussed in Sect.~\ref{Scgfs} as having excellent profiles but unsatisfactory oscillator strengths.

\subsection{Titanium}
\label{Tiresults}

Titanium is in principle an ideal case: a large number of good solar lines with accurate transition probabilities, very minor HFS and isotopic broadening, and extensive NLTE calculations available for the minority species (\alttii).  

Nonetheless, the derived 3D+NLTE Ti abundances from \tii ($\log \epsilon_\mathrm{Ti}=4.88\pm0.05$; $\pm$0.01 stat, $\pm$0.05 sys), and \tiii ($\log \epsilon_\mathrm{Ti}=4.97\pm0.04$; $\pm$0.01 stat, $\pm$0.03 sys) show a 0.09\,dex discrepancy. Even more puzzlingly, this discrepancy is not present in either the mean \citetalias{HM} or \oneDAV\ results.  Referring to Fig.~\ref{fig:ti}, 3D+NLTE abundances from \tii and \tiii show no perceptible trend with line strength or excitation potential, whereas the HM and \oneDAV\ results show a clear trend with line strength for both ionisation stages.  Given the rather large 3D corrections for \alttii, and the fact that we always \textit{assume} the NLTE corrections for the 3D model to be equal to those calculated for the \oneDAV\ model, we suspect that the NLTE corrections are somewhat underestimated in \oneDAV\ compared to full 3D.  Alternatively, our chosen efficiency of inelastic collisions with hydrogen ($S_\mathrm{H}=1$) may be somewhat too high in this case; had we instead adopted $S_\mathrm{H}=0.05$, the 3D+NLTE abundance from \tii would have been $\log \epsilon_\mathrm{Ti}=4.93$.

Considering the sensitivity of the \tii NTLE corrections to $S_{\rm H}$, for our final recommended Ti abundance we take a weighted mean of the \tii and \tiii results, with the weightings determined by the respective uncertainties of the two results.  This favours \tiii due to its smaller systematic uncertainty, resulting in
\begin{displaymath}
\log \epsilon_\mathrm{Ti}=4.93\pm0.04\ (\pm0.01\ \mathrm{stat},\ \pm0.04\ \mathrm{sys}).
\end{displaymath}
Here we have estimated the errors by considering all \tii and \tiii lines equally in a single list, rather than by using the same statitistical weighting procedure as for the mean.  Had we instead used the latter procedure, the final uncertainty would be just $0.03$\,dex; owing to the tension between \tii and \tiii, and for consistency with other elements where we include both ionisation stages in a single list, we think it more appropriate to adopt the larger estimate.  The final Ti abundance is in excellent agreement with the meteoritic value \citep[$4.91\pm0.03$;][]{Lodders09}, but the difference between the results returned by the two ionisation stages remains troubling.  

\citet{Bergemann11} found that Ti ionisation balance would be best satisfied with the MAFAGS-OS model if one were to adopt the \citet{BW06} and \citet{Pickering01} $gf$-values, giving $\log \epsilon_\mathrm{Ti}=4.94\pm0.05$ from \tii lines and $\log \epsilon_\mathrm{Ti}=4.95\pm0.06$ for \tiii lines.  Using more accurate $gf$-values (\citealt{Bizzarri93}) would lead to a larger discrepancy between \tii and \alttiii: $\log \epsilon_\mathrm{Ti}=4.93\pm0.04$ from \tii lines and $\log\epsilon_\mathrm{Ti}=4.98\pm0.04$ for \tiii lines.  As in this paper, \citet{Bergemann11} observed a strong dependence of \tii abundances upon the chosen model atmosphere and collisional efficiency parameters.

Compared to the result we adopted in \citetalias{AGSS} ($\log \epsilon_\mathrm{Ti}=4.95\pm0.05$), we now have dedicated NLTE calculations available in intensity for \tii based on the work of \citet{Bergemann11}.  We have also now dropped the \tii line at 522.4\,nm from our line list due to blending, adopted isotopic splitting and HFS data for \alttiii, and employed accurate new oscillator strengths for both ionisation stages \citep{Lawler13, Wood13}.

\subsection{Vanadium}
\label{Vresults}

Quality atomic data exists for both ionisation stages of V, but the only clear solar lines are of \altvi.  Unfortunately, only \vii lines are expected to form in LTE; the magnitude of NLTE corrections for \vi is unknown, so we apply an {\it ad hoc} correction of $+0.1$\,dex, as discussed in Sect.\ \ref{Vgfs}.

The minority species, \altvi, is more strongly affected by the temperature structure of the model atmosphere than \altvii.  The 3D LTE abundance from \vi lines is $\log \epsilon_\mathrm{V}=3.79\pm0.04$ ($1\sigma$ dispersion), whereas with the \citetalias{HM} model we obtain $\log \epsilon_\mathrm{V}=3.97\pm0.03$ ($1\sigma$).  For \altvii, the 3D and \citetalias{HM} LTE results agree well: $\log \epsilon_\mathrm{V}=4.00\pm0.05$ ($1\sigma$) in 3D, $\log \epsilon_\mathrm{V}=4.01\pm0.05$ ($1\sigma$) with \citetalias{HM}.  

These numbers are not a complete surprise: we know that \citetalias{HM} `includes' NLTE effects to some degree by way of its empirical temperature construction, as effects of departures from LTE can be partially mimicked by adjusting the spatially-averaged temperature structure of the model atmosphere, a phenomenon dubbed `NLTE masking' by \citet{1982A&A...115..104R}.  Similarly, the 3D \vi results exhibit a strong dependence upon excitation potential, whereas the results from \citetalias{HM} do not (Fig.~\ref{fig:v}). Because we applied the same NLTE correction to all \altvi\ lines, the trend also remains in NLTE. In reality we  expect more pronounced NLTE effects for lower-excitation lines, as these are sensitive to higher atmospheric layers, where lower densities and temperatures make LTE an increasingly poor approximation.  

Although low-excitation lines are most sensitive to the temperature structure, and therefore less reliable as abundance indicators, we have no way to know whether our universal NLTE correction of $+0.1$\,dex is more accurate at high or low excitation potential.  To avoid introducing any further systematic bias into our result, we therefore retain both the high- and low-excitation lines in our sample of \vi lines.  A dedicated NLTE study of V line formation in the Sun would clarify matters substantially.

Given the abysmal nature of the \vii lines in the solar spectrum, abundances from these lines are dominated by systematic and statistical errors in the determination of equivalent widths.  We therefore trust the absolute values of the \vii results even less than our \textit{ad hoc} \vi NLTE correction, and adopt the NLTE-corrected \vi 3D result as our recommended value:
\begin{displaymath}
\log \epsilon_\mathrm{V}=3.89\pm0.08\ (\pm0.01\ \mathrm{stat},\ \pm0.08\ \mathrm{sys}).
\end{displaymath}

Our result is significantly lower than the stated \citetalias{HM}-based abundances of \citet[][$3.99\pm0.01$]{Whaling85} and \citet[][$4.02\pm0.02$]{Biemont89}; their quoted uncertainties only consider statistical errors, not systematic errors stemming from, for example, the $gf$-values, model atmospheres or LTE line formation.  Our 3D result is also below the meteoritic value \citep[$3.96\pm0.02$;][]{Lodders09}, but the mutual uncertainties overlap.  Compared to \citetalias{AGSS} ($\log \epsilon_\mathrm{V}=3.93\pm0.08$), here we have added laboratory HFS data for \altvii.  We also discarded two \vi lines (619.9\,nm and 624.3\,nm), because we are suspicious as to the accuracy of the experimental branching fractions measured from their shared upper level.

As noted in Sect.\ \ref{Vgfs}, \citet{Wood14V2} have recently measured new experimental transition probabilities for a large number of \vii lines. Had we adopted their values for our five lines, the 3D-based \vii abundance would be 0.02\,dex lower, and thus in slightly better agreement with the \vi results. \citeauthor{Wood14V2} also performed spectrum synthesis (using the HM model) rather than fitting equivalent widths as we do here, which further reduces the inferred \vii abundance.  Comparing the results obtained with the HM model by both \citeauthor{Wood14V2} and us, and taking into account the differences in the adopted $gf$-values, we estimate that employing spectrum synthesis with our 3D models would have reduced the abundance by a further 0.02\,dex. Our final 3D \vii abundance would then have become $\log \epsilon_\mathrm{V}=3.96$, in perfect agreement with the meteoritic value. We note that \citeauthor{Wood14V2} employed a larger set of 15 \vii lines, resulting in a mean abundance of $\log \epsilon_\mathrm{V}=3.95\pm0.05$ ($1\sigma$), which should be contrasted with our HM-based value of $\log \epsilon_\mathrm{V}=4.01\pm0.05$ ($1\sigma$).  Adopting the oscillator strengths of \citeauthor{Wood14V2} and taking into account the $-0.02$\,dex impact of spectrum synthesis on our lines, our HM abundance would become $\log \epsilon_\mathrm{V}=3.97\pm0.04$, in perfect agreement with \citeauthor{Wood14V2} for the five lines in common.  In other words, there is a real possibility that our 3D \vii result should be decreased by about 0.04\,dex, bringing it into better agreement with \altvi. However, we still argue that \vi is a better indicator of the solar V abundance, in spite of the uncertainty in the NLTE effects.

\subsection{Chromium}
\label{Crresults}
For Cr\,\textsc{i}, we have a large number of very good solar lines (Sect.~\ref{CrNLTE}) and
very accurate $gf$-values (Sect.~\ref{Crgfs}).  Our derived NLTE abundance from \cri 
lines is $\log \epsilon_\mathrm{Cr}=5.60\pm0.04$ ($\pm$0.01 stat, $\pm$0.04 sys).  Using the very few recent experimental $gf$-values for Cr\,\textsc{ii} lines \citep{Nilsson06}, we found a very large scatter in
abundances. We therefore recommended (Sect.~\ref{Crgfs}) the theoretical $gf$-values of \citet{Kuruczweb} as the best currently available. With these data we find $\log \epsilon_\mathrm{Cr}=5.65\pm0.04$ ($\pm$0.02 stat, $\pm$0.04 sys) from \crii lines, in good agreement with both the \cri result and the meteoritic
abundance \citep[$5.64\pm0.01$;][]{Lodders09}.  We therefore adopt the mean result from all Cr lines
\begin{displaymath}
\log \epsilon_\mathrm{Cr}=5.62\pm0.04\ (\pm0.01\ \mathrm{stat},\ \pm0.03\ \mathrm{sys})
\end{displaymath}
as our final recommended solar Cr abundance.

No significant trends are visible with line strength or excitation potential in the results with any model (Fig.~\ref{fig:cr}, left panels). Our NLTE results with the \citetalias{HM} model agree very well with the LTE abundances derived from the Cr\,\textsc{i} lines by \citet[][$\log \epsilon_\mathrm{Cr}=5.67$]{Biemont78c}, \citet[][$\log \epsilon_\mathrm{Cr}=5.68$]{Blackwell87} and \citet[][$\log \epsilon_\mathrm{Cr}=5.64$]{Sobeck07}, which however is somewhat of a coincidence given the significant NLTE corrections.  The NLTE result of \citet{Bergemann11}, obtained by choosing $S_\mathrm{H}=0$ so as to impose ionisation balance with the MAFAGS-ODF model, is $0.12$ dex higher: $\log \epsilon_\mathrm{Cr}=5.74$.

For \altcrii, our results are much smaller than those of \citet[][$\log \epsilon_\mathrm{Cr}=5.67\pm0.13$ with \textsc{marcs}, $\log \epsilon_\mathrm{Cr}=5.77\pm0.13$ with \protect\citetalias{HM}]{Sobeck07} or \citet[][$\log \epsilon_\mathrm{Cr}=5.79\pm0.12$ with MAFAGS-ODF]{Bergemann10}.  Even using the theoretical $gf$-values of Kurucz, our dispersion is also much smaller ($\sigma=0.06$) than either of these results.  The substantially larger scatter seen with \citet{Nilsson06} $gf$-values \citep[as used by][]{Sobeck07,Bergemann10} than with semi-empirical Kurucz values is worrisome; it remains to be seen if the experimental measurements were affected by a systematic error of some kind.

The result we give here is slightly updated with respect to that in \citetalias{AGSS} ($\log \epsilon_\mathrm{Cr}=5.64\pm0.04$), as we now have dedicated NLTE intensity calculations for \cri (Sect.~\ref{CrNLTE}) for our specific 1D models with $S_\mathrm{H}=1$, based on the work of \citet{Bergemann10}.

\subsection{Manganese}
\label{Mnresults}

Following our consideration of the most reliable lines and oscillator strengths for \mni (Sect.~\ref{Mngfs}), we find a final NLTE Mn abundance of
\begin{displaymath}
\log \epsilon_\mathrm{Mn}=5.42\pm0.04\ (\pm0.01\ \mathrm{stat},\ \pm0.04\ \mathrm{sys}),
\end{displaymath}
slightly smaller than the meteoritic value \citep[$5.48\pm0.01$;][]{Lodders09}.  

Our result is somewhat larger than the earlier LTE \citetalias{HM} result of \citet[][$\log \epsilon_\mathrm{Mn}=5.39$]{Booth84b}. This shift can mainly be attributed to the positive NLTE abundance corrections and our more accurate oscillator strengths.  Our \citetalias{HM} and \textsc{marcs} abundances ($\log \epsilon_\mathrm{Mn}=5.47$ and 5.37, respectively) are in good agreement with the corresponding results of \citet[][$\log \epsilon_\mathrm{Mn}=5.46$ and 5.37, respectively]{BW07}.  No significant trends with equivalent width or excitation potential are visible in Fig.~\ref{fig:mn}.

Compared to the Mn abundance adopted in \citetalias{AGSS} ($\log \epsilon_\mathrm{Mn}=5.43\pm0.04$), we now have dedicated NLTE intensity calculations using the model atom of \citet{Bergemann07} for the individual Mn lines and 1D models we employ (instead of relying on the MAFAGS-ODF model), and adopted $S_\mathrm{H}=1$ (instead of $S_\mathrm{H}=0.05$).  We have also updated four oscillator strengths (Table~\ref{table:lines}) with the new data of \citet{DenHartog11}.

\subsection{Iron}
\label{Feresults}

Both \fei and \feii should be good indicators of the Fe abundance, as we have several clean solar lines, small NLTE corrections and accurate oscillator strengths.  Our derived Fe abundance from \fei lines ($\log \epsilon_\mathrm{Fe}=7.45\pm0.04$; $\pm$0.01 stat, $\pm$0.04 sys) overlaps the \feii result ($\log \epsilon_\mathrm{Fe}=7.51\pm0.04$; $\pm$0.01 stat, $\pm$0.04 sys) to within the mutual uncertainties, but the agreement is not perfect.  This may indicate a small error in the \feii oscillator strengths (as they are at least partially based on theoretical results, which are not always accurate), or perhaps slightly too high an adopted value of $S_\mathrm{H}$ (resulting in slightly too low NLTE corrections).  A similar size discrepancy exists in the \citetalias{HM} results (Table~\ref{table:abuns}), but reversed in sign: the neutral species returns an abundance $0.06$\,dex higher.  The only significant trend visible in Fig.~\ref{fig:fe} is in the difference between the 3D and \citetalias{HM} or \oneDAV\ results from \feii as a function of line strength: stronger \feii lines appear to show larger positive corrections due to 3D effects.

Considering all our adopted \fei and \feii lines, our final 3D+NLTE Fe abundance is
\begin{displaymath}
\log \epsilon_\mathrm{Fe}=7.47\pm0.04\ (\pm0.01\ \mathrm{stat},\ \pm0.04\ \mathrm{sys}),
\end{displaymath}
in very good agreement with the meteoritic value \citep[$7.45\pm0.01$;][]{Lodders09}.

Our derived Fe abundance from \feii lines is in perfect agreement with the 3D result of \citet[][$\log \epsilon_\mathrm{Fe}=7.51$]{Caffau11}, although the standard deviation of our result is smaller (0.04 vs.\ 0.06\,dex).  For lines in common, the equivalent widths employed in the two studies agree to within a few percent, so the difference in scatter presumably reflects a difference in the quality of the line selection.  Both our \fei and \feii abundances are consistent with those found in \citet[][$\log \epsilon_\mathrm{Fe}=7.44\pm0.05$ and $7.45\pm0.10$ respectively]{AspII}.  The difference in the scatter of the \feii result is in this case due to our use of the improved \citet{Melendez09} oscillator strengths.  The difference in the central value, whilst not statistically significant, probably reflects a slight difference in the temperature gradient between the two versions of the 3D model.  Our results are also in full agreement with \citet[][$\log \epsilon_\mathrm{Fe}=7.46\pm0.02$]{MB_fe}, who investigated NLTE line formation of Fe with the same \oneDAV\ solar model atmosphere as we employ here.

Recently, \citet{2010ApJ...724.1536F, 2012A&A...548A..35F} revisited the issue of the solar Fe abundance
in light of 3D magneto-hydrodynamic simulations of the solar atmosphere for different magnetic field strengths ($B_{\rm z} = 0-200$\,G). 
Their 3D models were calculated with the same {\sc stagger} code as we employ, but with less up-to-date
opacities and equation-of-state. 
They found quite substantial effects on the derived Fe abundance due to the presence of magnetic fields: in some
cases up to $+0.15$\,dex for the strongest magnetic fields. 
For typical \fei lines employed here and elsewhere, with small or negligible Land\'e factors, the effects are much more sedate: 
$\approx +0.04$\,dex for $B_{\rm z} = 200$\,G. Most of this is an indirect effect: it is not Zeeman broadening (which in 
any case would strengthen the line and thus lead to lower inferred Fe abundance), but the impact of magnetic fields
on the atmospheric temperature structure that matters most. With magnetic dissipation included, the higher atmospheric
layers are heated relative to the non-magnetic case, with the difference amounting to $\approx$$130$\,K at $\log \tau_{\rm 500} = -2$
for the $B_{\rm z} = 200$\,G case \citep{2012A&A...548A..35F}. As a consequence, the number density of \fei is decreased
and a higher Fe abundance is required to reproduce the observed \fei lines; although \citet{2012A&A...548A..35F} did not consider
typical \feii lines used for abundance purposes, the expectation is that those lines should be rather insensitive to the different
temperature structures, as they are formed in significantly deeper layers. At face value, the agreement between the \fei\ and \feii\ 
results would be improved, especially since observations of the quiet Sun suggest the presence of a ubiquitous mixed-polarity magnetic field 
with an average strength of $\approx 100$\,G \citep{2004Natur.430..326T}.

We intend to return to this important issue in the future, but in the meantime we note that the
case for a significant upward revision of the solar Fe abundance (and by consequence many other elements) 
due to the presence of magnetic fields is not as unequivocal as argued by \citet{2010ApJ...724.1536F, 2012A&A...548A..35F}.
Firstly, at magnetic fields of 100\,G, the effect is in fact rather minor: $\approx 0.02$\,dex for lines similar to those we use.
Secondly, our recommended Fe abundance is based on both \fei and \feii lines. Thirdly, 
\citet{2013A&A...554A.118P} found that 3D MHD models of the solar atmosphere perform worse
than simulations without magnetic fields against a number of key observational diagnostics, including
the continuum centre-to-limb variation; they thus conclude that current MHD solar models are in fact 
less realistic than the one employed by us. 
In view of these findings, we recommend our 3D+NLTE value based on a 3D hydrodynamic solar model, 
but caution that further studies into the importance of magnetic fields are needed. 

In \citetalias{AGSS}, we adopted the result from \feii ($\log \epsilon_\mathrm{Fe}=7.50\pm0.04$) as our reference abundance.  Here we also utilise \altfei, because we now have dedicated NLTE calculations available for our lines with the \oneDAV\ model.  Relative to the \citetalias{AGSS} analysis, we have dropped two \fei lines: 
657.4\,nm, because it sits in the wing of H$\alpha$, and 660.9\,nm, because of its relatively large line strength.

\subsection{Cobalt}
\label{Coresults}

From our selection of weak \coi lines, we find a mean NLTE Co abundance of
\begin{displaymath}
\log \epsilon_\mathrm{Co}=4.93\pm0.05\ (\pm0.01\ \mathrm{stat},\ \pm0.05\ \mathrm{sys}).
\end{displaymath}
This is somewhat higher than the meteoritic value \citep[$4.87\pm0.01$;][]{Lodders09}, but still marginally consistent to within the mutual errors.  Our result agrees well with that of \citet[][$4.95\pm0.04$]{Bergemann11}, although in that paper a different model atmosphere, flux spectra and $S_\mathrm{H}=0.05$ were used, resulting in larger NLTE corrections than we see here with $S_\mathrm{H}=1$ ($+0.14$ vs.\ $+0.08$\,dex).  Our mean LTE \citetalias{HM} result ($\log \epsilon_\mathrm{Co}=4.94$) is also consistent with the \citetalias{HM}-based abundance derived by \citet[][$\log \epsilon_\mathrm{Co}=4.92$]{Cardon82}.  Our result exhibits a smaller dispersion however, reflecting the care we took in our line selection: $\sigma=0.06$ in our \citetalias{HM} results, $\sigma=0.08$ in \citeauthor{Cardon82}'s.  The dispersions of our 3D LTE and NLTE results were $\sigma=0.05$\,dex, similar to those of \citet{Bergemann10Co}, which is indicative of the intrinsic uncertainty of the oscillator strengths.

No substantial trend in abundances with line strength can be seen in Fig.~\ref{fig:co}.  A weak trend with excitation potential is visible in the 3D results: lines with $\chi_\mathrm{exc}>3$\,eV lead to an abundance of $\log \epsilon_\mathrm{Co}=4.90\pm0.02$\,($1\sigma$), whereas lower-excitation lines return an abundance of $\log \epsilon_\mathrm{Co}=4.97\pm0.05$\,($1\sigma$).  This may be an effect of imperfect $gf$-values, NLTE corrections or the temperature structure of the model atmosphere.  Inspection of the lower right panel of Fig.~\ref{fig:co} reveals that the trend is more severe with the \oneDAV\ model than the full 3D model, and yet more severe again with the \citetalias{HM} model.  Using the \citetalias{HM} model, the high-excitation lines give $\log \epsilon_\mathrm{Co}=4.92\pm0.02$\,($1\sigma$), whereas the low-excitation lines return $\log \epsilon_\mathrm{Co}=5.04\pm0.05$\,($1\sigma$); the switch to 3D atmospheric modelling is a clear improvement for solar analysis of Co.

Relative to \citetalias{AGSS} ($\log \epsilon_\mathrm{Co}=4.99\pm0.07$), the main update to the Co abundance here is that we calculate NLTE intensity corrections \citep[based on][]{Bergemann10Co} specifically for our different 1D models rather than the MAFAGS-ODF model, and use $S_\mathrm{H}=1$ instead of $S_\mathrm{H}=0.05$.  This accounts for $0.03$\,dex of the reduction; the remaining $0.03$\,dex comes from the updated opacities, equation of state, ionisation potentials and partition functions.

\subsection{Nickel}
\label{Niresults}

The mean 3D nickel abundance from \nii lines
\begin{displaymath}
\log \epsilon_\mathrm{Ni}=6.20\pm0.04\ (\pm\!<\!0.01\ \mathrm{stat},\ \pm0.04\ \mathrm{sys})
\end{displaymath}
is in excellent agreement with the meteoritic value \citep[$6.20\pm0.01$;][]{Lodders09}.  \niii lines indicated widely varying abundances, though the mean values they return with each model are broadly consistent with \nii results.  Using the theoretical $gf$-values of \citet{Fritzsche00} results in a far lower abundance scatter than any other $gf$-values, leading us to believe that these are currently the most accurate oscillator strengths available for optical \niii lines.  Given the uncertainty in the mean \niii value, we adopt the 3D \nii result as the most reliable estimate of the solar abundance.  \niii is the most model-sensitive of our ionised species. 

No trends with line strength or excitation potential can be seen in the 3D results (Fig.~\ref{fig:ni}).  The abundance corrections due to 3D effects have a clear dependence upon line strength, and a smaller correlation with excitation potential.  In contrast, the effect of the mean temperature structure is 0.04\,dex regardless of line strength.

Our result is consistent with that presented in \citet[][$\log \epsilon_\mathrm{Ni}=6.17\pm0.05$]{Scott09Ni}, but slightly higher due to the improved temperature structure of the improved 3D model we use here.  Compared to the result we reported in \citetalias{AGSS} ($\log \epsilon_\mathrm{Ni}=6.22\pm0.04$), our result here is slightly lower because we employed new $gf$-values and an expanded set of isotopic separations from \citet{Wood14} (in \citetalias{AGSS} and \citealt{Scott09Ni} we used $gf$-values from \citealt{Wickliffe97} and isotopic separations from \citealt{Litzen93}).  Improvements in the overall opacity, equation of state, ionisation potential and partition function also play a small role in the difference from \citetalias{AGSS}.

\setcounter{table}{3}
	
\begin{table*}[tbp]
\centering
\caption[Abundances implied by all lines]{Average abundances implied by Sc\,\textsc{i}, Sc\,\textsc{ii}, Ti\,\textsc{i}, Ti\,\textsc{ii}, V\,\textsc{i}, V\,\textsc{ii}, Cr\,\textsc{i}, Cr\,\textsc{ii}, Mn\,\textsc{i}, Fe\,\textsc{i}, Fe\,\textsc{ii}, Co\,\textsc{i}, Ni\,\textsc{i} and Ni\,\textsc{ii} lines.  Abundances are given as the weighted mean across all lines in the given list, taking into account NLTE corrections for \altsci, \altscii, \alttii, \altvi, \altcri, \altmni, \fei and \altcoi.  \vii and \niii are shown in brackets because we do not consider these results reliable enough to include in our final adopted abundances.  We also give our final recommended solar photospheric abundance of each element, compared with the abundance in CI chondritic meteorites \protect\citep[][normalised to the silicon abundance determined in \protect\citetalias{AGSS_NaCa}]{Lodders09}.  Note that because all means were computed using abundances accurate to three decimal places, entries in columns 8 and 9 differ in some cases from the differences between the entries in columns 3--5.}
\label{table:abuns}
\begin{tabular}{l l c c c c c c c c c}
\hline
\hline
& Species & 3D & \oneDAV & HM & \textsc{marcs} & {\sc miss} & 3D$-$HM & 3D$-$\oneDAV & Recommended & Meteoritic\\
\hline
$\log \epsilon_\mathrm{Sc}$     & Sc\,\textsc{i}    & 3.14$\pm$0.09 & 3.21 & 3.28 & 3.18 & 3.23 & $-$0.14 & $-$0.07 & $3.16\pm0.04$ & $3.05\pm0.02$\\
                                & Sc\,\textsc{ii}   & 3.17$\pm$0.04 & 3.16 & 3.19 & 3.14 & 3.19 & $-$0.02 & \ph0.01 & \\
                                & Sc all            & 3.16$\pm$0.04 & 3.18 & 3.22 & 3.15 & 3.21 & $-$0.07 & $-$0.03 & \vspace{1mm}\\
$\log \epsilon_\mathrm{Ti}$     & Ti\,\textsc{i}    & 4.88$\pm$0.05 & 4.94 & 4.99 & 4.90 & 4.93 & $-$0.11 & $-$0.06 & $4.93\pm0.04$ & $4.91\pm0.03$\\
                                & Ti\,\textsc{ii}   & 4.97$\pm$0.04 & 4.94 & 4.97 & 4.91 & 4.97 & \ph0.00 & \ph0.02 & \\
                                & Ti all            & 4.90$\pm$0.04 & 4.94 & 4.99 & 4.90 & 4.94 & $-$0.08 & $-$0.04 & \vspace{1mm}\\
$\log \epsilon_\mathrm{V}$      & V\,\textsc{i}     & 3.89$\pm$0.08 & 3.99 & 4.07 & 3.96 & 4.00 & $-$0.18 & $-$0.10 & $3.89\pm0.08$ & $3.96\pm0.02$\\
                                & (V\,\textsc{ii})  & 4.00$\pm$0.04 & 3.98 & 4.01 & 3.95 & 4.01 & $-$0.01 & \ph0.02 & \vspace{1mm}\\
$\log \epsilon_\mathrm{Cr}$     & Cr\,\textsc{i}    & 5.60$\pm$0.04 & 5.62 & 5.66 & 5.57 & 5.63 & $-$0.06 & $-$0.02 & $5.62\pm0.04$ & $5.64\pm0.01$\\
                                & Cr\,\textsc{ii}   & 5.65$\pm$0.04 & 5.62 & 5.63 & 5.56 & 5.65 & $+$0.03 & \ph0.04 & \\
                                & Cr all            & 5.62$\pm$0.04 & 5.62 & 5.65 & 5.57 & 5.64 & $-$0.04 & $-$0.01 & \vspace{1mm}\\
$\log \epsilon_\mathrm{Mn}$     & Mn\,\textsc{i}    & 5.42$\pm$0.04 & 5.43 & 5.47 & 5.37 & 5.42 & $-$0.04 & $-$0.00 & $5.42\pm0.04$ & $5.48\pm0.01$\vspace{1mm}\\
$\log \epsilon_\mathrm{Fe}$     & Fe\,\textsc{i}    & 7.45$\pm$0.04 & 7.46 & 7.52 & 7.41 & 7.46 & $-$0.07 & $-$0.00 & $7.47\pm0.04$ & $7.45\pm0.01$ \\
                                & Fe\,\textsc{ii}   & 7.51$\pm$0.04 & 7.46 & 7.46 & 7.42 & 7.49 & $+$0.05 & \ph0.05 & \\
                                & Fe all            & 7.47$\pm$0.04 & 7.46 & 7.50 & 7.41 & 7.47 & $-$0.03 & \ph0.01 & \vspace{1mm}\\
$\log \epsilon_\mathrm{Co}$     & Co\,\textsc{i}    & 4.93$\pm$0.05 & 4.96 & 4.99 & 4.92 & 4.96 & $-$0.06 & $-$0.03 & $4.93\pm0.05$ & $4.87\pm0.01$\vspace{1mm}\\
$\log \epsilon_\mathrm{Ni}$     & Ni\,\textsc{i}    & 6.20$\pm$0.04 & 6.20 & 6.24 & 6.15 & 6.23 & $-$0.04 & \ph0.00 & $6.20\pm0.04$ & $6.20\pm0.01$\\
                                & (Ni\,\textsc{ii}) & 6.30$\pm$0.10 & 6.23 & 6.24 & 6.19 & 6.26 & $+$0.06 & \ph0.08 & \vspace{0.5mm}\\
\hline							  
\end{tabular}						  
\end{table*}

\section{Comments and discussion}
\label{discussion}

\subsection{Sensitivity to temperature: 3D vs.\ \oneDAV\ vs.\ HM}
\label{sensitivity}

Table~\ref{table:abuns} shows that the results for the once-ionised species 
are typically less model-dependent than those of the neutral species; this is to be expected for these 
dominant species.  We notice that the model-dependence of the abundances of neutral species increases 
with decreasing ionisation potential, whereas the model-dependence of abundances from ionised lines 
increases with ionisation potential.  This reflects the general rule that 
the more in majority a species is, the less sensitive its lines will be to the ionisation balance,
and therefore less affected by the temperature structure of the model atmosphere.
   
From Table~\ref{table:abuns}, we see that the differences 3D$-$\oneDAV\ vary widely between different neutral species. From values of 0.06--0.10\,dex for \altsci, \tii and \altvi, they decrease to 0.00--0.03\,dex for the rest of the neutrals, reflecting the lower ionisation energies and therefore more severe minority status of \altsci, \tii and \vi compared to the other neutrals.

When looking at the plots in Figs.~\ref{fig:sc}--\ref{fig:ni}, we clearly see that the 3D$-$\oneDAV\ abundance
difference is also related to the excitation potentials of individual lines, or
more precisely, the difference between the ionisation and excitation
energies ($E_{\rm ion}-E_{\rm exc}$).  This difference is the most important parameter for the
temperature sensitivity of lines of minor species like the
neutral iron group elements. Lines with lower excitation energies are typically more sensitive than higher-excitation
lines to 
 higher atmospheric layers and the presence of atmospheric inhomogeneities, as seen in our 3D$-$\citetalias{HM} and 3D$-$\oneDAV\ results, respectively. 
A similar argument holds also for line strengths: stronger lines are typically formed higher, so 
show larger sensitivity to both the mean structure and horizontal inhomogeneities.

\subsection{Sensitivity to collisional broadening and HFS}
\label{controls}

Collisional broadening is now well determined for neutral species and \feii \citep{Anstee95, Barklem97, 
Barklem98, Barklem00, 2005A&A...435..373B}.  Even extreme collisional sensitivity should therefore not be a major source of error when using neutral lines and \feii in the current analysis.  For other ionic lines, the enhancement factor used with the classical \citet{Unsold} broadening recipe is a potential source of error.  As \altscii, \tiii and \vii lines were mostly insensitive to its variation though, even with ionic lines the broadening treatment should contribute very little to our uncertainties.

The sensitivity of derived abundances to hyperfine (and by
implication, isotopic) structure varies greatly with different
lines, species and abundance-determination techniques.  Clearly,
lines with large HFS (i.e.~transitions between
levels with large $A$, $B$, and/or $J$ values, or in nuclei with
large $I$) will be most affected.  If one finds abundances using
equivalent widths, the strongest lines are those most sensitive to
the HFS treatment.  This is because the spreading of a strong line
into multiple components causes it to become either partially or
wholly desaturated, whereas a single component would be more
saturated.  This means that completely neglecting HFS or isotopic
structure often leads to overestimated abundances, a common concern
in past 1D analyses \citep[e.g.~][]{vonderHeide68, Holweger71,
Kurucz93, Prochaska00}.  Broadening by HFS or isotopic structure 
modifies the depth of line formation in general for all lines 
(pushing them deeper into the photosphere), so it can play a 
role even for fainter lines, even when equivalent widths are used for fitting
rather than profile fits. Furthermore, it is very important 
in combination with NLTE line formation, especially when NLTE 
abundance corrections are computed from differences between 
LTE and NLTE equivalent widths.

To ascertain the overall impact of HFS on our abundances, we also computed all \oneDAV\ abundances with HFS neglected, and calculated the mean HFS correction $\Delta_\mathrm{HFS}\equiv\log\epsilon_\mathrm{no\,HFS}-\log\epsilon_\mathrm{HFS}$ for our sample of lines. We found that \mni was by far the species most affected, with $\Delta_\mathrm{HFS}=0.16$\,dex.  \coi was the next most strongly affected ($\Delta_\mathrm{HFS}=0.05$\,dex), followed by \vii ($\Delta_\mathrm{HFS}=0.04$\,dex), \vi ($\Delta_\mathrm{HFS}=0.03$\,dex) and \altsci/\scanii (both $\Delta_\mathrm{HFS}=0.01$\,dex).  HFS in Ti and Cr had virtually no effect.

\section{Previous Solar Abundance Compilations}
\label{compilations}

Table \ref{table:compilations} compares the values we recommend here with those adopted in some of the most commonly-used compilations:
\citetalias{AG89}, \citetalias{GS98}, \citetalias{AGS05}, \citetalias{AGSS} and \citet{Lodders09}. 
We note however that with the exception of \citetalias{AGSS}, all of the others are in fact compilations
of results from the literature, all with their own methodologies, spectrum synthesis codes, model atmospheres,
and error estimation procedures, which makes the recommended solar values a rather inhomogeneous mixture. 
In particular, none of the previous studies have attempted to account for systematic errors in 
the quoted abundance uncertainties.

Not surprisingly, the solar abundances that we present here are quite similar to those of \citetalias{AGSS}.  As outlined in detail in Sect. \ref{results} however, we have updated them following a complete re-assessment of all analysis ingredients, including continuous opacities, equation-of-state, line selection, atomic data and NLTE abundance corrections. In most cases this has resulted in very minor changes.  Cobalt ($-0.06$\,dex, see Sect.~\ref{Coresults}) is the notable exception, explained mostly by improved NLTE calculations.

For the Fe-peak elements, \citetalias{AGS05} only included 1D-based analyses with the 
exception of Fe \citep{AspII}, although it still updated the recommended values for a few elements relative to \citetalias{GS98}.  \citetalias{GS98} was in turn primarily based on \citetalias{AG89}. The main difference between the latter two is the adopted Fe value, where \citetalias{AG89} still preferred a high value (0.2\,dex larger than derived here); see \citet{GS99} for a detailed description of the reasons for the long-standing debate on the solar Fe abundance. Since then the preferred Fe value has not changed drastically, in spite of the advent of 3D hydrodynamic model atmospheres, more complete NLTE calculations and improved $gf$-values -- which is reassuring.

Compared with \citet{Lodders09}, our solar abundances for the Fe-peak elements are
similar overall, but there are some rather large isolated differences.  These include
Sc ($+0.06$\,dex), V ($-0.11$\,dex) and Mn ($+0.05$\,dex). As outlined in Sect. \ref{results},
we are confident that our analysis is the most reliable and accurate possible today.

\begin{table}
\centering
\caption{The present-day solar photospheric abundances for the Fe-peak elements Sc to Ni that we recommend here, compared with oft-used solar abundance compilations: \citetalias{AG89}, \citetalias{GS98}, \citetalias{AGS05}, \citetalias{AGSS} and \citet{Lodders09} (LPG09)}
\label{table:compilations} 
\begin{tabular}{c@{\hspace{3mm}}l@{\hspace{3mm}}c@{\hspace{3mm}}c@{\hspace{3mm}}c@{\hspace{3mm}}c@{\hspace{2mm}}c@{\hspace{2mm}}c}
\hline Z & el. & This work & AG89 & GS98 & AGS05 & AGSS09 & LPG09 \\
\hline
\hline
  21 &  Sc& $    3.16\pm 0.04 $ &    3.10 &    3.17 &    3.05 &    3.15 &    3.10\\
  22 &  Ti& $    4.93\pm 0.04 $ &    4.99 &    5.02 &    4.90 &    4.95 &    4.90\\
  23 &   V& $    3.89\pm 0.08 $ &    4.00 &    4.00 &    4.00 &    3.93 &    4.00\\
  24 &  Cr& $    5.62\pm 0.04 $ &    5.67 &    5.67 &    5.64 &    5.64 &    5.64\\
  25 &  Mn& $    5.42\pm 0.04 $ &    5.39 &    5.39 &    5.39 &    5.43 &    5.37\\
  26 &  Fe& $    7.47\pm 0.04 $ &    7.67 &    7.50 &    7.45 &    7.50 &    7.45\\
  27 &  Co& $    4.93\pm 0.05 $ &    4.92 &    4.92 &    4.92 &    4.99 &    4.92\\
  28 &  Ni& $    6.20\pm 0.04 $ &    6.25 &    6.25 &    6.23 &    6.22 &    6.23\\
\hline
\end{tabular}
\end{table}

\section{Conclusions}
\label{conclusions}

We have determined the abundances of all the iron group elements in the Sun. For our analysis, we have
carefully assessed all relevant atomic data, made very stringent line selections, employed a
highly realistic 3D model for the solar atmosphere and accounted for departures from LTE.  
We have attempted to quantify the remaining systematic uncertainties stemming from 
possible errors in atmospheric and line-formation modelling, and to properly account for statistical errors. 

Our final recommended abundances of Sc, Ti, V, Cr, Mn, Fe, Co and Ni are given in Table~\ref{table:abuns}.
The derived abundances generally show good agreement with the meteoritic values, and 
between different ionisation stages, but some discrepancies remain.  Trends in abundances with 
excitation potential or line strength are largely absent in the 3D results, but are visible in a number of results 
from 1D models.  The level of agreement between theoretical and observed line profiles with the 3D model is clearly satisfactory.  
Nonetheless, theoretical profiles computed in 3D systematically underestimate the line width by a small amount, suggesting 
that some additional work on improving the atmospheric velocity field or NLTE effects is still required before perfect 
agreement can be claimed. Nevertheless, we are confident that the solar photospheric abundances that we present here are the most accurate possible by today's standards.

\begin{acknowledgements}
We thank Dan Bayliss, Mike Bessell, Remo Collet, Peter Hannaford, Wolfgang Hayek, Lyudmila Mashonkina, Tiago Pereira, Chris Sneden and Regner Trampedach for helpful discussions, and the referee for constructive feedback.  PS, NG and MA variously thank the Max Planck Institut f\"ur Astrophysik, Garching, the Centre Spatial de Li\`ege, the Department of Astrophysics, Geophysics and Oceanography, University of Li\`ege and Mount Stromlo Observatory for support and hospitality during the production of this paper.  We acknowledge further support from IAU Commission 46, the Lorne Trottier Chair in Astrophysics, the (Canadian) Institute for Particle Physics, the Banting Fellowship scheme as administered by the Natural Science and Engineering Research Council of Canada, the UK Science \& Technology Facilities Council (PS), the Australian Research Council (MA) and the Royal Belgian Observatory (NG).
\end{acknowledgements}

\bibliography{CObiblio,AbuGen,FePeakGeneral,Sc,Ti,V,Cr,Mn,Co,Ni,Others,MA}

\Online

\onecolumn
\setcounter{table}{0}
\begin{center}
\topcaption{\label{table:lines} Lines retained in this analysis: atomic and solar data, line weightings, LTE abundance results for the 5 models used in this analysis, NLTE corrections to the LTE result (when available), and the corresponding 3D+NLTE abundance result.  Asterisks (*) indicate lines for which the weighting has been reduced by 1 due to a large uncertainty in the $gf$ value.}
\tablefirsthead{%
  \hline
  $\lambda$ & & \multicolumn{4}{c}{Atomic levels} & $E_{\rm exc}$ & $\log$ $gf$ & $gf$ & $W_\lambda$ & Wt.  &\multicolumn{5}{c}{LTE Abundances}& $\Delta_{\rm NLTE}$ & 3D \\
  (nm) & & \multicolumn{2}{c}{Lower} & \multicolumn{2}{c}{Upper} & (eV) & & ref. & (pm) &  & 3D & $\langle 3{\rm D}\rangle$ & HM & \marcs & {\sc miss} & (3D) & NLTE\\
  \hline
   & & & & & & & & & & & & & & & & & \\
  }
\tablehead{%
  \hline 
  $\lambda$ & & \multicolumn{4}{c}{Atomic levels} & $E_{\rm exc}$ & $\log$ $gf$ & $gf$ & $W_\lambda$ & Wt.  &\multicolumn{5}{c}{LTE Abundances}& $\Delta_{\rm NLTE}$ & 3D \\
  (nm) & & \multicolumn{2}{c}{Lower} & \multicolumn{2}{c}{Upper} & (eV) & & ref. & (pm) &  & 3D & $\langle 3{\rm D}\rangle$ & HM & \marcs & {\sc miss} & (3D) & NLTE\\
  \hline
  \multicolumn{18}{r}{continued.}\\
  \hline
   & & & & & & & & & & & & & & & & & \\
  }
\tabletail{%
  \hline
  \multicolumn{18}{r}{continued on next page}\\
  \hline
}
\tablelasttail{\hline}
\begin{mpsupertabular}{r@{}c@{\ \ \ }r@{~}l@{\ }r@{~}l@{\ \ \ }c@{\ \ \ }c@{\ \ \ }c@{\ \ \ }c@{\ \ \ }c@{\ \ \ }c@{\ \ \ }c@{\ \ \ }c@{\ \ \ }c@{\ \ \ }c@{\ \ \ }r@{\ \ \ }c}
\multicolumn{18}{c} {\scani} \\
474.3821 & &  \footnotesize($^3$F)4s &\footnotesize $^4$F$_\frac{9}{2}$  &\footnotesize ($^3$F)4p &\footnotesize $^4$D$_\frac{7}{2}$  & 1.448 & \ph0.422 &  1 & 0.82 & 1\pst      & 2.974 & 3.039 & 3.104 & 3.006 & 3.065 & $+$0.15\phz & 3.124 \\	      
508.1561 & &  \footnotesize($^3$F)4s &\footnotesize $^4$F$_\frac{9}{2}$  &\footnotesize ($^3$F)4p &\footnotesize $^4$F$_\frac{9}{2}$  & 1.448 & \ph0.469 &  1 & 1.00 & 2\pst      & 2.989 & 3.056 & 3.121 & 3.022 & 3.080 & $+$0.16\phz & 3.149 \\	      
535.6097 & &  \footnotesize($^3$F)4s &\footnotesize $^2$F$_\frac{7}{2}$  &\footnotesize ($^3$F)4p &\footnotesize $^2$D$_\frac{5}{2}$  & 1.865 & \ph0.168 &  1 & 0.20 & 3\pst      & 2.974 & 3.030 & 3.089 & 2.994 & 3.056 & $+$0.15\phz & 3.124 \\	      
567.1828 & &  \footnotesize($^3$F)4s &\footnotesize $^4$F$_\frac{9}{2}$  &\footnotesize ($^3$F)4p &\footnotesize $^4$G$_\frac{11}{2}$ & 1.448 & \ph0.495 &  1 & 1.24 & 2\pst      & 3.024 & 3.092 & 3.159 & 3.058 & 3.114 & $+$0.15\phz & 3.174 \\	      
623.9800 & &  \footnotesize4s$^2$    &\footnotesize $^2$D$_\frac{3}{2}$  &\footnotesize ($^3$D)4sp&\footnotesize $^4$D$_\frac{3}{2}$  & 0.000 & $-$1.780 &  1 & 0.22 & 1\pst      & 2.985 & 3.133 & 3.224 & 3.108 & 3.146 & $+$0.15\phz & 3.135 \\     
   & & & & & & & & & & & & & & &\\                           			                            													                                          
\multicolumn{18}{c} {\scanii} \\    																							                    
442.0661 & &  \footnotesize   3d$^2$ &\footnotesize $^3$F$_4$            &\footnotesize   4p      &\footnotesize $^3$F$_3$            & 0.618 & $-$2.273 &  1 & 1.51 & 2\pst      & 3.109 & 3.109 & 3.140 & 3.093 & 3.143 & $-$0.01\phz & 3.099 \\
443.1362 & &  \footnotesize   3d$^2$ &\footnotesize $^3$F$_3$            &\footnotesize   4p      &\footnotesize $^3$F$_2$            & 0.605 & $-$1.969 &  1 & 2.92 & 1\pst      & 3.165 & 3.162 & 3.193 & 3.142 & 3.195 & $-$0.01\phz & 3.155 \\
535.7202 & &  \footnotesize   3d$^2$ &\footnotesize $^3$P$_2$            &\footnotesize   4p      &\footnotesize $^1$P$_1$            & 1.507 & $-$2.111 &  1 & 0.43 & 2\pst      & 3.131 & 3.130 & 3.153 & 3.110 & 3.164 & \ph0.00\phz & 3.131 \\
564.1000 & &  \footnotesize   3d$^2$ &\footnotesize $^3$P$_1$            &\footnotesize   4p      &\footnotesize $^3$P$_2$            & 1.500 & $-$1.131 &  1 & 3.65 & 1\pst      & 3.246 & 3.236 & 3.255 & 3.205 & 3.264 & $-$0.02\phz & 3.226 \\
565.8362 & &  \footnotesize   3d$^2$ &\footnotesize $^3$P$_0$            &\footnotesize   4p      &\footnotesize $^3$P$_1$            & 1.497 & $-$1.208 &  1 & 3.14 & 1\pst      & 3.221 & 3.212 & 3.232 & 3.183 & 3.241 & $-$0.01\phz & 3.211 \\
566.7164 & &  \footnotesize   3d$^2$ &\footnotesize $^3$P$_1$            &\footnotesize   4p      &\footnotesize $^3$P$_1$            & 1.500 & $-$1.309 &  1 & 2.81 & 1\pst      & 3.245 & 3.238 & 3.258 & 3.212 & 3.268 & $-$0.01\phz & 3.235 \\
566.9055 & &  \footnotesize   3d$^2$ &\footnotesize $^3$P$_1$            &\footnotesize   4p      &\footnotesize $^3$P$_0$            & 1.500 & $-$1.200 &  1 & 3.26 & 1\pst      & 3.256 & 3.243 & 3.263 & 3.214 & 3.272 & $-$0.01\phz & 3.246 \\
568.4214 & &  \footnotesize   3d$^2$ &\footnotesize $^3$P$_2$            &\footnotesize   4p      &\footnotesize $^3$P$_1$            & 1.507 & $-$1.074 &  1 & 3.56 & 2\pst      & 3.174 & 3.164 & 3.183 & 3.133 & 3.193 & $-$0.02\phz & 3.154 \\
660.4578 & &  \footnotesize   3d$^2$ &\footnotesize $^1$D$_2$            &\footnotesize   4p      &\footnotesize $^1$D$_2$            & 1.357 & $-$1.309 &  1 & 3.54 & 1*         & 3.214 & 3.202 & 3.219 & 3.173 & 3.227 & $-$0.01\phz & 3.204 \\
   & & & & & & & & & & & & & & & \\

\multicolumn{18}{c} {\tii} \\
428.1363 & & \footnotesize ($^4$F)4s &\footnotesize$^5$F$_1$             &\footnotesize($^4$F)4p  &\footnotesize$^5$D$_2$             & 0.813 & $-$1.260 & 2  & 2.40 & 1\pst      & 4.787 & 4.856 & 4.938 & 4.825 & 4.878 & $+$0.056    & 4.843 \\
446.5805 & & \footnotesize ($^4$P)4s &\footnotesize$^5$P$_2$             &\footnotesize($^4$P)4p  &\footnotesize$^5$P$_3$             & 1.739 & $-$0.130 & 2  & 3.56 & 2\pst      & 4.829 & 4.862 & 4.933 & 4.822 & 4.886 & $+$0.050    & 4.879 \\
475.8118 & & \footnotesize ($^2$H)4s &\footnotesize$^3$H$_5$             &\footnotesize($^2$H)4p  &\footnotesize$^3$H$_5$             & 2.249 & \ph0.510 & 2  & 4.18 & 3\pst      & 4.792 & 4.809 & 4.876 & 4.763 & 4.832 & $+$0.053    & 4.845 \\
475.9269 & & \footnotesize ($^2$H)4s &\footnotesize$^3$H$_6$             &\footnotesize($^2$H)4p  &\footnotesize$^3$H$_6$             & 2.256 & \ph0.590 & 2  & 4.60 & 2\pst      & 4.810 & 4.821 & 4.889 & 4.772 & 4.842 & $+$0.053    & 4.863 \\
496.4715 & & \footnotesize ($^3$F)4sp&\footnotesize$^5$G$_2$             &\footnotesize4s($^4$F)5s&\footnotesize$^5$F$_2$             & 1.969 & $-$0.820 & 3  & 0.77 & 2\pst      & 4.826 & 4.880 & 4.942 & 4.845 & 4.905 & $+$0.060    & 4.886 \\
502.2866 & & \footnotesize ($^4$F)4s &\footnotesize$^5$F$_3$             &\footnotesize($^4$F)4p  &\footnotesize$^5$G$_3$             & 0.826 & $-$0.330 & 2  & 6.99 & 1\pst      & 4.817 & 4.804 & 4.896 & 4.753 & 4.808 & $+$0.065    & 4.882 \\
511.3439 & & \footnotesize ($^4$F)4s &\footnotesize$^3$F$_3$             &\footnotesize($^3$P)4sp &\footnotesize$^3$D$_2$             & 1.443 & $-$0.700 & 2  & 2.45 & 2\pst      & 4.782 & 4.838 & 4.912 & 4.802 & 4.860 & $+$0.049    & 4.831 \\
514.5459 & & \footnotesize ($^4$F)4s &\footnotesize$^3$F$_4$             &\footnotesize($^3$P)4sp &\footnotesize$^3$D$_3$             & 1.460 & $-$0.540 & 2  & 3.31 & 1\pst      & 4.838 & 4.884 & 4.959 & 4.845 & 4.904 & $+$0.048    & 4.886 \\
514.7477 & & \footnotesize 4s$^2$    &\footnotesize$^3$F$_2$             &\footnotesize($^3$F)4sp &\footnotesize$^3$F$_3$             & 0.000 & $-$1.940 & 2  & 3.46 & 1\pst      & 4.781 & 4.871 & 4.968 & 4.840 & 4.882 & $+$0.086    & 4.867 \\
515.2184 & & \footnotesize 4s$^2$    &\footnotesize$^3$F$_3$             &\footnotesize($^3$F)4sp &\footnotesize$^3$F$_4$             & 0.021 & $-$1.950 & 2  & 3.32 & 2\pst      & 4.782 & 4.873 & 4.970 & 4.843 & 4.885 & $+$0.084    & 4.866 \\
521.9699 & & \footnotesize 4s$^2$    &\footnotesize$^3$F$_3$             &\footnotesize($^3$F)4sp &\footnotesize$^3$F$_2$             & 0.021 & $-$2.220 & 2  & 2.24 & 2\pst      & 4.783 & 4.896 & 4.991 & 4.868 & 4.909 & $+$0.081    & 4.864 \\
522.3620 & & \footnotesize ($^3$F)4sp&\footnotesize$^5$F$_2$             &\footnotesize4s($^4$F)5s&\footnotesize$^5$F$_2$             & 2.092 & $-$0.490 & 2  & 1.22 & 1\pst      & 4.824 & 4.874 & 4.935 & 4.836 & 4.899 & $+$0.054    & 4.878 \\
524.7288 & & \footnotesize ($^3$F)4sp&\footnotesize$^5$F$_3$             &\footnotesize4s($^4$F)5s&\footnotesize$^5$F$_2$             & 2.103 & $-$0.640 & 3  & 0.85 & 1\pst      & 4.802 & 4.853 & 4.914 & 4.817 & 4.879 & $+$0.057    & 4.859 \\
525.2098 & & \footnotesize 4s$^2$    &\footnotesize$^3$F$_4$             &\footnotesize($^3$F)4sp &\footnotesize$^3$F$_3$             & 0.048 & $-$2.360 & 4  & 1.64 & 1\pst      & 4.768 & 4.890 & 4.984 & 4.863 & 4.905 & $+$0.079    & 4.847 \\
529.5774 & & \footnotesize 4s$^2$    &\footnotesize$^3$P$_2$             &\footnotesize($^1$D)4sp &\footnotesize$^3$D$_3$             & 1.067 & $-$1.590 & 2  & 1.06 & 2\pst      & 4.824 & 4.907 & 4.984 & 4.875 & 4.927 & $+$0.065    & 4.889 \\
549.0147 & & \footnotesize ($^4$F)4s &\footnotesize$^3$F$_4$             &\footnotesize($^4$F)4p  &\footnotesize$^5$D$_3$             & 1.460 & $-$0.840 & 2  & 2.03 & 1\pst      & 4.795 & 4.857 & 4.929 & 4.821 & 4.878 & $+$0.051    & 4.846 \\
566.2147 & & \footnotesize ($^3$F)4sp&\footnotesize$^5$D$_4$             &\footnotesize4s($^4$F)5s&\footnotesize$^5$F$_5$             & 2.318 & \ph0.010 & 3  & 2.12 & 1\pst      & 4.802 & 4.843 & 4.903 & 4.802 & 4.866 & $+$0.057    & 4.859 \\
568.9459 & & \footnotesize ($^3$F)4sp&\footnotesize$^5$D$_2$             &\footnotesize4s($^4$F)5s&\footnotesize$^5$F$_3$             & 2.297 & $-$0.360 & 3  & 1.13 & 1\pst      & 4.824 & 4.871 & 4.930 & 4.833 & 4.896 & $+$0.058    & 4.882 \\
570.2658 & & \footnotesize ($^3$F)4sp&\footnotesize$^5$D$_1$             &\footnotesize4s($^4$F)5s&\footnotesize$^5$F$_2$             & 2.292 & $-$0.590 & 3  & 0.70 & 1\pst      & 4.815 & 4.864 & 4.922 & 4.827 & 4.889 & $+$0.059    & 4.874 \\
571.6441 & & \footnotesize ($^3$F)4sp&\footnotesize$^5$D$_2$             &\footnotesize4s($^4$F)5s&\footnotesize$^5$F$_2$             & 2.297 & $-$0.720 & 3  & 0.54 & 1\pst      & 4.828 & 4.878 & 4.935 & 4.841 & 4.903 & $+$0.058    & 4.886 \\
586.6429 & \footnote[1]{Isotopic splitting included (see Table \protect\ref{table:hfs}); wavelength corresponds to $^{50}$Ti component.\protect\label{Tiisonote}}		     	             	                                       
           & \footnotesize 4s$^2$    &\footnotesize$^3$P$_2$             &\footnotesize($^4$F)4p  &\footnotesize$^3$D$_3$             & 1.067 & $-$0.790 & 2  & 4.46 & 2\pst      & 4.833 & 4.887 & 4.971 & 4.849 & 4.899 & $+$0.042    & 4.875 \\
592.2088 & \color{BrickRed}{$^a$}																							                       
           & \footnotesize 4s$^2$    &\footnotesize$^3$P$_0$             &\footnotesize($^4$F)4p  &\footnotesize$^3$D$_1$             & 1.046 & $-$1.380 & 2  & 1.79 & 2\pst      & 4.802 & 4.886 & 4.964 & 4.853 & 4.903 & $+$0.048    & 4.850 \\
609.2789 & & \footnotesize ($^2$G)4s &\footnotesize$^3$G$_5$             &\footnotesize($^4$F)4p  &\footnotesize$^3$G$_5$             & 1.887 & $-$1.380 & 2  & 0.36 & 2\pst      & 4.881 & 4.942 & 5.008 & 4.909 & 4.964 & $+$0.057    & 4.938 \\
625.8099 & & \footnotesize ($^4$F)4s &\footnotesize$^3$F$_3$             &\footnotesize($^3$F)4sp &\footnotesize$^3$G$_4$             & 1.443 & $-$0.390 & 2  & 5.05 & 3\pst      & 4.912 & 4.943 & 5.023 & 4.901 & 4.952 & $+$0.045    & 4.957 \\
630.3753 & & \footnotesize ($^4$F)4s &\footnotesize$^3$F$_3$             &\footnotesize($^3$F)4sp &\footnotesize$^3$G$_3$             & 1.443 & $-$1.580 & 2  & 0.68 & 1\pst      & 4.912 & 4.986 & 5.061 & 4.955 & 5.004 & $+$0.056    & 4.968 \\
631.2234 & & \footnotesize ($^4$F)4s &\footnotesize$^3$F$_4$             &\footnotesize($^3$F)4sp &\footnotesize$^3$G$_4$             & 1.460 & $-$1.550 & 2  & 0.68 & 2\pst      & 4.892 & 4.966 & 5.040 & 4.935 & 4.985 & $+$0.055    & 4.947 \\
659.9104 & & \footnotesize 4s$^2$    &\footnotesize$^1$D$_2$             &\footnotesize($^3$F)4sp &\footnotesize$^1$F$_3$             & 0.900 & $-$2.029 & 5  & 0.80 & 2\pst      & 4.857 & 4.955 & 5.038 & 4.927 & 4.970 & $+$0.083    & 4.940 \\
735.7726 & & \footnotesize ($^4$F)4s &\footnotesize$^3$F$_3$             &\footnotesize($^3$F)4sp &\footnotesize$^3$F$_3$             & 1.443 & $-$1.020 & 2  & 1.99 & 2\pst      & 4.814 & 4.886 & 4.960 & 4.854 & 4.901 & $+$0.054    & 4.868 \\
842.6504 & & \footnotesize ($^4$F)4s &\footnotesize$^5$F$_3$             &\footnotesize($^3$F)4sp &\footnotesize$^5$D$_2$             & 0.826 & $-$1.197 & 6  & 4.66 & 2\pst      & 4.832 & 4.903 & 4.992 & 4.871 & 4.904 & $+$0.071    & 4.903 \\
843.5648 & & \footnotesize ($^4$F)4s &\footnotesize$^5$F$_4$             &\footnotesize($^3$F)4sp &\footnotesize$^5$D$_3$             & 0.836 & $-$0.967 & 6  & 5.76 & 1\pst      & 4.805 & 4.859 & 4.951 & 4.825 & 4.855 & $+$0.073    & 4.878 \\
867.5371 & & \footnotesize 4s$^2$    &\footnotesize$^3$P$_2$             &\footnotesize($^3$F)4sp &\footnotesize$^3$D$_3$             & 1.067 & $-$1.500 & 2  & 1.85 & 2\pst      & 4.774 & 4.869 & 4.948 & 4.838 & 4.880 & $+$0.075    & 4.849 \\
868.2979 & & \footnotesize 4s$^2$    &\footnotesize$^3$P$_1$             &\footnotesize($^3$F)4sp &\footnotesize$^3$D$_2$             & 1.053 & $-$1.790 & 2  & 1.07 & 2\pst      & 4.773 & 4.876 & 4.954 & 4.846 & 4.888 & $+$0.075    & 4.848 \\
869.2328 & & \footnotesize 4s$^2$    &\footnotesize$^3$P$_0$             &\footnotesize($^3$F)4sp &\footnotesize$^3$D$_1$             & 1.046 & $-$2.130 & 2  & 0.52 & 2\pst      & 4.765 & 4.874 & 4.951 & 4.845 & 4.887 & $+$0.074    & 4.839 \\
873.4711 & & \footnotesize 4s$^2$    &\footnotesize$^3$P$_1$             &\footnotesize($^3$F)4sp &\footnotesize$^3$D$_1$             & 1.053 & $-$2.240 & 2  & 0.41 & 1\pst      & 4.771 & 4.880 & 4.957 & 4.851 & 4.894 & $+$0.075    & 4.846 \\
   & & & & & & & & & & & & & & & \\

\multicolumn{18}{c} {\tiii} \\									                         												                       
440.9520 & & \footnotesize ($^3$P)4s &\footnotesize$^4$P$_\frac{3}{2}$   &\footnotesize($^3$F)4p  &\footnotesize$^4$D$_\frac{3}{2}$   & 1.231 & $-$2.530 & 7  & 3.81 & 2\pst      & 4.938 & 4.923 & 4.951 & 4.894 & 4.956 & & \\
444.4524 & \footnote[1]{Isotopic splitting included (see Table \protect\ref{table:hfs}); wavelength corresponds to $^{50}$Ti component.}					                                                  
           & \footnotesize 3d$^3$    &\footnotesize$^2$G$_\frac{7}{2}$   &\footnotesize($^3$F)4p  &\footnotesize$^2$F$_\frac{7}{2}$   & 1.116 & $-$2.200 & 7  & 5.99 & 1\pst      & 4.960 & 4.923 & 4.952 & 4.881 & 4.951 & & \\
449.3525 & \color{BrickRed}{$^a$}	          																                                         		                       
           & \footnotesize ($^1$D)4s &\footnotesize$^2$D$_\frac{3}{2}$   &\footnotesize($^3$F)4p  &\footnotesize$^4$F$_\frac{5}{2}$   & 1.080 & $-$2.780 & 7  & 3.18 & 1\pst      & 4.897 & 4.887 & 4.916 & 4.862 & 4.920 & & \\
458.3396 & \color{BrickRed}{$^a$}	          																                                         		                       
           & \footnotesize 3d$^3$    &\footnotesize$^4$P$_\frac{3}{2}$   &\footnotesize($^3$F)4p  &\footnotesize$^2$F$_\frac{5}{2}$   & 1.165 & $-$2.840 & 7  & 3.02 & 2\pst      & 4.985 & 4.977 & 5.005 & 4.953 & 5.010 & & \\
460.9253 & \color{BrickRed}{$^a$}																		                                         		                       
           & \footnotesize 3d$^3$    &\footnotesize$^4$P$_\frac{5}{2}$   &\footnotesize($^3$F)4p  &\footnotesize$^2$F$_\frac{5}{2}$   & 1.180 & $-$3.320 & 7  & 1.16 & 1\pst      & 4.933 & 4.932 & 4.959 & 4.914 & 4.966 & & \\
																						                                         
465.7212 & & \footnotesize ($^3$P)4s &\footnotesize$^4$P$_\frac{5}{2}$   &\footnotesize($^3$F)4p  &\footnotesize$^2$F$_\frac{7}{2}$   & 1.243 & $-$2.290 & 7  & 5.18 & 1\pst      & 4.988 & 4.956 & 4.983 & 4.917 & 4.986 & & \\
470.8656 & \color{BrickRed}{$^a$}																		                                         		                       
           & \footnotesize 3d$^3$    &\footnotesize$^2$P$_\frac{3}{2}$   &\footnotesize($^3$F)4p  &\footnotesize$^2$F$_\frac{5}{2}$   & 1.237 & $-$2.350 & 7  & 5.06 & 1\pst      & 4.951 & 4.923 & 4.949 & 4.885 & 4.953 & & \\
																						                                         
471.9533 & & \footnotesize ($^3$P)4s &\footnotesize$^4$P$_\frac{5}{2}$   &\footnotesize($^3$F)4p  &\footnotesize$^2$F$_\frac{5}{2}$   & 1.243 & $-$3.320 & 7  & 1.22 & 1\pst      & 5.013 & 5.011 & 5.038 & 4.993 & 5.043 & & \\
476.4518 & \color{BrickRed}{$^a$}					                                 									                                         		                       
           & \footnotesize 3d$^3$    &\footnotesize$^2$P$_\frac{3}{2}$   &\footnotesize($^3$F)4p  &\footnotesize$^4$F$_\frac{5}{2}$   & 1.237 & $-$2.690 & 7  & 3.35 & 1\pst      & 4.968 & 4.956 & 4.982 & 4.928 & 4.989 & & \\
479.8535 & \color{BrickRed}{$^a$}					                          										                                         		                       
           & \footnotesize ($^1$D)4s &\footnotesize$^2$D$_\frac{3}{2}$   &\footnotesize($^3$F)4p  &\footnotesize$^4$G$_\frac{5}{2}$   & 1.080 & $-$2.660 & 7  & 4.29 & 1\pst      & 4.990 & 4.970 & 4.997 & 4.937 & 5.001 & & \\
486.5597 & \color{BrickRed}{$^a$}									  									                                         		                       
           & \footnotesize 3d$^3$    &\footnotesize$^2$G$_\frac{7}{2}$   &\footnotesize($^3$F)4p  &\footnotesize$^4$G$_\frac{5}{2}$   & 1.116 & $-$2.700 & 7  & 3.50 & 1\pst      & 4.877 & 4.866 & 4.893 & 4.838 & 4.898 & & \\
533.6770 & \color{BrickRed}{$^a$}			                             												                                         		                       
           & \footnotesize 3d$^3$    &\footnotesize$^2$D2$_\frac{5}{2}$  &\footnotesize($^3$F)4p  &\footnotesize$^2$F$_\frac{7}{2}$   & 1.582 & $-$1.600 & 7  & 7.20 & 2\pst      & 4.991 & 4.922 & 4.944 & 4.870 & 4.942 & & \\
538.1013 & \color{BrickRed}{$^a$}											                 					                                         		                       
           & \footnotesize 3d$^3$    &\footnotesize$^2$D2$_\frac{3}{2}$  &\footnotesize($^3$F)4p  &\footnotesize$^2$F$_\frac{5}{2}$   & 1.566 & $-$1.970 & 7  & 5.66 & 1\pst      & 5.004 & 4.963 & 4.983 & 4.918 & 4.988 & & \\
541.8760 & \color{BrickRed}{$^a$}				                        											                                         		                       
           & \footnotesize 3d$^3$    &\footnotesize$^2$D2$_\frac{5}{2}$  &\footnotesize($^3$F)4p  &\footnotesize$^2$F$_\frac{5}{2}$   & 1.582 & $-$2.130 & 7  & 4.81 & 3\pst      & 4.999 & 4.970 & 4.990 & 4.930 & 4.997 & & \\
   & & & & & & & & & & & & & & & \\

\multicolumn{18}{c} {\vi} \\
458.6370 & &  \footnotesize 4s$^2$   &\footnotesize $^4$F$_\frac{7}{2}$  &\footnotesize ($^4$F)4sp&\footnotesize$^4$G$_\frac{9}{2}$   & 0.040 & $-$0.793 & 8  & 4.14 & 2\pst      & 3.750 & 3.849 & 3.943 & 3.820 & 3.865 & \phz$+$0.1\phz & 3.850 \\ 
459.4119 & &  \footnotesize 4s$^2$   &\footnotesize $^4$F$_\frac{9}{2}$  &\footnotesize ($^4$F)4sp&\footnotesize$^4$G$_\frac{11}{2}$  & 0.069 & $-$0.672 & 8  & 5.26 & 2\pst      & 3.762 & 3.862 & 3.956 & 3.833 & 3.878 & \phz$+$0.1\phz & 3.862 \\  
463.5172 & &  \footnotesize 4s$^2$   &\footnotesize $^4$F$_\frac{9}{2}$  &\footnotesize ($^4$F)4sp&\footnotesize$^4$G$_\frac{9}{2}$   & 0.069 & $-$1.924 & 8  & 0.45 & 1\pst      & 3.753 & 3.886 & 3.975 & 3.861 & 3.905 & \phz$+$0.1\phz & 3.853 \\  
482.7452 & &  \footnotesize 4s$^2$   &\footnotesize $^4$F$_\frac{7}{2}$  &\footnotesize ($^4$F)4sp&\footnotesize$^4$D$_\frac{7}{2}$   & 0.040 & $-$1.478 & 8  & 1.30 & 1\pst      & 3.742 & 3.875 & 3.965 & 3.849 & 3.892 & \phz$+$0.1\phz & 3.842 \\  
487.5486 & &  \footnotesize 4s$^2$   &\footnotesize $^4$F$_\frac{7}{2}$  &\footnotesize ($^4$D)4sp&\footnotesize$^4$G$_\frac{5}{2}$   & 0.040 & $-$0.806 & 8  & 4.22 & 1\pst      & 3.754 & 3.853 & 3.948 & 3.822 & 3.866 & \phz$+$0.1\phz & 3.854 \\  
488.1555 & &  \footnotesize 4s$^2$   &\footnotesize $^4$F$_\frac{9}{2}$  &\footnotesize ($^4$F)4sp&\footnotesize$^4$D$_\frac{7}{2}$   & 0.069 & $-$0.657 & 8  & 5.39 & 1\pst      & 3.749 & 3.848 & 3.943 & 3.818 & 3.861 & \phz$+$0.1\phz & 3.849 \\  
562.6019 & &  \footnotesize ($^5$D)4s&\footnotesize$^4$D$_\frac{1}{2}$   &\footnotesize ($^5$D)4p &\footnotesize$^4$D$_\frac{1}{2}$   & 1.043 & $-$1.252 & 8  & 0.31 & 1\pst      & 3.834 & 3.923 & 3.999 & 3.894 & 3.943 & \phz$+$0.1\phz & 3.934 \\ 
564.6108 & &  \footnotesize ($^5$D)4s&\footnotesize$^4$D$_\frac{3}{2}$   &\footnotesize ($^5$D)4p &\footnotesize$^4$D$_\frac{1}{2}$   & 1.051 & $-$1.187 & 8  & 0.37 & 2\pst      & 3.855 & 3.941 & 4.019 & 3.912 & 3.961 & \phz$+$0.1\phz & 3.955 \\  
565.7438 & &  \footnotesize ($^5$D)4s&\footnotesize$^4$D$_\frac{5}{2}$   &\footnotesize ($^5$D)4p &\footnotesize$^4$D$_\frac{3}{2}$   & 1.064 & $-$1.018 & 8  & 0.50 & 3\pst      & 3.842 & 3.928 & 4.004 & 3.898 & 3.947 & \phz$+$0.1\phz & 3.942 \\  
566.8361 & &  \footnotesize ($^5$D)4s&\footnotesize$^4$D$_\frac{7}{2}$   &\footnotesize ($^5$D)4p &\footnotesize$^4$D$_\frac{5}{2}$   & 1.081 & $-$1.021 & 8  & 0.49 & 1\pst      & 3.849 & 3.934 & 4.011 & 3.905 & 3.954 & \phz$+$0.1\phz & 3.949 \\  
567.0847 & &  \footnotesize ($^5$D)4s&\footnotesize$^4$D$_\frac{7}{2}$   &\footnotesize ($^4$F)4sp&\footnotesize$^2$G$_\frac{9}{2}$   & 1.081 & $-$0.425 & 8  & 1.69 & 3\pst      & 3.823 & 3.907 & 3.983 & 3.876 & 3.926 & \phz$+$0.1\phz & 3.923 \\  
570.3586 & &  \footnotesize ($^5$D)4s&\footnotesize$^4$D$_\frac{3}{2}$   &\footnotesize ($^5$D)4p &\footnotesize$^4$F$_\frac{5}{2}$   & 1.051 & $-$0.212 & 8  & 2.57 & 2\pst      & 3.823 & 3.900 & 3.978 & 3.867 & 3.917 & \phz$+$0.1\phz & 3.923 \\  
572.7046 & &  \footnotesize ($^5$D)4s&\footnotesize$^4$D$_\frac{7}{2}$   &\footnotesize ($^5$D)4p &\footnotesize$^4$F$_\frac{9}{2}$   & 1.081 & $-$0.012 & 8  & 3.62 & 3\pst      & 3.810 & 3.888 & 3.966 & 3.855 & 3.905 & \phz$+$0.1\phz & 3.910 \\ 
572.7655 & &  \footnotesize ($^5$D)4s&\footnotesize$^4$D$_\frac{3}{2}$   &\footnotesize ($^5$D)4p &\footnotesize$^4$F$_\frac{3}{2}$   & 1.051 & $-$0.875 & 8  & 0.78 & 2\pst      & 3.876 & 3.961 & 4.041 & 3.933 & 3.980 & \phz$+$0.1\phz & 3.976 \\  
573.1249 & &  \footnotesize ($^5$D)4s&\footnotesize$^4$D$_\frac{5}{2}$   &\footnotesize ($^4$F)4sp&\footnotesize$^2$G$_\frac{7}{2}$   & 1.064 & $-$0.732 & 8  & 0.97 & 2\pst      & 3.836 & 3.924 & 4.000 & 3.894 & 3.943 & \phz$+$0.1\phz & 3.936 \\  
573.7065 & &  \footnotesize ($^5$D)4s&\footnotesize$^4$D$_\frac{5}{2}$   &\footnotesize ($^5$D)4p &\footnotesize$^4$F$_\frac{5}{2}$   & 1.064 & $-$0.736 & 8  & 0.92 & 2\pst      & 3.828 & 3.915 & 3.991 & 3.885 & 3.934 & \phz$+$0.1\phz & 3.928 \\  
600.2294 & &  \footnotesize 4s$^2$   &\footnotesize$^4$P$_\frac{5}{2}$   &\footnotesize ($^5$D)4p &\footnotesize$^4$D$_\frac{7}{2}$   & 1.218 & $-$1.773 & 8  & 0.07 & 1\pst      & 3.843 & 3.926 & 4.000 & 3.897 & 3.946 & \phz$+$0.1\phz & 3.943 \\  
603.9728 & &  \footnotesize ($^5$D)4s&\footnotesize$^4$D$_\frac{5}{2}$   &\footnotesize ($^5$D)4p &\footnotesize$^4$P$_\frac{5}{2}$   & 1.064 & $-$0.652 & 8  & 1.12 & 3\pst      & 3.834 & 3.917 & 3.994 & 3.886 & 3.934 & \phz$+$0.1\phz & 3.934 \\  
608.1441 & &  \footnotesize ($^5$D)4s&\footnotesize$^4$D$_\frac{3}{2}$   &\footnotesize ($^5$D)4p &\footnotesize$^4$P$_\frac{3}{2}$   & 1.051 & $-$0.579 & 8  & 1.26 & 3\pst      & 3.788 & 3.875 & 3.951 & 3.844 & 3.893 & \phz$+$0.1\phz & 3.888 \\ 
609.0208 & &  \footnotesize ($^5$D)4s&\footnotesize$^4$D$_\frac{7}{2}$   &\footnotesize ($^5$D)4p &\footnotesize$^4$P$_\frac{5}{2}$   & 1.081 & $-$0.062 & 8  & 3.07 & 3\pst      & 3.799 & 3.869 & 3.947 & 3.835 & 3.884 & \phz$+$0.1\phz & 3.899 \\  
611.1650 & &  \footnotesize ($^5$D)4s&\footnotesize$^4$D$_\frac{1}{2}$   &\footnotesize ($^5$D)4p &\footnotesize$^4$P$_\frac{1}{2}$   & 1.043 & $-$0.714 & 8  & 0.99 & 3\pst      & 3.784 & 3.876 & 3.951 & 3.846 & 3.895 & \phz$+$0.1\phz & 3.884 \\  
611.9528 & &  \footnotesize ($^5$D)4s&\footnotesize$^4$D$_\frac{5}{2}$   &\footnotesize ($^5$D)4p &\footnotesize$^4$P$_\frac{3}{2}$   & 1.064 & $-$0.320 & 8  & 1.95 & 2\pst      & 3.779 & 3.857 & 3.933 & 3.824 & 3.874 & \phz$+$0.1\phz & 3.879 \\  
613.5363 & &  \footnotesize ($^5$D)4s&\footnotesize$^4$D$_\frac{3}{2}$   &\footnotesize ($^5$D)4p &\footnotesize$^4$P$_\frac{1}{2}$   & 1.051 & $-$0.746 & 8  & 0.94 & 1\pst      & 3.807 & 3.896 & 3.972 & 3.866 & 3.914 & \phz$+$0.1\phz & 3.907 \\  
624.2828 & &  \footnotesize ($^5$D)4s&\footnotesize$^6$D$_\frac{1}{2}$   &\footnotesize ($^4$F)4sp&\footnotesize$^6$D$_\frac{3}{2}$   & 0.262 & $-$1.552 & 8  & 0.83 & 3\pst      & 3.716 & 3.850 & 3.936 & 3.823 & 3.863 & \phz$+$0.1\phz & 3.816 \\ 
625.1823 & &  \footnotesize ($^5$D)4s&\footnotesize$^6$D$_\frac{7}{2}$   &\footnotesize ($^4$F)4sp&\footnotesize$^6$D$_\frac{7}{2}$   & 0.287 & $-$1.342 & 8  & 1.29 & 3\pst      & 3.745 & 3.874 & 3.962 & 3.847 & 3.887 & \phz$+$0.1\phz & 3.845 \\  
625.6903 & &  \footnotesize ($^5$D)4s&\footnotesize$^6$D$_\frac{5}{2}$   &\footnotesize ($^4$F)4sp&\footnotesize$^6$D$_\frac{5}{2}$   & 0.275 & $-$2.006 & 8  & 0.31 & 2\pst      & 3.740 & 3.878 & 3.964 & 3.851 & 3.892 & \phz$+$0.1\phz & 3.840 \\  
627.4653 & &  \footnotesize ($^5$D)4s&\footnotesize$^6$D$_\frac{3}{2}$   &\footnotesize ($^4$F)4sp&\footnotesize$^6$D$_\frac{1}{2}$   & 0.267 & $-$1.673 & 8  & 0.71 & 1\pst      & 3.774 & 3.906 & 3.995 & 3.880 & 3.919 & \phz$+$0.1\phz & 3.874 \\  
628.5160 & &  \footnotesize ($^5$D)4s&\footnotesize$^6$D$_\frac{5}{2}$   &\footnotesize ($^4$F)4sp&\footnotesize$^6$D$_\frac{3}{2}$   & 0.275 & $-$1.512 & 8  & 0.88 & 3\pst      & 3.724 & 3.854 & 3.940 & 3.826 & 3.867 & \phz$+$0.1\phz & 3.824 \\  
629.2824 & &  \footnotesize ($^5$D)4s&\footnotesize$^6$D$_\frac{7}{2}$   &\footnotesize ($^4$F)4sp&\footnotesize$^6$D$_\frac{5}{2}$   & 0.287 & $-$1.471 & 8  & 1.02 & 1\pst      & 3.768 & 3.895 & 3.983 & 3.869 & 3.908 & \phz$+$0.1\phz & 3.868 \\ 
653.1401 & &  \footnotesize 4s$^2$   &\footnotesize$^4$P$_\frac{5}{2}$   &\footnotesize ($^5$D)4p &\footnotesize$^4$P$_\frac{5}{2}$   & 1.218 & $-$0.836 & 8  & 0.57 & 2\pst      & 3.809 & 3.895 & 3.967 & 3.864 & 3.913 & \phz$+$0.1\phz & 3.909 \\
   & & & & & & & & & & & & & & & \\
     	       
\multicolumn{18}{c} {\vii} \\				  												           	       
376.0222 & &  \footnotesize 3d$^4$   &\footnotesize $^3$F$_4$            &\footnotesize ($^4$F)4p &\footnotesize $^3$F$_3$            & 1.687 & $-$1.153 & 9  & 3.64 & 1\pst      & 3.958 & 3.942 & 3.970 & 3.917 & 3.975 &             & \\  
386.6740 & &  \footnotesize 3d$^4$   &\footnotesize $^3$P$_1$            &\footnotesize ($^4$F)4p &\footnotesize $^5$D$_2$            & 1.428 & $-$1.550 & 9  & 3.29 & 1\pst      & 4.015 & 4.000 & 4.030 & 3.978 & 4.034 &             & \\ 
395.1960 & &  \footnotesize 3d$^4$   &\footnotesize $^3$P$_2$            &\footnotesize ($^4$F)4p &\footnotesize $^3$D$_3$            & 1.476 & $-$0.740 &10  & 6.47 & 1\pst      & 3.950 & 3.904 & 3.933 & 3.861 & 3.932 &             & \\ 
399.7117 & &  \footnotesize 3d$^4$   &\footnotesize $^3$P$_2$            &\footnotesize ($^4$F)4p &\footnotesize $^5$F$_3$            & 1.476 & $-$1.230 & 9  & 5.01 & 1\pst      & 4.062 & 4.041 & 4.069 & 4.008 & 4.073 &             & \\ 
403.6777 & &  \footnotesize 3d$^4$   &\footnotesize $^3$P$_2$            &\footnotesize ($^4$F)4p &\footnotesize $^5$F$_2$            & 1.476 & $-$1.594 & 9  & 3.17 & 1\pst      & 4.015 & 4.006 & 4.036 & 3.986 & 4.041 &             & \\ 
   & & & & & & & & & & & & & & & \\

\multicolumn{18}{c} {\cri} \\
437.3259 & & \footnotesize 4s$^2$    &\footnotesize$^5$D$_2$             &\footnotesize($^5$D)4sp &\footnotesize$^5$F$_1$             & 0.983 & $-$2.323 &11  & 3.76 & 1\pst      & 5.584 & 5.615 & 5.698 & 5.578 & 5.636 & $+$0.031    & 5.615 \\
452.9838 &  \footnote[2]{Isotopic splitting included (see Table \protect\ref{table:hfs}); wavelength corresponds to $^{50}$Cr component.}										         	       
           & \footnotesize ($^4$G)4s &\footnotesize$^5$G$_6$             &\footnotesize($^4$G)4p  &\footnotesize$^5$G$_5$             & 2.544 & $-$1.380 &12  & 1.77 & 1\pst      & 5.598 & 5.626 & 5.685 & 5.589 & 5.653 & $+$0.023    & 5.621 \\
453.5127 & & \footnotesize ($^4$G)4s &\footnotesize$^5$G$_3$             &\footnotesize($^4$G)4p  &\footnotesize$^5$G$_4$             & 2.544 & $-$0.993 &13  & 3.15 & 2\pst      & 5.576 & 5.593 & 5.653 & 5.550 & 5.619 & $+$0.023    & 5.599 \\
454.1060 & & \footnotesize ($^4$G)4s &\footnotesize$^5$G$_4$             &\footnotesize($^4$G)4p  &\footnotesize$^5$G$_3$             & 2.545 & $-$1.143 &13  & 2.50 & 1\pst      & 5.578 & 5.600 & 5.660 & 5.560 & 5.627 & $+$0.023    & 5.601 \\
463.3259 & & \footnotesize ($^5$D)4sp&\footnotesize$^7$F$_3$             &\footnotesize     4s5s  &\footnotesize$^7$D$_4$             & 3.125 & $-$1.110 &14  & 0.93 & 2\pst      & 5.534 & 5.560 & 5.610 & 5.522 & 5.589 & $+$0.046    & 5.580 \\
470.0599 & & \footnotesize ($^4$P)4s &\footnotesize$^5$P$_1$             &\footnotesize($^3$P2)4sp&\footnotesize$^5$S$_2$             & 2.710 & $-$1.255 &15  & 1.51 & 2\pst      & 5.581 & 5.608 & 5.665 & 5.570 & 5.636 & $+$0.027    & 5.608 \\
470.8017 & & \footnotesize ($^5$D)4sp&\footnotesize$^7$F$_5$             &\footnotesize     4s5s  &\footnotesize$^7$D$_4$             & 3.168 & \ph0.090 &16  & 5.67 & 1\pst      & 5.595 & 5.578 & 5.633 & 5.516 & 5.599 & $+$0.066    & 5.661 \\
474.5270 & & \footnotesize ($^4$P)4s &\footnotesize$^5$P$_3$             &\footnotesize($^3$P2)4sp&\footnotesize$^5$D$_4$             & 2.708 & $-$1.380 &14  & 1.22 & 2\pst      & 5.536 & 5.566 & 5.622 & 5.528 & 5.595 & $+$0.026    & 5.562 \\
478.9340 & & \footnotesize ($^4$G)4s &\footnotesize$^5$G$_6$             &\footnotesize($^5$D)4sp &\footnotesize$^5$F$_5$             & 2.544 & $-$0.348 &16  & 5.99 & 2\pst      & 5.528 & 5.501 & 5.567 & 5.443 & 5.519 & $+$0.019    & 5.547 \\
480.1048 & & \footnotesize 4s$^2$    &\footnotesize$^3$F$_4$             &\footnotesize($^4$G)4p  &\footnotesize$^3$F$_3$             & 3.122 & $-$0.131 &15  & 4.79 & 2\pst      & 5.613 & 5.600 & 5.657 & 5.544 & 5.623 & $+$0.045    & 5.658 \\
488.5733 & & \footnotesize ($^4$G)4s &\footnotesize$^5$G$_3$             &\footnotesize($^5$D)4sp &\footnotesize$^5$P$_2$             & 2.544 & $-$1.055 &15  & 2.82 & 2\pst      & 5.558 & 5.579 & 5.639 & 5.536 & 5.604 & $+$0.025    & 5.583 \\
493.6336 & & \footnotesize 4s$^2$    &\footnotesize$^3$F$_4$             &\footnotesize($^4$G)4p  &\footnotesize$^3$H$_4$             & 3.113 & $-$0.237 &16  & 4.27 & 1\pst      & 5.591 & 5.586 & 5.642 & 5.533 & 5.610 & $+$0.037    & 5.628 \\
495.3714 & & \footnotesize 4s$^2$    &\footnotesize$^3$F$_4$             &\footnotesize($^4$G)4p  &\footnotesize$^3$H$_4$             & 3.122 & $-$1.480 &14  & 0.47 & 1\pst      & 5.562 & 5.590 & 5.639 & 5.551 & 5.619 & $+$0.035    & 5.597 \\
522.0913 & & \footnotesize ($^5$D)4sp&\footnotesize$^7$D$_1$             &\footnotesize     4s5s  &\footnotesize$^7$D$_1$             & 3.385 & $-$0.890 &14  & 1.09 & 2\pst      & 5.599 & 5.622 & 5.669 & 5.581 & 5.650 & $+$0.025    & 5.624 \\
524.1454 & & \footnotesize ($^4$P)4s &\footnotesize$^5$P$_1$             &\footnotesize($^5$D)4sp &\footnotesize$^5$P$_1$             & 2.710 & $-$1.920 &14  & 0.35 & 3\pst      & 5.450 & 5.484 & 5.540 & 5.449 & 5.512 & $+$0.025    & 5.475 \\
527.2008 & & \footnotesize ($^5$D)4sp&\footnotesize$^7$P$_3$             &\footnotesize     4s5s  &\footnotesize$^7$D$_4$             & 3.449 & $-$0.421 &16  & 2.29 & 1\pst      & 5.603 & 5.620 & 5.667 & 5.573 & 5.646 & $+$0.026    & 5.629 \\
528.7201 & & \footnotesize ($^5$D)4sp&\footnotesize$^7$P$_2$             &\footnotesize     4s5s  &\footnotesize$^7$D$_3$             & 3.438 & $-$0.888 &16  & 1.00 & 2\pst      & 5.608 & 5.630 & 5.677 & 5.589 & 5.657 & $+$0.025    & 5.633 \\
530.0743 & & \footnotesize 4s$^2$    &\footnotesize$^5$D$_2$             &\footnotesize($^6$S)4p  &\footnotesize$^5$P$_3$             & 0.983 & $-$2.083 &17  & 5.48 & 2\pst      & 5.558 & 5.566 & 5.652 & 5.522 & 5.575 & $+$0.035    & 5.593 \\
530.4184 & & \footnotesize ($^5$D)4sp&\footnotesize$^7$P$_4$             &\footnotesize     4s5s  &\footnotesize$^7$D$_4$             & 3.464 & $-$0.681 &16  & 1.45 & 2\pst      & 5.612 & 5.633 & 5.679 & 5.590 & 5.659 & $+$0.025    & 5.637 \\
531.2871 & & \footnotesize ($^5$D)4sp&\footnotesize$^7$P$_3$             &\footnotesize     4s5s  &\footnotesize$^7$D$_3$             & 3.449 & $-$0.556 &16  & 1.85 & 1\pst      & 5.599 & 5.618 & 5.665 & 5.573 & 5.644 & $+$0.026    & 5.625 \\
531.8810 & & \footnotesize ($^5$D)4sp&\footnotesize$^7$P$_2$             &\footnotesize     4s5s  &\footnotesize$^7$D$_2$             & 3.438 & $-$0.679 &16  & 1.49 & 3\pst      & 5.596 & 5.617 & 5.663 & 5.573 & 5.643 & $+$0.026    & 5.622 \\
534.0474 & & \footnotesize ($^5$D)4sp&\footnotesize$^7$P$_2$             &\footnotesize     4s5s  &\footnotesize$^7$D$_1$             & 3.438 & $-$0.730 &16  & 1.48 & 1\pst      & 5.642 & 5.662 & 5.710 & 5.620 & 5.688 & $+$0.026    & 5.668 \\
562.8621 & & \footnotesize 4s$^2$    &\footnotesize$^3$G$_3$             &\footnotesize($^4$G)4p  &\footnotesize$^3$H$_4$             & 3.422 & $-$0.756 &16  & 1.36 & 2\pst      & 5.597 & 5.618 & 5.665 & 5.575 & 5.644 & $+$0.033    & 5.630 \\
571.9809 & & \footnotesize ($^4$D)4s &\footnotesize$^5$D$_3$             &\footnotesize($^5$D)4sp &\footnotesize$^5$D$_4$             & 3.013 & $-$1.620 &16  & 0.43 & 1\pst      & 5.503 & 5.535 & 5.586 & 5.496 & 5.562 & $+$0.026    & 5.529 \\
578.1163 & & \footnotesize ($^4$D)4s &\footnotesize$^5$D$_4$             &\footnotesize($^5$D)4sp &\footnotesize$^5$D$_3$             & 3.011 & $-$1.000 &14  & 1.56 & 1\pst      & 5.498 & 5.525 & 5.577 & 5.483 & 5.551 & $+$0.002    & 5.500 \\
578.5024 & & \footnotesize ($^6$S)4p &\footnotesize$^5$P$_3$             &\footnotesize($^6$S)4d  &\footnotesize$^5$D$_3$             & 3.321 & $-$0.380 &15  & 3.13 & 1\pst      & 5.595 & 5.610 & 5.659 & 5.560 & 5.633 & $+$0.029    & 5.624 \\
584.4592 & & \footnotesize ($^4$D)4s &\footnotesize$^5$D$_3$             &\footnotesize($^5$D)4sp &\footnotesize$^5$D$_2$             & 3.013 & $-$1.770 &14  & 0.40 & 2\pst      & 5.616 & 5.647 & 5.698 & 5.609 & 5.673 & $+$0.026    & 5.642 \\
688.2477 & & \footnotesize ($^5$D)4sp&\footnotesize$^7$P$_2$             &\footnotesize($^6$S)4d  &\footnotesize$^7$D$_2$             & 3.438 & $-$0.375 &15  & 3.13 & 1\pst      & 5.626 & 5.640 & 5.687 & 5.592 & 5.659 & $+$0.025    & 5.651 \\
688.2997 & & \footnotesize ($^5$D)4sp&\footnotesize$^7$P$_2$             &\footnotesize($^6$S)4d  &\footnotesize$^7$D$_1$             & 3.438 & $-$0.420 &15  & 2.96 & 3\pst      & 5.636 & 5.651 & 5.698 & 5.604 & 5.671 & $+$0.025    & 5.661 \\
   & & & & & & & & & & & & & & & \\

\multicolumn{18}{c} {\crii} \\
455.4990 & & \footnotesize d$^5$     &\footnotesize$^4$F$_\frac{7}{2}$   &\footnotesize($^5$D)4p  &\footnotesize$^4$D$_\frac{7}{2}$   & 4.071 & $-$1.249 &18  & 4.66 & 1\pst      & 5.632 & 5.596 & 5.606 & 5.538 & 5.627 & & \\
458.8200 & & \footnotesize d$^5$     &\footnotesize$^4$F$_\frac{7}{2}$   &\footnotesize($^5$D)4p  &\footnotesize$^4$D$_\frac{5}{2}$   & 4.071 & $-$0.594 &18  & 7.47 & 2\pst      & 5.648 & 5.565 & 5.576 & 5.492 & 5.588 & & \\
484.8237 & & \footnotesize ($^3$F)4s &\footnotesize$^4$F$_\frac{7}{2}$   &\footnotesize($^5$D)4p  &\footnotesize$^4$F$_\frac{7}{2}$   & 3.864 & $-$1.160 &18  & 6.11 & 3\pst      & 5.689 & 5.621 & 5.629 & 5.555 & 5.648 & & \\
523.7328 & & \footnotesize d$^5$     &\footnotesize$^4$F$_\frac{9}{2}$   &\footnotesize($^5$D)4p  &\footnotesize$^4$F$_\frac{9}{2}$   & 4.073 & $-$1.087 &18  & 5.36 & 2\pst      & 5.610 & 5.554 & 5.557 & 5.490 & 5.581 & & \\
524.6768 & & \footnotesize ($^3$F)4s &\footnotesize$^4$P$_\frac{1}{2}$   &\footnotesize($^5$D)4p  &\footnotesize$^4$P$_\frac{3}{2}$   & 3.714 & $-$2.436 &18  & 1.62 & 1\pst      & 5.654 & 5.652 & 5.665 & 5.620 & 5.684 & & \\
527.9877 & & \footnotesize d$^5$     &\footnotesize$^4$F$_\frac{9}{2}$   &\footnotesize($^5$D)4p  &\footnotesize$^4$F$_\frac{7}{2}$   & 4.073 & $-$1.909 &18  & 2.02 & 1\pst      & 5.580 & 5.576 & 5.587 & 5.539 & 5.608 & & \\
531.0686 & & \footnotesize d$^5$     &\footnotesize$^4$F$_\frac{3}{2}$   &\footnotesize($^5$D)4p  &\footnotesize$^4$F$_\frac{5}{2}$   & 4.072 & $-$2.144 &18  & 1.32 & 1\pst      & 5.564 & 5.566 & 5.578 & 5.534 & 5.597 & & \\
531.3561 & & \footnotesize d$^5$     &\footnotesize$^4$F$_\frac{5}{2}$   &\footnotesize($^5$D)4p  &\footnotesize$^4$F$_\frac{5}{2}$   & 4.073 & $-$1.473 &18  & 3.49 & 1\pst      & 5.544 & 5.523 & 5.529 & 5.473 & 5.553 & & \\
550.2068 & & \footnotesize ($^3$G)4s &\footnotesize$^4$G$_\frac{9}{2}$   &\footnotesize($^5$D)4p  &\footnotesize$^4$F$_\frac{7}{2}$   & 4.168 & $-$2.049 &18  & 1.84 & 2\pst      & 5.743 & 5.741 & 5.750 & 5.704 & 5.770 & & \\
612.9226 & & \footnotesize ($^3$G)4s &\footnotesize$^4$D$_\frac{5}{2}$   &\footnotesize($^5$D)4p  &\footnotesize$^4$D$_\frac{5}{2}$   & 4.750 & $-$2.478 &18  & 0.29 & 1\pst      & 5.749 & 5.759 & 5.768 & 5.731 & 5.785 & & \\
   & & & & & & & & & & & & & & & \\

\multicolumn{18}{c} {\mni} \\
408.2945 & & \footnotesize ($^5$D)4s &\footnotesize$^6$D$_\frac{3}{2}$   &\footnotesize($^5$D)4p  &\footnotesize$^6$D$_\frac{5}{2}$   & 2.178 & $-$0.365 &19  & 8.97 & 2\pst      & 5.380 & 5.316 & 5.392 & 5.256 & 5.329 & $+$0.016    & 5.396 \\
426.5928 & & \footnotesize ($^5$D)4s &\footnotesize$^4$D$_\frac{3}{2}$   &\footnotesize($^5$D)4p  &\footnotesize$^4$P$_\frac{3}{2}$   & 2.941 & $-$0.400 &20  & 5.85 & 1*         & 5.376 & 5.354 & 5.420 & 5.301 & 5.376 & $+$0.076    & 5.452 \\     
445.3013 & & \footnotesize ($^5$D)4s &\footnotesize$^4$D$_\frac{3}{2}$   &\footnotesize($^5$D)4p  &\footnotesize$^4$D$_\frac{1}{2}$   & 2.941 & $-$0.620 &21  & 5.19 & 2\pst      & 5.368 & 5.371 & 5.436 & 5.322 & 5.395 & $+$0.070    & 5.438 \\
445.7041 & & \footnotesize ($^6$S)4sp&\footnotesize$^6$P$_\frac{5}{2}$   &\footnotesize($^7$S)4sd &\footnotesize$^6$D$_\frac{3}{2}$   & 3.073 & $-$0.685 &20  & 4.33 & 1*         & 5.392 & 5.400 & 5.460 & 5.353 & 5.426 & $+$0.065    & 5.457 \\
447.0142 & & \footnotesize ($^5$D)4s &\footnotesize$^4$D$_\frac{3}{2}$   &\footnotesize($^5$D)4p  &\footnotesize$^4$D$_\frac{3}{2}$   & 2.941 & $-$0.560 &21  & 5.22 & 2\pst      & 5.411 & 5.389 & 5.454 & 5.336 & 5.411 & $+$0.061    & 5.472 \\
449.8897 & & \footnotesize ($^5$D)4s &\footnotesize$^4$D$_\frac{3}{2}$   &\footnotesize($^5$D)4p  &\footnotesize$^4$D$_\frac{5}{2}$   & 2.941 & $-$0.460 &21  & 5.54 & 1\pst      & 5.398 & 5.364 & 5.430 & 5.309 & 5.385 & $+$0.054    & 5.452 \\
450.2223 & & \footnotesize ($^5$D)4s &\footnotesize$^4$D$_\frac{5}{2}$   &\footnotesize($^5$D)4p  &\footnotesize$^4$D$_\frac{7}{2}$   & 2.920 & $-$0.430 &21  & 5.81 & 2\pst      & 5.349 & 5.328 & 5.393 & 5.273 & 5.348 & $+$0.055    & 5.404 \\
467.1688 & & \footnotesize ($^5$D)4s &\footnotesize$^4$D$_\frac{7}{2}$   &\footnotesize($^5$D)4p  &\footnotesize$^4$F$_\frac{5}{2}$   & 2.888 & $-$1.660 &21  & 1.23 & 1\pst      & 5.379 & 5.412 & 5.472 & 5.377 & 5.440 & $+$0.061    & 5.440 \\
470.9710 & & \footnotesize ($^5$D)4s &\footnotesize$^4$D$_\frac{7}{2}$   &\footnotesize($^5$D)4p  &\footnotesize$^4$F$_\frac{7}{2}$   & 2.888 & $-$0.487 &19  & 6.88 & 2\pst      & 5.353 & 5.360 & 5.427 & 5.306 & 5.379 & $+$0.065    & 5.418 \\
473.9110 & & \footnotesize ($^5$D)4s &\footnotesize$^4$D$_\frac{3}{2}$   &\footnotesize($^5$D)4p  &\footnotesize$^4$F$_\frac{3}{2}$   & 2.941 & $-$0.604 &19  & 5.82 & 3\pst      & 5.346 & 5.359 & 5.424 & 5.310 & 5.382 & $+$0.065    & 5.411 \\
500.4891 & & \footnotesize ($^5$D)4s &\footnotesize$^4$D$_\frac{5}{2}$   &\footnotesize($^5$D)4p  &\footnotesize$^6$F$_\frac{7}{2}$   & 2.920 & $-$1.636 &22  & 1.31 & 2\pst      & 5.404 & 5.434 & 5.493 & 5.398 & 5.461 & $+$0.064    & 5.468 \\
525.5330 & & \footnotesize 4s$^2$    &\footnotesize$^4$G$_\frac{11}{2}$  &\footnotesize($^5$D)4p  &\footnotesize$^4$F$_\frac{9}{2}$   & 3.133 & $-$0.858 &19  & 3.69 & 2\pst      & 5.326 & 5.355 & 5.413 & 5.313 & 5.381 & $+$0.069    & 5.395 \\
538.8538 & & \footnotesize 4s$^2$    &\footnotesize$^4$P$_\frac{5}{2}$   &\footnotesize($^5$D)4p  &\footnotesize$^4$D$_\frac{7}{2}$   & 3.373 & $-$1.620 &21  & 0.49 & 1\pst      & 5.334 & 5.361 & 5.416 & 5.322 & 5.389 & $+$0.064    & 5.398 \\
542.0368 & & \footnotesize ($^5$D)4s &\footnotesize$^6$D$_\frac{7}{2}$   &\footnotesize($^6$S)4sp &\footnotesize$^6$P$_\frac{5}{2}$   & 2.143 & $-$1.462 &23  & 7.86 & 3\pst      & 5.315 & 5.361 & 5.434 & 5.323 & 5.382 & $+$0.072    & 5.387 \\
   & & & & & & & & & & & & & & & \\

\multicolumn{18}{c} {\fei} \\
444.5472 & & \footnotesize4s$^2$     &\footnotesize$^5$D$_2$             &\footnotesize4s4p($^3$P)&\footnotesize$^7$F$_2$             & 0.087 & $-$5.412 & 24 & 3.80 & 2\pst      & 7.419 & 7.463 & 7.568 & 7.436 & 7.474 & $+$0.016    & 7.435 \\  
524.7050 & & \footnotesize4s$^2$     &\footnotesize$^5$D$_2$             &\footnotesize4s4p($^3$P)&\footnotesize$^7$D$_3$             & 0.087 & $-$4.961 & 24 & 6.40 & 3\pst      & 7.472 & 7.449 & 7.559 & 7.412 & 7.440 & $+$0.022    & 7.494 \\ 
549.1832 & & \footnotesize3d$^8$     &\footnotesize$^3$F$_2$             &\footnotesize($^2$P)4p  &\footnotesize$^3$D$_3$             & 4.186 & $-$2.188 & 25 & 1.23 & 1\pst      & 7.441 & 7.452 & 7.500 & 7.411 & 7.481 & $+$0.006    & 7.447 \\ 
560.0224 & & \footnotesize4s4p($^3$P)&\footnotesize$^3$P$_1$             &\footnotesize4s($^4$D)5s&\footnotesize$^5$D$_1$             & 4.260 & $-$1.420 & 25 & 3.65 & 1\pst      & 7.369 & 7.367 & 7.412 & 7.316 & 7.390 & $+$0.007    & 7.376 \\ 
566.1346 & & \footnotesize4s4p($^3$P)&\footnotesize$^3$P$_0$             &\footnotesize4s($^4$D)5s&\footnotesize$^5$D$_1$             & 4.284 & $-$1.756 & 25 & 2.22 & 2\pst      & 7.414 & 7.419 & 7.465 & 7.374 & 7.445 & $+$0.006    & 7.420 \\ 
570.5465 & & \footnotesize($^4$F)4p  &\footnotesize$^5$F$_1$             &\footnotesize4s($^4$D)5s&\footnotesize$^5$D$_1$             & 4.301 & $-$1.355 & 25 & 3.96 & 2\pst      & 7.418 & 7.409 & 7.455 & 7.356 & 7.431 & $+$0.005    & 7.423 \\ 
577.8453 & & \footnotesize4s$^2$     &\footnotesize$^3$F2$_3$            &\footnotesize($^4$F)4p  &\footnotesize$^3$D$_3$             & 2.588 & $-$3.440 & 25 & 2.04 & 2\pst      & 7.403 & 7.427 & 7.495 & 7.391 & 7.450 & $+$0.005    & 7.408 \\ 
578.4658 & & \footnotesize4s4p($^3$P)&\footnotesize$^5$F$_3$             &\footnotesize4s($^6$D)5s&\footnotesize$^5$D$_4$             & 3.396 & $-$2.532 & 25 & 2.58 & 2\pst      & 7.418 & 7.429 & 7.486 & 7.387 & 7.453 & $+$0.007    & 7.425 \\ 
585.5077 & & \footnotesize($^4$F)4p  &\footnotesize$^3$F$_3$             &\footnotesize($^4$F)4d  &\footnotesize$^5$H$_4$             & 4.608 & $-$1.478 & 25 & 2.20 & 1\pst      & 7.422 & 7.426 & 7.468 & 7.379 & 7.452 & $+$0.006    & 7.428 \\ 
595.6694 & & \footnotesize($^4$F)4s  &\footnotesize$^5$F$_5$             &\footnotesize4s4p($^3$P)&\footnotesize$^7$P$_4$             & 0.859 & $-$4.552 & 24 & 5.02 & 3\pst      & 7.430 & 7.441 & 7.538 & 7.408 & 7.443 & $+$0.017    & 7.447 \\ 
615.1618 & & \footnotesize($^4$P)4s  &\footnotesize$^5$P$_3$             &\footnotesize($^4$F)4p  &\footnotesize$^5$D$_2$             & 2.176 & $-$3.282 & 26 & 4.88 & 3\pst      & 7.445 & 7.437 & 7.514 & 7.397 & 7.447 & $+$0.012    & 7.457 \\ 
624.0646 & & \footnotesize($^4$P)4s  &\footnotesize$^5$P$_1$             &\footnotesize4s4p($^3$P)&\footnotesize$^3$P$_2$             & 2.223 & $-$3.287 & 27 & 4.76 & 3\pst      & 7.469 & 7.461 & 7.538 & 7.421 & 7.472 & $+$0.012    & 7.481 \\ 
631.1500 & & \footnotesize($^4$P)4s  &\footnotesize$^3$P$_2$             &\footnotesize ($^4$F)4p &\footnotesize$^3$D$_2$             & 2.831 & $-$3.141 & 25 & 2.66 & 1\pst      & 7.470 & 7.485 & 7.549 & 7.447 & 7.505 & $+$0.008    & 7.478 \\ 
649.8939 & & \footnotesize($^4$F)4s  &\footnotesize$^5$F$_3$             &\footnotesize4s4p($^3$P)&\footnotesize$^7$F$_3$             & 0.958 & $-$4.695 & 24 & 4.39 & 3\pst      & 7.488 & 7.516 & 7.610 & 7.486 & 7.519 & $+$0.015    & 7.503 \\ 
651.8367 & & \footnotesize($^4$P)4s  &\footnotesize$^3$P$_2$             &\footnotesize($^4$F)4p  &\footnotesize$^3$D$_3$             & 2.831 & $-$2.448 & 27 & 5.72 & 2\pst      & 7.429 & 7.389 & 7.459 & 7.343 & 7.396 & $+$0.012    & 7.441 \\ 
669.9142 & & \footnotesize($^2$F)4s  &\footnotesize$^3$F$_4$             &\footnotesize($^2$P)4p  &\footnotesize$^3$D$_3$             & 4.593 & $-$2.101 & 25 & 0.81 & 2\pst      & 7.515 & 7.469 & 7.543 & 7.422 & 7.471 & $+$0.006    & 7.489 \\ 
679.3259 & & \footnotesize3d$^8$     &\footnotesize$^3$F$_4$             &\footnotesize4s4p($^3$P)&\footnotesize$^5$G$_4$             & 4.076 & $-$2.326 & 25 & 1.25 & 1\pst      & 7.420 & 7.431 & 7.479 & 7.390 & 7.455 & $+$0.006    & 7.426 \\ 
683.7006 & & \footnotesize($^2$F)4s  &\footnotesize$^3$F$_4$             &\footnotesize($^2$H)4p  &\footnotesize$^3$G$_4$             & 4.593 & $-$1.687 & 25 & 1.77 & 1\pst      & 7.466 & 7.468 & 7.509 & 7.422 & 7.492 & $+$0.006    & 7.472 \\ 
685.4823 & & \footnotesize($^2$F)4s  &\footnotesize$^3$F$_4$             &\footnotesize4s4p($^3$P)&\footnotesize$^1$H$_5$             & 4.593 & $-$1.926 & 25 & 1.22 & 1\pst      & 7.506 & 7.512 & 7.553 & 7.469 & 7.536 & $+$0.006    & 7.512 \\ 
740.1685 & & \footnotesize3d$^8$     &\footnotesize$^3$F$_2$             &\footnotesize($^4$P)4p  &\footnotesize$^3$D$_1$             & 4.186 & $-$1.500 & 25 & 4.16 & 3\pst      & 7.381 & 7.371 & 7.417 & 7.323 & 7.387 & $+$0.008    & 7.389 \\ 
791.2867 & & \footnotesize($^4$F)4s  &\footnotesize$^5$F$_5$             &\footnotesize4s4p($^3$P)&\footnotesize$^7$D$_4$             & 0.859 & $-$4.848 & 29 & 4.57 & 2\pst      & 7.451 & 7.489 & 7.586 & 7.462 & 7.486 & $+$0.017    & 7.468 \\ 
829.3515 & & \footnotesize($^2$D)4s  &\footnotesize$^3$D$_2$             &\footnotesize($^4$F)4p  &\footnotesize$^3$D$_2$             & 3.301 & $-$2.203 & 30 & 5.85 & 1*         & 7.471 & 7.448 & 7.509 & 7.401 & 7.453 & $+$0.011    & 7.482 \\ 
   & & & & & & & & & & & & & & & \\		

\multicolumn{18}{c} {\feii} \\	
462.0513 & & \footnotesize4s         &\footnotesize$^4$F$_\frac{7}{2}$   &\footnotesize4p         &\footnotesize$^4$D$_\frac{7}{2}$   & 2.828 & $-$3.210 & 31 & 5.40 & 1\pst      & 7.474 & 7.405 & 7.416 & 7.350 & 7.436 & & \\ 
526.4804 & & \footnotesize4s         &\footnotesize$^4$G$_\frac{5}{2}$   &\footnotesize4p         &\footnotesize$^4$D$_\frac{3}{2}$   & 3.230 & $-$3.130 & 31 & 4.74 & 3\pst      & 7.556 & 7.500 & 7.503 & 7.445 & 7.530 & & \\ 
541.4072 & & \footnotesize4s         &\footnotesize$^4$G$_\frac{7}{2}$   &\footnotesize4p         &\footnotesize$^4$D$_\frac{7}{2}$   & 3.221 & $-$3.580 & 31 & 2.73 & 2\pst      & 7.483 & 7.464 & 7.471 & 7.424 & 7.496 & & \\ 
643.2676 & & \footnotesize4s$^2$     &\footnotesize$^6$S$_\frac{5}{2}$   &\footnotesize4p         &\footnotesize$^6$D$_\frac{5}{2}$   & 2.891 & $-$3.570 & 31 & 4.30 & 3\pst      & 7.515 & 7.463 & 7.462 & 7.416 & 7.488 & & \\ 
651.6077 & & \footnotesize4s$^2$     &\footnotesize$^6$S$_\frac{5}{2}$   &\footnotesize4p         &\footnotesize$^6$D$_\frac{7}{2}$   & 2.891 & $-$3.310 & 31 & 5.69 & 3\pst      & 7.569 & 7.485 & 7.482 & 7.432 & 7.504 & & \\ 
722.2392 & & \footnotesize4s         &\footnotesize$^4$D$_\frac{3}{2}$   &\footnotesize4p         &\footnotesize$^4$D$_\frac{1}{2}$   & 3.889 & $-$3.260 & 31 & 2.03 & 1\pst      & 7.519 & 7.504 & 7.501 & 7.466 & 7.530 & & \\ 
722.4479 & & \footnotesize4s         &\footnotesize$^4$D$_\frac{1}{2}$   &\footnotesize4p         &\footnotesize$^4$D$_\frac{1}{2}$   & 3.889 & $-$3.200 & 31 & 2.10 & 1\pst      & 7.480 & 7.464 & 7.461 & 7.425 & 7.490 & & \\ 
751.5831 & & \footnotesize4s         &\footnotesize$^4$D$_\frac{7}{2}$   &\footnotesize4p         &\footnotesize$^4$D$_\frac{5}{2}$   & 3.903 & $-$3.390 & 31 & 1.47 & 2\pst      & 7.455 & 7.445 & 7.444 & 7.411 & 7.472 & & \\ 
771.1721 & & \footnotesize4s         &\footnotesize$^4$D$_\frac{7}{2}$   &\footnotesize4p         &\footnotesize$^4$D$_\frac{7}{2}$   & 3.903 & $-$2.500 & 31 & 5.04 & 3\pst      & 7.500 & 7.431 & 7.417 & 7.378 & 7.448 & & \\ 
   & & & & & & & & & & & & & & & \\

\multicolumn{18}{c} {\coi} \\			   
521.2688 & &\footnotesize($^4$F)4sp  &\footnotesize$^4$F$_\frac{9}{2}$   &\footnotesize4s($^5$F)5s&\footnotesize$^4$F$_\frac{9}{2}$   & 3.514 & $-$0.110 & 32 & 1.91 & 3\pst      & 4.807 & 4.822 & 4.873 & 4.785 & 4.851 & $+$0.072    & 4.879 \\
528.0627 & &\footnotesize($^4$F)4sp  &\footnotesize$^4$G$_\frac{9}{2}$   &\footnotesize4s($^5$F)5s&\footnotesize$^4$F$_\frac{7}{2}$   & 3.629 & $-$0.030 & 32 & 1.78 & 2\pst      & 4.820 & 4.833 & 4.883 & 4.795 & 4.862 & $+$0.077    & 4.897 \\
530.1044 & &\footnotesize4s$^2$      &\footnotesize$^4$P$_\frac{5}{2}$   &\footnotesize($^3$F)4p  &\footnotesize$^4$D$_\frac{5}{2}$   & 1.710 & $-$1.940 & 33 & 1.79 & 1\pst      & 4.859 & 4.899 & 4.973 & 4.869 & 4.922 & $+$0.100    & 4.959 \\
535.2041 & &\footnotesize($^4$F)4sp  &\footnotesize$^4$G$_\frac{11}{2}$  &\footnotesize4s($^5$F)5s&\footnotesize$^4$F$_\frac{9}{2}$   & 3.576 & \ph0.060 & 32 & 2.38 & 2\pst      & 4.823 & 4.836 & 4.886 & 4.796 & 4.864 & $+$0.082    & 4.905 \\
548.3353 & &\footnotesize4s$^2$      &\footnotesize$^4$P$_\frac{5}{2}$   &\footnotesize($^3$F)4p  &\footnotesize$^4$D$_\frac{7}{2}$   & 1.710 & $-$1.410 & 33 & 4.64 & 3\pst      & 4.800 & 4.837 & 4.913 & 4.804 & 4.857 & $+$0.099    & 4.899 \\
564.7233 & &\footnotesize($^3$P)4s   &\footnotesize$^2$P$_\frac{3}{2}$   &\footnotesize($^3$F)4p  &\footnotesize$^2$D$_\frac{5}{2}$   & 2.280 & $-$1.560 & 32 & 1.24 & 2*         & 4.837 & 4.869 & 4.936 & 4.836 & 4.894 & $+$0.084    & 4.921 \\
593.5390 & &\footnotesize($^3$P)4s   &\footnotesize$^4$P$_\frac{5}{2}$   &\footnotesize($^3$F)4p  &\footnotesize$^4$D$_\frac{7}{2}$   & 1.883 & $-$2.610 & 33 & 0.32 & 1\pst      & 4.849 & 4.891 & 4.963 & 4.861 & 4.914 & $+$0.087    & 4.936 \\
608.2423 & &\footnotesize($^4$F)4sp  &\footnotesize$^4$F$_\frac{9}{2}$   &\footnotesize($^3$F)5s  &\footnotesize$^4$F$_\frac{9}{2}$   & 3.514 & $-$0.520 & 32 & 1.00 & 3\pst      & 4.857 & 4.873 & 4.924 & 4.836 & 4.901 & $+$0.072    & 4.929 \\
609.3141 & &\footnotesize4s$^2$      &\footnotesize$^4$P$_\frac{3}{2}$   &\footnotesize($^4$F)4sp &\footnotesize$^4$D$_\frac{3}{2}$   & 1.740 & $-$2.440 & 32 & 0.79 & 2*         & 4.935 & 4.978 & 5.050 & 4.950 & 4.998 & $+$0.085    & 5.020 \\
618.9005 & &\footnotesize4s$^2$      &\footnotesize$^4$P$_\frac{5}{2}$   &\footnotesize($^4$F)4sp &\footnotesize$^4$D$_\frac{5}{2}$   & 1.710 & $-$2.450 & 32 & 0.89 & 2*         & 4.956 & 4.998 & 5.071 & 4.971 & 5.018 & $+$0.085    & 5.041 \\
642.9913 & &\footnotesize4s$^2$      &\footnotesize$^2$G$_\frac{7}{2}$   &\footnotesize($^4$F)4sp &\footnotesize$^2$F$_\frac{5}{2}$   & 2.137 & $-$2.410 & 32 & 0.31 & 2*         & 4.854 & 4.892 & 4.960 & 4.861 & 4.914 & $+$0.087    & 4.941 \\
645.4995 & &\footnotesize($^4$F)4sp  &\footnotesize$^4$D$_\frac{7}{2}$   &\footnotesize($^3$F)5s  &\footnotesize$^4$F$_\frac{9}{2}$   & 3.632 & $-$0.250 & 32 & 1.34 & 2\pst      & 4.826 & 4.840 & 4.889 & 4.803 & 4.867 & $+$0.084    & 4.910 \\
741.7386 & &\footnotesize($^1$D)4s   &\footnotesize$^2$D$_\frac{3}{2}$   &\footnotesize($^4$F)4sp &\footnotesize$^4$D$_\frac{5}{2}$   & 2.042 & $-$2.070 & 32 & 1.00 & 2*         & 4.880 & 4.920 & 4.989 & 4.892 & 4.939 & $+$0.090    & 4.970 \\
   & & & & & & & & & & & & & & & \\             	                                                                             	                        
\multicolumn{18}{c} {\nii} \\
474.0166 & \footnote[3]{\protect\label{Niisonote}Isotopic splitting included (see Table \protect\ref{table:hfs}); wavelength corresponds to $^{58}$Ni component.}&\footnotesize($^3$F)4sp  &\footnotesize$^5$G$_4$             &\footnotesize($^2$D)4d  &\footnotesize$^3$G$_5$             & 3.480 & $-$1.720 & 34 & 1.60 & 1\pst      & 6.192 & 6.197 & 6.246 & 6.160 & 6.228 &             &   \\
481.1977 & \color{BrickRed}{$^c$}	  	          	
           &\footnotesize($^2$D)4p   &\footnotesize$^3$P$_1$             &\footnotesize($^2$D)4d  &\footnotesize$^3$P$_0$             & 3.658 & $-$1.450 & 35 & 2.13 & 1\pst      & 6.238 & 6.241 & 6.287 & 6.201 & 6.271 &             &   \\   
481.4598 & &\footnotesize($^3$F)4sp  &\footnotesize$^5$G$_2$             &\footnotesize4s($^4$F)5s&\footnotesize$^5$F$_3$             & 3.597 & $-$1.630 & 34 & 1.59 & 1\pst      & 6.216 & 6.218 & 6.265 & 6.179 & 6.248 &             &   \\   
487.4793 & &\footnotesize($^3$F)4sp  &\footnotesize$^5$G$_3$             &\footnotesize4s($^4$F)5s&\footnotesize$^5$F$_4$             & 3.543 & $-$1.440 & 34 & 2.35 & 1\pst      & 6.177 & 6.178 & 6.225 & 6.137 & 6.207 &             &   \\   
488.6711 & &\footnotesize($^2$D)4p   &\footnotesize$^3$D$_2$             &\footnotesize4s($^4$F)5s&\footnotesize$^5$F$_2$             & 3.706 & $-$1.810 & 34 & 0.90 & 1\pst      & 6.201 & 6.207 & 6.253 & 6.170 & 6.238 &             &   \\   
490.0971 & &\footnotesize($^3$F)4sp  &\footnotesize$^5$G$_4$             &\footnotesize4s($^4$F)5s&\footnotesize$^5$F$_5$             & 3.480 & $-$1.660 & 34 & 1.79 & 1\pst      & 6.195 & 6.196 & 6.244 & 6.156 & 6.226 &             &   \\   
497.6135 & \color{BrickRed}{$^c$}	  	          	
           &\footnotesize($^3$F)4sp  &\footnotesize$^5$F$_4$             &\footnotesize($^2$D)4d  &\footnotesize$^3$G$_4$             & 3.606 & $-$1.260 & 34 & 2.86 & 2\pst      & 6.179 & 6.176 & 6.222 & 6.132 & 6.204 &             &   \\   
515.7981 & &\footnotesize($^3$F)4sp  &\footnotesize$^5$F$_4$             &\footnotesize4s($^4$F)5s&\footnotesize$^5$F$_5$             & 3.606 & $-$1.510 & 34 & 1.86 & 3\pst      & 6.169 & 6.169 & 6.215 & 6.128 & 6.198 &             &   \\   
550.4095 & &\footnotesize($^3$F)4sp  &\footnotesize$^3$G$_5$             &\footnotesize4s($^4$F)5s&\footnotesize$^5$F$_4$             & 3.834 & $-$1.690 & 34 & 0.97 & 1\pst      & 6.207 & 6.211 & 6.254 & 6.170 & 6.240 &             &   \\   
551.0009 & \color{BrickRed}{$^c$}	  	          	
           &\footnotesize($^2$D)4p   &\footnotesize$^1$F$_3$             &\footnotesize($^2$D)4d  &\footnotesize$^3$G$_4$             & 3.847 & $-$0.880 & 34 & 3.75 & 2\pst      & 6.189 & 6.176 & 6.219 & 6.125 & 6.201 &             &   \\   
553.7105 & &\footnotesize($^2$D)4p   &\footnotesize$^1$F$_3$             &\footnotesize4s($^4$F)5s&\footnotesize$^5$F$_4$             & 3.847 & $-$2.220 & 34 & 0.31 & 3\pst      & 6.213 & 6.220 & 6.263 & 6.182 & 6.250 &             &   \\   
574.9280 & \color{BrickRed}{$^c$}	  	          	
           &\footnotesize($^3$F)4sp  &\footnotesize$^3$G$_3$             &\footnotesize($^2$D)4d  &\footnotesize$^3$G$_4$             & 3.941 & $-$1.920 & 34 & 0.44 & 2\pst      & 6.145 & 6.152 & 6.193 & 6.113 & 6.181 &             &   \\   
617.6820 & \color{BrickRed}{$^c$}					    		  	          	
           &\footnotesize($^3$F)4sp  &\footnotesize$^3$F$_4$             &\footnotesize($^2$D)4d  &\footnotesize$^3$G$_5$             & 4.088 & $-$0.260 & 34 & 6.64 & 2\pst      & 6.225 & 6.193 & 6.234 & 6.130 & 6.208 &             &   \\   
620.4605 & &\footnotesize($^3$F)4sp  &\footnotesize$^3$F$_4$             &\footnotesize4s($^4$F)5s&\footnotesize$^5$F$_4$             & 4.088 & $-$1.080 & 34 & 2.11 & 3\pst      & 6.211 & 6.206 & 6.245 & 6.161 & 6.232 &             &   \\   
622.3991 & \color{BrickRed}{$^c$}	  	          	
           &\footnotesize($^3$F)4sp  &\footnotesize$^3$F$_3$             &\footnotesize($^2$D)4d  &\footnotesize$^3$G$_4$             & 4.105 & $-$0.910 & 34 & 2.79 & 3\pst      & 6.197 & 6.194 & 6.233 & 6.148 & 6.220 &             &   \\   
637.8258 & \color{BrickRed}{$^c$}								   	  	          	
           &\footnotesize($^3$F)4sp  &\footnotesize$^3$D$_3$             &\footnotesize($^2$D)4d  &\footnotesize$^3$G$_4$             & 4.154 & $-$0.820 & 34 & 3.20 & 3\pst      & 6.225 & 6.221 & 6.259 & 6.173 & 6.245 &             &   \\   
641.4588 & &\footnotesize($^3$F)4sp  &\footnotesize$^3$D$_3$             &\footnotesize4s($^4$F)5s&\footnotesize$^5$F$_4$             & 4.154 & $-$1.160 & 34 & 1.68 & 2\pst      & 6.215 & 6.213 & 6.251 & 6.169 & 6.239 &             &   \\

\end{mpsupertabular}
\end{center}
\vspace{3mm}
\textbf{References:}\nopagebreak\\
\begin{minipage}[t]{0.43\linewidth}
\vspace{0pt}
\begin{enumerate}
\item\citet{Lawler89}
\item\citet{Lawler13}
\item\citet{Nitz98}
\item\citet{Blackwell1}, as corrected by \citet{Grevesse89}
\item\citet{Blackwell3}, as corrected by \citet{Grevesse89}
\item\citet{Blackwell2}, as corrected by \citet{Grevesse89}
\item\citet{Wood13}
\item\citet{Whaling85}, with \mbox{572.77\,nm} corrected for arithmetic error in converting from BFs to $A$ values as per \citet{Martin88}
\item\citet{Biemont89}
\item\citet{Karamatskos86}
\item mean of \citet{Sobeck07} and \citet{Blackwell84}
\item mean of \citet{Sobeck07} and \citet{Tozzi85}
\item mean of \citet{Sobeck07}, \citet{Tozzi85} and \citet{Blackwell86}
\item\citet{Sobeck07}
\item\citet{Blackwell86}
\item mean of \citet{Sobeck07} and \citet{Blackwell86}
\item mean of \citet{Sobeck07}, \citet{Tozzi85} and \citet{Blackwell84}
\end{enumerate}
\end{minipage}%
\begin{minipage}[t]{0.05\linewidth}
\vspace{0pt}
\hspace{3mm}
\end{minipage}%
\begin{minipage}[t]{0.43\linewidth}
\vspace{0pt}
\begin{enumerate}
\setcounter{enumi}{17}
\item\citet{Kuruczweb}
\item\citet{DenHartog11}
\item\citet{Booth84c}, renormalised to the absolute scale of \citet{BW07} (lines with excitation $\mathrm{potential}\approx3$\,eV; see Sect.~\protect\ref{Mngfs})
\item\citet{BW07}
\item derived from BFs of \citet{Greenlee79} and lifetimes of \citet{Schnabel95}
\item\citet{Booth84c}
\item mean of Oxford data \citep{Oxford1,Oxford2,Oxford3,Oxford4,Oxford5,Oxford6} and \citet{OBrian91}
\item Hannover data \citep{Bard91, Bard94}
\item mean of Oxford (see Ref.~24) and Hannover data (see Ref.~25)
\item mean of Hannover data (see Ref.~25) and \citet{OBrian91}, with double weight to Hannover
\item mean of Oxford data (see Ref.~24) and \citet{OBrian91}, with double weight to Oxford
\item Oxford data (see Ref.~24)
\item mean of Hannover data (see Ref.~25) and \citet{OBrian91}
\item\citet{Melendez09}
\item\citet{Cardon82}
\item\citet{Nitz99}
\item\citet{Wood14}
\item\citet{Johansson03}
\end{enumerate}
\end{minipage}
\vspace{6mm}

\newpage
\begin{center}
\topcaption{\label{table:hfs} HFS and isotopic splitting data for the lines retained in this analysis.}
\tablefirsthead{%
  \hline
   & & \multicolumn{4}{c}{Lower level} && \multicolumn{4}{c}{Upper level} \\
  \cline{3-6}
  \cline{8-11}
  $\lambda$ & Iso. & $J$ & $A$   & $B$   & HFS  && $J$ & $A$   & $B$   & HFS \\
  (nm)      &      &     & (MHz) & (MHz) & ref. &&     & (MHz) & (MHz) & ref. \\
  \hline
   & & & & & & & & & \\
  }
\tablehead{%
  \hline 
   & & \multicolumn{4}{c}{Lower level} && \multicolumn{4}{c}{Upper level} \\
  \cline{3-6}
  \cline{8-11}
  $\lambda$ & Iso. & $J$ & $A$   & $B$   & HFS  && $J$ & $A$   & $B$   & HFS \\
  (nm)      &      &     & (MHz) & (MHz) & ref. &&     & (MHz) & (MHz) & ref. \\
  \hline
  \multicolumn{11}{r}{continued.}\\
  \hline
   & & & & & & & & & \\
  }
\tabletail{%
  \hline
  \multicolumn{11}{r}{continued on next page}\\
  \hline
}
\tablelasttail{\hline}
\begin{mpsupertabular}{r c@{\hspace{10mm}}c r r ccc r r c}
\multicolumn{11}{c} {\scani: 100\% $^{45}$Sc ($I=\frac72$)} \vspace{2mm}\\
474.3821 & $^{45}$Sc & $9/2$ &     285.967  &  $-$15.460    &  1 && $7/2$  &          &               &    \\ 
508.1561 & $^{45}$Sc & $9/2$ &     285.967  &  $-$15.460    &  1 && $9/2$  &          &               &    \\ 
535.6097 & $^{45}$Sc & $7/2$ &   $-$25.000  &               &  2 && $5/2$  &          &               &    \\ 
567.1828 & $^{45}$Sc & $9/2$ &     285.967  &  $-$15.460    &  1 && $11/2$ &   55.000 &     25.000    &  3 \\       
623.9800 & $^{45}$Sc & $3/2$ &     269.556  &  $-$26.346    &  4 && $3/2$  &  348.320 &               &  5 \\  
   & & & & & & \hspace{6mm} & & & & \\

\multicolumn{11}{c} {\scanii} \vspace{2mm}\\ 			    		  			    		 
442.0661 & $^{45}$Sc &   4   &      38.357  &  $-$16.456    &  6 && 3      &  205.400 &  $-$70.000    &  7 \\            
443.1362 & $^{45}$Sc &   3   &     113.674  &  $-$12.615    &  6 && 2      &  366.800 &  $-$40.000    &  7 \\            
535.7202 & $^{45}$Sc &   2   &   $-$27.732  &     22.127    &  8 && 1      &          &               &    \\         
564.1000 & $^{45}$Sc &   1   &  $-$107.501  &  $-$12.300    &  8 && 2      &  106.117 &  $-$20.200    &  8 \\            
565.8362 & $^{45}$Sc &   0   &       0.000  &      0.000    &  N && 1      &  255.155 &     11.753    &  8 \\            
566.7164 & $^{45}$Sc &   1   &  $-$107.501  &  $-$12.300    &  8 && 1      &  255.155 &     11.753    &  8 \\            
566.9055 & $^{45}$Sc &   1   &  $-$107.501  &  $-$12.300    &  8 && 0      &    0.000 &      0.000    &  N \\            
568.4214 & $^{45}$Sc &   2   &   $-$27.732  &     22.127    &  8 && 1      &  255.155 &     11.753    &  8 \\            
624.5641 & $^{45}$Sc &   2   &   $-$27.732  &     22.127    &  8 && 3      &   99.730 &     21.495    &  8 \\            
630.0746 & $^{45}$Sc &   2   &   $-$27.732  &     22.127    &  8 && 2      &  125.423 &      8.769    &  8 \\            
632.0843 & $^{45}$Sc &   1   &  $-$107.501  &  $-$12.300    &  8 && 1      &  304.788 &      3.824    &  8 \\            
660.4578 & $^{45}$Sc &   2   &     149.361  &      7.818    &  6 && 2      &  215.700 &     18.000    &  9 \\           
   & & & & & & \hspace{6mm} & & & & \\

\multicolumn{11}{c} {\tii: 8.25\% $^{46}$Ti ($I=0$), 7.44\% $^{47}$Ti ($I=\frac{5}{2}$), 73.72\% $^{48}$Ti ($I=0$), 5.41\% $^{49}$Ti ($I=\frac{7}{2}$), 5.18\% $^{50}$Ti ($I=0$)}\\
\multicolumn{11}{c} {Isotopic separations from \citet{Gangrsky95}} \vspace{2mm}\\           	   	    					   	   
586.6429 & $^{50}$Ti & 2     &      0.000   &     0.000     &  N && 3      &     0.000&      0.000    &  N \\
586.6439 & $^{49}$Ti & 2     &  $-$25.216   & $-$39.202     & 10 && 3      &          &               &    \\
586.6448 & $^{48}$Ti & 2     &      0.000   &     0.000     &  N && 3      &     0.000&      0.000    &  N \\
586.6458 & $^{47}$Ti & 2     &  $-$25.216   & $-$47.826     & 10 && 3      &          &               &    \\ 
586.6468 & $^{46}$Ti & 2     &      0.000   &     0.000     &  N && 3      &     0.000&      0.000    &  N \\
592.2088 & $^{50}$Ti & 0     &      0.000   &     0.000     &  N && 1      &     0.000&      0.000    &  N \\
592.2097 & $^{49}$Ti & 0     &      0.000   &     0.000     &  N && 1      &$-$140.600&      0.000    & 11 \\
592.2107 & $^{48}$Ti & 0     &      0.000   &     0.000     &  N && 1      &     0.000&      0.000    &  N \\
592.2117 & $^{47}$Ti & 0     &      0.000   &     0.000     &  N && 1      &$-$140.700&      0.000    & 11 \\
592.2128 & $^{46}$Ti & 0     &      0.000   &     0.000     &  N && 1      &     0.000&      0.000    &  N \\
   & & & & & & \hspace{6mm} & & & & \\

\multicolumn{11}{c} {\tiii:  Isotopic separations from \citet{Nouri10}} \vspace{2mm}\\	
4444.524 & $^{50}$Ti & $7/2$ &      0.000   &     0.000     &  N && $7/2$  &     0.000&      0.000    &  N \\
4444.530 & $^{49}$Ti & $7/2$ &  $-$54.374   &    26.422     & 12 && $7/2$  & $-$31.500&  $-$14.000    & 12 \\
4444.536 & $^{48}$Ti & $7/2$ &      0.000   &     0.000     &  N && $7/2$  &     0.000&      0.000    &  N \\
4444.542 & $^{47}$Ti & $7/2$ &  $-$54.374   &    32.235     & 12 && $7/2$  & $-$31.500&  $-$17.080    & 12 \\
4444.547 & $^{46}$Ti & $7/2$ &      0.000   &     0.000     &  N && $7/2$  &     0.000&      0.000    &  N \\
4493.520 & $^{46}$Ti & $3/2$ &      0.000   &     0.000     &  N && $5/2$  &     0.000&      0.000    &  N \\
4493.521 & $^{47}$Ti & $3/2$ &     97.013   & $-$19.453     & 12 && $5/2$  &          &               &    \\
4493.523 & $^{48}$Ti & $3/2$ &      0.000   &     0.000     &  N && $5/2$  &     0.000&      0.000    &  N \\
4493.524 & $^{49}$Ti & $3/2$ &     97.013   & $-$23.733     & 12 && $5/2$  &          &               &    \\
4493.525 & $^{50}$Ti & $3/2$ &      0.000   &     0.000     &  N && $5/2$  &     0.000&      0.000    &  N \\
4583.396 & $^{50}$Ti & $3/2$ &      0.000   &     0.000     &  N && $5/2$  &     0.000&      0.000    &  N \\
4583.403 & $^{49}$Ti & $3/2$ &   $-$6.630   & $-$24.100     & 13 && $5/2$  & $-$84.210&  $-$44.000    & 12 \\
4583.409 & $^{48}$Ti & $3/2$ &      0.000   &     0.000     &  N && $5/2$  &     0.000&      0.000    &  N \\
4583.415 & $^{47}$Ti & $3/2$ &   $-$6.630   & $-$29.402     & 13 && $5/2$  & $-$84.210&  $-$53.680    & 12 \\
4583.421 & $^{46}$Ti & $3/2$ &      0.000   &     0.000     &  N && $5/2$  &     0.000&      0.000    &  N \\
4609.253 & $^{50}$Ti & $5/2$ &      0.000   &     0.000     &  N && $5/2$  &     0.000&      0.000    &  N \\
4609.259 & $^{49}$Ti & $5/2$ &     11.520   &    38.400     & 13 && $5/2$  & $-$84.210&  $-$44.000    & 12 \\
4609.265 & $^{48}$Ti & $5/2$ &      0.000   &     0.000     &  N && $5/2$  &     0.000&      0.000    &  N \\
4609.271 & $^{47}$Ti & $5/2$ &     11.520   &    46.848     & 13 && $5/2$  & $-$84.210&  $-$53.680    & 12 \\
4609.277 & $^{46}$Ti & $5/2$ &      0.000   &     0.000     &  N && $5/2$  &     0.000&      0.000    &  N \\
4708.656 & $^{50}$Ti & $3/2$ &      0.000   &     0.000     &  N && $5/2$  &     0.000&      0.000    &  N \\
4708.659 & $^{49}$Ti & $3/2$ &     53.334   & $-$23.471     & 12 && $5/2$  & $-$84.210&  $-$44.000    & 12 \\
4708.662 & $^{48}$Ti & $3/2$ &      0.000   &     0.000     &  N && $5/2$  &     0.000&      0.000    &  N \\
4708.665 & $^{47}$Ti & $3/2$ &     53.334   & $-$28.635     & 12 && $5/2$  & $-$84.210&  $-$53.680    & 12 \\
4708.668 & $^{46}$Ti & $3/2$ &      0.000   &     0.000     &  N && $5/2$  &     0.000&      0.000    &  N \\
4764.518 & $^{50}$Ti & $3/2$ &      0.000   &     0.000     &  N && $5/2$  &     0.000&      0.000    &  N \\
4764.521 & $^{49}$Ti & $3/2$ &     53.334   & $-$23.471     & 12 && $5/2$  &          &               &    \\
4764.524 & $^{48}$Ti & $3/2$ &      0.000   &     0.000     &  N && $5/2$  &     0.000&      0.000    &  N \\
4764.527 & $^{47}$Ti & $3/2$ &     53.334   & $-$28.635     & 12 && $5/2$  &          &               &    \\
4764.530 & $^{46}$Ti & $3/2$ &      0.000   &     0.000     &  N && $5/2$  &     0.000&      0.000    &  N \\
4798.529 & $^{46}$Ti & $3/2$ &      0.000   &     0.000     &  N && $5/2$  &     0.000&      0.000    &  N \\
4798.530 & $^{47}$Ti & $3/2$ &     97.013   & $-$19.453     & 12 && $5/2$  &          &               &    \\
4798.532 & $^{48}$Ti & $3/2$ &      0.000   &     0.000     &  N && $5/2$  &     0.000&      0.000    &  N \\
4798.533 & $^{49}$Ti & $3/2$ &     97.013   & $-$23.733     & 12 && $5/2$  &          &               &    \\
4798.535 & $^{50}$Ti & $3/2$ &      0.000   &     0.000     &  N && $5/2$  &     0.000&      0.000    &  N \\
4865.597 & $^{50}$Ti & $7/2$ &      0.000   &     0.000     &  N && $5/2$  &     0.000&      0.000    &  N \\
4865.605 & $^{49}$Ti & $7/2$ &  $-$54.374   &    26.422     & 12 && $5/2$  &          &               &    \\
4865.611 & $^{48}$Ti & $7/2$ &      0.000   &     0.000     &  N && $5/2$  &     0.000&      0.000    &  N \\
4865.618 & $^{47}$Ti & $7/2$ &  $-$54.374   &    32.235     & 12 && $5/2$  &          &               &    \\
4865.625 & $^{46}$Ti & $7/2$ &      0.000   &     0.000     &  N && $5/2$  &     0.000&      0.000    &  N \\  
5336.770 & $^{50}$Ti & $5/2$ &      0.000   &     0.000     &  N && $7/2$  &     0.000&      0.000    &  N \\
5336.774 & $^{49}$Ti & $5/2$ &              &               &    && $7/2$  & $-$31.500&  $-$14.000    & 12 \\
5336.778 & $^{48}$Ti & $5/2$ &      0.000   &     0.000     &  N && $7/2$  &     0.000&      0.000    &  N \\
5336.782 & $^{47}$Ti & $5/2$ &              &               &    && $7/2$  & $-$31.500&  $-$17.080    & 12 \\
5336.786 & $^{46}$Ti & $5/2$ &      0.000   &     0.000     &  N && $7/2$  &     0.000&      0.000    &  N \\       
5381.013 & $^{50}$Ti & $3/2$ &      0.000   &     0.000     &  N && $5/2$  &     0.000&      0.000    &  N \\
5381.017 & $^{49}$Ti & $3/2$ &              &               &    && $5/2$  & $-$84.210&  $-$44.000    & 12 \\
5381.021 & $^{48}$Ti & $3/2$ &      0.000   &     0.000     &  N && $5/2$  &     0.000&      0.000    &  N \\
5381.025 & $^{47}$Ti & $3/2$ &              &               &    && $5/2$  & $-$84.210&  $-$53.680    & 12 \\
5381.029 & $^{46}$Ti & $3/2$ &      0.000   &     0.000     &  N && $5/2$  &     0.000&      0.000    &  N \\       
5418.760 & $^{50}$Ti & $5/2$ &      0.000   &     0.000     &  N && $5/2$  &     0.000&      0.000    &  N \\
5418.764 & $^{49}$Ti & $5/2$ &              &               &    && $5/2$  & $-$84.210&  $-$44.000    & 12 \\
5418.768 & $^{48}$Ti & $5/2$ &      0.000   &     0.000     &  N && $5/2$  &     0.000&      0.000    &  N \\
5418.771 & $^{47}$Ti & $5/2$ &              &               &    && $5/2$  & $-$84.210&  $-$53.680    & 12 \\
5418.775 & $^{46}$Ti & $5/2$ &      0.000   &     0.000     &  N && $5/2$  &     0.000&      0.000    &  N \\

   & & & & & & \hspace{6mm} & & & & \\

\multicolumn{11}{c} {\vi: 99.75\% $^{51}$V ($I=\frac72$)} \vspace{2mm}\\
458.6370 & $^{51}$V  & $7/2$ &     249.739  &      5.081    & 14 && $9/2$  &   408.197&               & 15 \\
459.4119 & $^{51}$V  & $9/2$ &     227.132  &      7.822    & 14 && $11/2$ &   448.669&               & 15 \\

463.5172 & $^{51}$V  & $9/2$ &     227.132  &      7.822    & 14 && $9/2$  &   408.197&               & 15 \\
482.7452 & $^{51}$V  & $7/2$ &     249.739  &      5.081    & 14 && $7/2$  &   606.150&               & 15 \\
487.5486 & $^{51}$V  & $7/2$ &     249.739  &      5.081    & 14 && $5/2$  &   611.067&               & 15 \\
488.1555 & $^{51}$V  & $9/2$ &     227.132  &      7.822    & 14 && $7/2$  &   606.150&               & 15 \\
562.6019 & $^{51}$V  & $1/2$ &    1276.000  &               & 16 && $1/2$  &  1100.238&               & 15 \\
564.6108 & $^{51}$V  & $3/2$ &       6.966  &  $-$10.854    & 16 && $1/2$  &  1100.238&               & 15 \\
565.7438 & $^{51}$V  & $5/2$ &  $-$143.432  &   $-$1.196    & 16 && $3/2$  &   141.202&               & 15 \\
566.8361 & $^{51}$V  & $7/2$ &  $-$160.219  &     10.229    & 16 && $5/2$  &    15.289&               & 15 \\
567.0847 & $^{51}$V  & $7/2$ &  $-$160.219  &     10.229    & 16 && $9/2$  &    94.644&               & 15 \\
570.3586 & $^{51}$V  & $3/2$ &       6.966  &  $-$10.854    & 16 && $5/2$  &   215.851&               & 17 \\
572.7046 & $^{51}$V  & $7/2$ &  $-$160.219  &     10.229    & 16 && $9/2$  &    89.038&               & 15 \\
572.7655 & $^{51}$V  & $3/2$ &       6.966  &  $-$10.854    & 16 && $3/2$  &   634.361&               & 15 \\
573.1249 & $^{51}$V  & $5/2$ &  $-$143.432  &   $-$1.196    & 16 && $7/2$  &   431.551&               & 15 \\
573.7065 & $^{51}$V  & $5/2$ &  $-$143.432  &   $-$1.196    & 16 && $5/2$  &   215.851&               & 17 \\
600.2294 & $^{51}$V  & $5/2$ &     112.835  &               & 18 && $7/2$  & $-$17.088&               & 15 \\
603.9728 & $^{51}$V  & $5/2$ &  $-$143.432  &   $-$1.196    & 16 && $5/2$  & $-$89.800&    8.000      & 16 \\
608.1441 & $^{51}$V  & $3/2$ &       6.966  &  $-$10.854    & 16 && $3/2$  &$-$286.400& $-$6.000      & 16 \\
609.0208 & $^{51}$V  & $7/2$ &  $-$160.219  &     10.229    & 16 && $5/2$  & $-$89.800&    8.000      & 16 \\
611.1650 & $^{51}$V  & $1/2$ &    1276.000  &               & 16 && $1/2$  &$-$795.200&               & 16 \\
611.9528 & $^{51}$V  & $5/2$ &  $-$143.432  &   $-$1.196    & 16 && $3/2$  &$-$286.400& $-$6.000      & 16 \\
613.5363 & $^{51}$V  & $3/2$ &       6.966  &  $-$10.854    & 16 && $1/2$  &$-$795.200&               & 16 \\
619.9191 & $^{51}$V  & $7/2$ &     382.367  &      2.268    & 14 && $9/2$  &   503.460&    3.300      & 19 \\
624.2828 & $^{51}$V  & $1/2$ &     751.478  &      3.337    & 14 && $3/2$  &   594.690& $-$4.400      & 19 \\
624.3110 & $^{51}$V  & $9/2$ &     406.851  &     14.324    & 14 && $9/2$  &   503.460&    3.300      & 19 \\
625.1823 & $^{51}$V  & $7/2$ &     382.367  &      2.268    & 14 && $7/2$  &   514.350& $-$1.200      & 19 \\
625.6903 & $^{51}$V  & $5/2$ &     373.518  &   $-$5.459    & 14 && $5/2$  &   537.440& $-$4.000      & 19 \\
627.4653 & $^{51}$V  & $3/2$ &     405.604  &   $-$8.107    & 14 && $1/2$  &   939.940&    0.000      & 19 \\
628.5160 & $^{51}$V  & $5/2$ &     373.518  &   $-$5.459    & 14 && $3/2$  &   594.690& $-$4.400      & 19 \\
629.2824 & $^{51}$V  & $7/2$ &     382.367  &      2.268    & 14 && $5/2$  &   537.440& $-$4.000      & 19 \\
653.1401 & $^{51}$V  & $5/2$ &     112.835  &               & 18 && $5/2$  & $-$89.800&    8.000      & 17 \\
   & & & & & & \hspace{6mm} & & & & \\

\multicolumn{11}{c} {\vii} \vspace{2mm}\\ 			    		  			    		 
371.8152 & $^{51}$V  & 3     &     250.910  &               & 20 && 4      &   178.223&               & 20 \\	             
376.0222 & $^{51}$V  & 4     &     171.400  &               & 20 && 3      &   301.130&               & 20 \\
386.6740 & $^{51}$V  & 1     &   $-$73.330  &               & 20 && 2      &          &               &    \\
395.1960 & $^{51}$V  & 2     &       0.000  &               & 20 && 3      &   160.220&               & 20 \\
399.7117 & $^{51}$V  & 2     &      50.000  &               & 21 && 3      &   200.000&               & 21 \\
403.6777 & $^{51}$V  & 2     &       0.000  &               & 20 && 2      &   239.500&               & 20 \\
   & & & & & & \hspace{6mm} & & & & \\

\multicolumn{11}{c} {\cri: 4.35\% $^{50}$Cr ($I=0$), 83.79\% $^{52}$Cr ($I=0$), 9.50\% $^{53}$Cr ($I=\frac32$), 2.37\% $^{54}$Cr ($I=0$)}\\
\multicolumn{11}{c} {Isotopic separations from \citet{Furmann05}} \vspace{2mm}\\
452.98384& $^{50}$Cr & 6     &       0.000  &      0.000    &  N && 5      &     0.000&    0.000      &  N \\
452.98396& $^{52}$Cr & 6     &       0.000  &      0.000    &  N && 5      &     0.000&    0.000      &  N \\
452.98404& $^{53}$Cr & 6     &  $−$112.000  &      8.300    & 22 && 5      &     0.000&    0.000      &  N \\
452.98412& $^{54}$Cr & 6     &       0.000  &      0.000    &  N && 5      &     0.000&    0.000      &  N \\
   & & & & & & \hspace{6mm} & & & & \\

\multicolumn{11}{c} {\mni: 100\% $^{55}$Mn ($I=\frac52$)} \vspace{2mm}\\
408.2945 & $^{55}$Mn & $3/2$ &     469.391  &  $-$65.091    & 23 && $5/2$  & $-$26.981&               & 24 \\
426.5928 & $^{55}$Mn & $3/2$ &      50.965  &               & 24 && $3/2$  &$-$293.797&               & 24 \\
445.3013 & $^{55}$Mn & $3/2$ &      50.965  &               & 24 && $1/2$  &  1067.261&               & 24 \\
445.7041 & $^{55}$Mn & $5/2$ &     467.410  &  $-$73.460    & 25 && $3/2$  &   683.527&  224.844      & 26 \\
447.0142 & $^{55}$Mn & $3/2$ &      50.965  &               & 24 && $3/2$  &   191.867&               & 24 \\
449.8897 & $^{55}$Mn & $3/2$ &      50.965  &               & 24 && $5/2$  &    92.936&               & 24 \\
450.2223 & $^{55}$Mn & $5/2$ &  $-$137.905  &               & 24 && $7/2$  &    44.969&               & 24 \\
467.1688 & $^{55}$Mn & $7/2$ &  $-$161.888  &               & 24 && $5/2$  &   284.803&               & 24 \\
470.9710 & $^{55}$Mn & $7/2$ &  $-$161.888  &               & 24 && $7/2$  &   170.882&               & 24 \\
473.9110 & $^{55}$Mn & $3/2$ &      50.965  &               & 24 && $3/2$  &   668.537&               & 24 \\
500.4891 & $^{55}$Mn & $5/2$ &  $-$137.905  &               & 24 && $7/2$  &   137.905&               & 27 \\
525.5330 & $^{55}$Mn & $11/2$&     405.265  &               & 28 && $9/2$  &   131.909&               & 24 \\
538.8538 & $^{55}$Mn & $5/2$ &      89.938  &               & 24 && $7/2$  &    44.969&               & 24 \\
542.0368 & $^{55}$Mn & $7/2$ &     458.930  &     21.701    & 23 && $5/2$  &$-$549.000&               & 23 \\
   & & & & & & \hspace{6mm} & & & & \\
	
\multicolumn{11}{c} {\coi: 100\% $^{59}$Co ($I=\frac72$)} \vspace{2mm}\\
521.2688 & $^{59}$Co & $9/2$ &     810.039  &  $-$59.958    & 29 && $9/2$  &  1076.855&  149.896      & 29 \\
528.0627 & $^{59}$Co & $9/2$ &     517.142  &    179.875    & 29 && $7/2$  &   846.914&   89.938      & 29 \\
530.1044 & $^{59}$Co & $5/2$ &     178.900  & $-$170.000    & 30 && $5/2$  &   464.678&               & 29 \\
535.2041 & $^{59}$Co & $11/2$&     771.966  &    209.855    & 29 && $9/2$  &  1076.855&  149.896      & 29 \\
548.3353 & $^{59}$Co & $5/2$ &     178.900  & $-$170.000    & 30 && $7/2$  &   478.169&  149.896      & 29 \\
564.7233 & $^{59}$Co & $3/2$ &     332.000  &    101.000    & 30 && $5/2$  &   491.660&    0.000      & 29 \\
593.5390 & $^{59}$Co & $5/2$ &    1124.800  &    144.000    & 30 && $7/2$  &   478.169&  149.896      & 29 \\
608.2423 & $^{59}$Co & $9/2$ &     810.039  &  $-$59.958    & 29 && $9/2$  &   401.722&               & 29 \\
609.3141 & $^{59}$Co & $3/2$ &     317.780  &    119.917    & 29 && $3/2$  &   702.400&$-$15.000      & 31 \\
618.9005 & $^{59}$Co & $5/2$ &     178.900  & $-$170.000    & 30 && $5/2$  &   696.118&   29.979      & 29 \\
642.9913 & $^{59}$Co & $7/2$ &     839.400  &  $-$97.000    & 30 && $5/2$  &  1046.276&               & 29 \\
645.4995 & $^{59}$Co & $7/2$ &     751.500  &     31.000    & 30 && $9/2$  &   401.722&               & 29 \\
741.7386 & $^{59}$Co & $3/2$ &     389.730  &               & 29 && $5/2$  &   696.118&   29.979      & 29 \\
   & & & & & & \hspace{6mm} & & & & \\

\multicolumn{11}{c} {\nii: 68.08\% $^{58}$Ni ($I=0$), 26.22\% $^{60}$Ni ($I=0$), 5.7\% $^{61,62,64}$Ni} \\
\multicolumn{11}{c} {Isotopic separations from \citeauthor{Johansson03} (\citeyear{Johansson03}; 481.2\,nm) and \citeauthor{Wood14} (\citeyear{Wood14}; all except 481.2\,nm)} \vspace{2mm}\\
474.0134 & $^{61,62,64}$Ni           & 4     &      0.000   &     0.000     &  N && 5      &     0.000&      0.000    &  N \\
474.0150 & $^{\phantom{61,62,}60}$Ni & 4     &      0.000   &     0.000     &  N && 5      &     0.000&      0.000    &  N \\
474.0166 & $^{\phantom{61,62,}58}$Ni & 4     &      0.000   &     0.000     &  N && 5      &     0.000&      0.000    &  N \\
481.1977 & $^{\phantom{61,62,}58}$Ni & 1     &      0.000   &     0.000     &  N && 0      &     0.000&      0.000    &  N \\
481.1993 & $^{\phantom{61,62,}60}$Ni & 1     &      0.000   &     0.000     &  N && 0      &     0.000&      0.000    &  N \\
481.1993 & $^{61,62,64}$Ni           & 1     &      0.000   &     0.000     &  N && 0      &     0.000&      0.000    &  N \\
497.6114 & $^{61,62,64}$Ni           & 4     &      0.000   &     0.000     &  N && 4      &     0.000&      0.000    &  N \\
497.6125 & $^{\phantom{61,62,}60}$Ni & 4     &      0.000   &     0.000     &  N && 4      &     0.000&      0.000    &  N \\
497.6135 & $^{\phantom{61,62,}58}$Ni & 4     &      0.000   &     0.000     &  N && 4      &     0.000&      0.000    &  N \\
550.9994 & $^{61,62,64}$Ni           & 3     &      0.000   &     0.000     &  N && 4      &     0.000&      0.000    &  N \\
551.0002 & $^{\phantom{61,62,}60}$Ni & 3     &      0.000   &     0.000     &  N && 4      &     0.000&      0.000    &  N \\
551.0009 & $^{\phantom{61,62,}58}$Ni & 3     &      0.000   &     0.000     &  N && 4      &     0.000&      0.000    &  N \\
574.9257 & $^{61,62,64}$Ni           & 3     &      0.000   &     0.000     &  N && 4      &     0.000&      0.000    &  N \\
574.9280 & $^{\phantom{61,62,}60}$Ni & 3     &      0.000   &     0.000     &  N && 4      &     0.000&      0.000    &  N \\
574.9304 & $^{\phantom{61,62,}58}$Ni & 3     &      0.000   &     0.000     &  N && 4      &     0.000&      0.000    &  N \\
617.6777 & $^{61,62,64}$Ni           & 4     &      0.000   &     0.000     &  N && 5      &     0.000&      0.000    &  N \\
617.6798 & $^{\phantom{61,62,}60}$Ni & 4     &      0.000   &     0.000     &  N && 5      &     0.000&      0.000    &  N \\
617.6820 & $^{\phantom{61,62,}58}$Ni & 4     &      0.000   &     0.000     &  N && 5      &     0.000&      0.000    &  N \\
622.3949 & $^{61,62,64}$Ni           & 3     &      0.000   &     0.000     &  N && 4      &     0.000&      0.000    &  N \\
622.3971 & $^{\phantom{61,62,}60}$Ni & 3     &      0.000   &     0.000     &  N && 4      &     0.000&      0.000    &  N \\
622.3991 & $^{\phantom{61,62,}58}$Ni & 3     &      0.000   &     0.000     &  N && 4      &     0.000&      0.000    &  N \\
637.8206 & $^{61,62,60}$Ni           & 3     &      0.000   &     0.000     &  N && 4      &     0.000&      0.000    &  N \\
637.8233 & $^{\phantom{61,62,}60}$Ni & 3     &      0.000   &     0.000     &  N && 4      &     0.000&      0.000    &  N \\
637.8258 & $^{\phantom{61,62,}58}$Ni & 3     &      0.000   &     0.000     &  N && 4      &     0.000&      0.000    &  N \\
   & & & & & & \hspace{6mm} & & & & \\

\end{mpsupertabular}
\end{center}
\vspace{3mm}
\textbf{References:}\nopagebreak\\
\begin{minipage}[t]{0.43\linewidth}
\vspace{0pt}
\begin{enumerate}
\item\citet[][ABMR]{Ertmer76}
\item\citet[theoretical calculations]{Basar04}
\item\citet{Singh91}
\item\citet{Childs71}
\item\citet{Aboussaid96}
\item\citet{Mansour89}
\item\citet{Young88}
\item statistically-weighted average of \citet{Mansour89} and \citet{Villemoes92}
\item\citet{Arnesen82}
\item\citet{Aydin90}
\item\citet{Gangrsky95}
\item\citet{Berrah92}
\item\citet{Nouri10}
\item\citet{Childs67b}
\item\citet{Palmeri95}
\item\citet{Childs79}
\item\citet{Lefebvre02}
\end{enumerate}
\end{minipage}%
\begin{minipage}[t]{0.05\linewidth}
\hspace{3mm}
\end{minipage}%
\begin{minipage}[t]{0.43\linewidth}
\begin{enumerate}
\setcounter{enumi}{17}
\item\citet{Johann81}
\item\citet{Cochrane98}
\item\citet{Armstrong11}
\item estimated from solar profiles by trial and error with a single snapshot of an earlier version of the 3D model \citep{AspI}
\item\citet{Jarosz07}
\item\citet{D79}
\item\citet{BW05c}
\item\citet{Handrich69}
\item\citet{Luc72}
\item\citet{Lefebvre03}
\item\citet{Johann81}
\item\citet{Pickering96a}
\item\citet{Guthlorien90}
\item unpublished work of R. Wenzel, reproduced in \citet{Guthlorien90}
\item[N.]no HFS because $J=0$ or $I=0$
\end{enumerate}
\end{minipage}

\begin{table}
\caption{Our adopted ionisation energies $E_{\rm ion}$ and partition functions $U(T)$ for relevant ionisation stages of the iron group elements.}
\label{table:partition} 
\centering
\begin{tabular}{l r r r r r}
\hline
\hline Species & $E_{\rm ion}$ (eV) &  \multicolumn{4}{c}{$U(T)$} \\
               &                    & 3000\,K & 5000\,K & 8000\,K & 12000\,K \\
\hline
  \scani&   6.562 &      9.65 &     11.95 &     21.54 &     49.08 \\
  \scanii& 12.800 &     17.80 &     22.74 &     29.45 &     37.65 \\
  \tii  &   6.828 &     20.88 &     29.61 &     54.83 &    114.24 \\
  \tiii &  13.580 &     44.03 &     55.45 &     72.24 &     95.29 \\
  \vi   &   6.746 &     34.67 &     47.50 &     79.09 &    152.62 \\
  \vii  &  14.660 &     31.77 &     43.36 &     64.24 &     98.87 \\
  \cri  &   6.766 &      7.61 &     10.26 &     20.20 &     52.33 \\
  \crii &  16.500 &      6.03 &      7.09 &     12.15 &     27.16 \\
  \mni  &   7.434 &      5.99 &      6.34 &      9.95 &     24.10 \\
  \fei  &   7.902 &     22.01 &     27.78 &     43.05 &     81.05 \\
  \feii &  16.190 &     34.20 &     43.21 &     56.33 &     78.50 \\
  \coi  &   7.881 &     24.44 &     33.43 &     48.16 &     76.50 \\
  \nii  &   7.640 &     26.34 &     30.76 &     36.35 &     48.84 \\
  \niii &  18.170 &      8.30 &     10.83 &     15.73 &     23.21 \\
\hline
\end{tabular}
\end{table}

\end{document}